\documentclass[prd,twocolumn,nofootinbib,superscriptaddress,amsmath,amssymb]{revtex4-1}
\usepackage{mathtools}
\usepackage{graphics}
\usepackage{graphicx}
\usepackage{dcolumn}
\usepackage{bm}
\usepackage{dsfont} 
\usepackage{amsmath,amssymb}
\usepackage{hyperref}
\usepackage{tabularx}
\usepackage{epstopdf}
\usepackage[normalem]{ulem}
\usepackage[usenames]{color}
\usepackage{multirow}
\usepackage{makecell}
\usepackage{diagbox}
\usepackage[abs]{overpic}
\usepackage{soul,xcolor}
\allowdisplaybreaks
\hypersetup{
    colorlinks=true,
    linkcolor=blue,
    filecolor=magenta,      
    urlcolor=blue,
    citecolor=blue
}
\newcommand{\A}{{\mbox{\tiny A}}}

\definecolor{red(ncs)}{rgb}{0.77, 0.01, 0.2}

\setstcolor{magenta}

\usepackage{array}
\newcolumntype{C}[1]{>{\centering\let\newline\\\arraybackslash\hspace{0pt}}m{#1}}

\newcolumntype{C}[1]{>{\centering\arraybackslash}m{#1}}

\begin{document}

\title{Future Prospects for Constraining Nuclear Matter Parameters with \\ Gravitational Waves}

\author{Zack Carson}
\affiliation{%
 Department of Physics, University of Virginia, Charlottesville, Virginia 22904, USA
}%

\author{Andrew W. Steiner}
\affiliation{%
 Department of Physics and Astronomy, University of Tennessee, Knoxville, TN 37996, USA
}%
\affiliation{%
 Physics Division, Oak Ridge National Laboratory, Oak Ridge, TN 37831, USA
}%

\author{Kent Yagi}
\affiliation{%
 Department of Physics, University of Virginia, Charlottesville, Virginia 22904, USA
}%

\date{\today}


\begin{abstract}
The gravitational wave emission from the merging binary neutron star system GW170817 arrived full of tidal information which can be used to probe the fundamental ultra-dense nuclear physics residing in these stars.
In previous work, we used two-dimensional correlations between nuclear matter parameters and tidal deformabilities of neutron stars applying specifically to GW170817 to derive constraints on the former.
Here, we extend this analysis by finding similar correlations for varying chirp masses, the dominant determining factor in the frequency evolution of the inspiral, such that one can apply the same method to future detections.
We estimate how accurately one can measure nuclear parameters with future gravitational wave interferometers and show how such measurements can be improved by combining multiple events.
We find that bounds on the nuclear parameters with future observations can improve from the current one with GW170817 only by $\sim 30\%$ due to the existence of systematic errors caused mainly by the remaining uncertainty in the equation of state near and
just above the nuclear saturation density.
We show that such systematic errors can be reduced by considering multidimensional correlations among nuclear parameters and tidal deformabilities with various neutron star masses.
\end{abstract}

\maketitle


\section{Introduction}\label{sec:intro}

Neutron stars (NSs) exist in one of the most extreme states of matter found in the universe.
However, the determination of the equation of state (EoS) of ultra-dense matter found exclusively in such compact objects remains to be one of the largest unsolved mysteries in both nuclear physics and astrophysics to date.
The nuclear matter EoS determines many important stellar properties, such as the mass, radius, and tidal deformability, and is vital to the further study of supranuclear matter.
Independent measurements of certain macroscopic NS observables determined by the EoS, such as the mass and radius, can be used to constrain the EoS, as was indeed done in Refs.~\cite{guver,ozel-baym-guver,steiner-lattimer-brown,Lattimer2014,Ozel:2016oaf} via x-ray observations of the mass-radius relationship.
However, such measurements potentially suffer from large systematic errors due to uncertainties in the astrophysical modeling of x-ray bursts.

Recent observations of gravitational waves (GWs) from a merging binary NS system (GW170817~\cite{TheLIGOScientific:2017qsa}) have been used to probe the interior nuclear structure via imprinted tidal effects~\cite{Abbott2018,Abbott:2018exr,Paschalidis2018,Burgio2018,Malik2018,Landry:2018prl}, which offers us a cleaner method of determining the nuclear matter EoS than electromagnetic wave observations of neutron stars.
As the NSs lose energy via GW emission, they inspiral towards each other and become increasingly tidally deformed in response to the companion stars' tidal field.
This deformation is characterized by the \emph{tidal deformability}~\cite{Flanagan2008} of the NS, and is strongly dependent on the underlying EoS.
Further, the mass-weighted combination of such tidal deformabilities from each star is the leading tidal parameter in the gravitational waveform, which has been constrained by the LIGO and VIRGO Collaboration to a 90\% credible bound of $70 \leq \tilde{\Lambda} \leq 720$~\cite{Abbott2018,De:2018uhw}.
Such observations have also been mapped to the NS radius in Refs.~\cite{Annala:2017llu,Abbott:2018exr,Lim:2018bkq,Bauswein:2017vtn,De:2018uhw,Most:2018hfd}.

While all currently-proposed EoSs to date utilize various different
approximations, one way to effectively study them is by measuring the
nuclear parameters which parameterize the EoSs using a model-independent formalism.
One such method for doing this\footnote{Piecewise polytropic constructions~\cite{Read2009,Lackey:2014fwa,Carney:2018sdv} and spectral EoSs~\cite{Lindblom:2010bb,Lindblom:2012zi,Lindblom:2013kra,Lindblom:2018rfr,Abbott:2018exr} similarly parameterize nuclear matter EoSs in a model-independent way. See also~\cite{Landry:2018prl,Kumar:2019xgp} for related works on piecewise unified EoSs.} is to Taylor expand the energy per nucleon of asymmetric nuclear matter about the saturation density of symmetric matter~\cite{Myers:1969zz,Vidana2009}.
The resulting coefficients are known as the ``nuclear matter parameters" and consist of the following: the slope of the symmetry energy $L_0$; the nuclear incompressibility $K_0$; the slope of the incompressibility $M_0$; the curvature of the symmetry energy $K_{\text{sym},0}$; and higher orders, each evaluated at the nuclear saturation density.

Previous important analyses by Alam \emph{et al.}~\cite{Alam2016} found approximately universal relations between the NS radius at a given mass and the nuclear parameters mentioned above (similar work can be found in Refs.~\cite{Sotani:2013dga,Silva:2016myw}).
Further, Malik \emph{et al.}~\cite{Malik2018} found that certain linear combinations of nuclear parameters (such as $K_0+\alpha L_0$ with $\alpha$ chosen to give maximal correlation) gave way to heightened correlations with the individual tidal deformabilities evaluated at a given mass.
By assuming individual masses for GW170817 and taking the approximate universal relations to be exact, the authors utilized prior constraints on the tidal deformability from GW170817~\cite{Abbott2017,Radice2018} and $L_0$~\cite{Abbott2018,Oertel2017,Lattimer2014} to derive new constraints on the nuclear parameters.

In Ref.~\cite{Zack:nuclearConstraints}, we expanded upon the previous work of Malik \emph{et al.}~\cite{Malik2018} and found improved constraints on $K_0$, $M_0$, and $K_{\text{sym},0}$.
We made various improvements upon the original analysis, such as (i) using an expanded set of EoSs in order to fully take into account the EoS-variation systematic errors; (ii) considering the correlations with the mass-weighted tidal deformability $\tilde\Lambda$ (rather than $\Lambda_{1.4}$, the tidal deformability at $1.4\text{ M}_\odot$) which was directly measured by GW observations; (iii) fully considering the EoS variation scatter uncertainty in the estimation of constraints; and (iv) adopting the full posterior distribution on $\tilde\Lambda$ as measured from GW170817.
Further, we found that by using our method of computing the posterior distribution on nuclear parameters, high degrees of correlation between $\tilde\Lambda$ and such nuclear parameters was not necessary, as any covariances between the two was taken into account by the multi-variate Gaussian probability distribution between them.
On the other hand, the authors of Ref.~\cite{Malik2018} absolutely required high correlations with nuclear parameters in order to assume that the relationship between the tidal deformability and the nuclear parameters lie exactly on the best-fit line between the two.
The resulting 90\% confidence interval on the curvature of symmetry energy was found to be $-259 \text{ MeV} \leq K_{\text{sym},0} \leq 32 \text{ MeV}$, which was more conservative than that derived in Ref.~\cite{Malik2018}.

Here, we extend the work of Ref.~\cite{Zack:nuclearConstraints} into the future of GW astronomy.
While every future merger event will be composed of NSs with varying individual masses which are difficult to measure, we can categorize them by the \emph{chirp mass} $\mathcal{M}$, which is the dominant driving factor in the frequency evolution of the inspiral event given by a certain combination of individual NS masses.
In this investigation, we repeat the analysis of Ref.~\cite{Zack:nuclearConstraints} as a function of chirp mass, applicable to any future event.
Further, we restrict the set of EoSs to those that obey the nuclear parameter correlations of Ref.~\cite{Tews2017}, and consider the implications of observations using future GW interferometers: Advanced LIGO (aLIGO)~\cite{aLIGO}, LIGO A\texttt{+} (A\texttt{+})~\cite{Ap_Voyager_CE}, Voyager~\cite{Ap_Voyager_CE}, Cosmic Explorer (CE)~\cite{Ap_Voyager_CE}, and Einstein Telescope (ET)~\cite{ET}.
We will consider not only the increased sensitivities from current detectors but also the combined uncertainties from multiple-event detections (relevant for future detectors with expanded horizon volumes). 


\subsection{Executive summary}

Here we summarize our results for busy readers.
In the current analysis, we attempt to find constraints on the nuclear matter parameters as a function of the binary systems' chirp mass.
We begin by finding the correlations between the mass-weighted tidal deformability $\tilde\Lambda$ and various nuclear parameters, and combinations thereof.
Similar to our results in Ref.~\cite{Zack:nuclearConstraints}, we find that the low-order nuclear parameters $K_0$ and $M_0$ observe small correlations with $\tilde\Lambda$, while the correlations for higher-order parameter $K_{\text{sym},0}$ remains high at $\sim 80\%$.
For this reason, we consider constraints on the curvature $K_{\text{sym},0}$ of the symmetry energy which is one of the most uncertain parts of the EoS of dense nucleonic matter~\cite{Li:2019xxz}.

In this analysis, we compute the posterior probability distribution for the curvature of symmetry energy $K_{\text{sym},0}$, for 22 different values of chirp mass between $0.94 \text{ M}_\odot$ and $1.6 \text{ M}_\odot$.
For each value of chirp mass $\mathcal{M}_i$ considered, we compute the single-event $\tilde\Lambda$ uncertainties using Fisher analysis techniques.
Approximated as a Gaussian prior, the uncertainty in $\tilde\Lambda$ may be used to estimate the posterior probability distribution on $K_{\text{sym},0}$, by multiplication with the one-dimensional conditional probability distribution on $K_{\text{sym},0}$ given an observation of $\tilde\Lambda$, and then integrating over all $\tilde\Lambda$ values.
The process is then repeated for each value of chirp mass $\mathcal{M}_i$, resulting in a relationship between the uncertainties in $K_{\text{sym},0}$ and the chirp mass $\mathcal{M}$.

\begin{figure}
\begin{center} 
\begin{overpic}[width=1\linewidth]{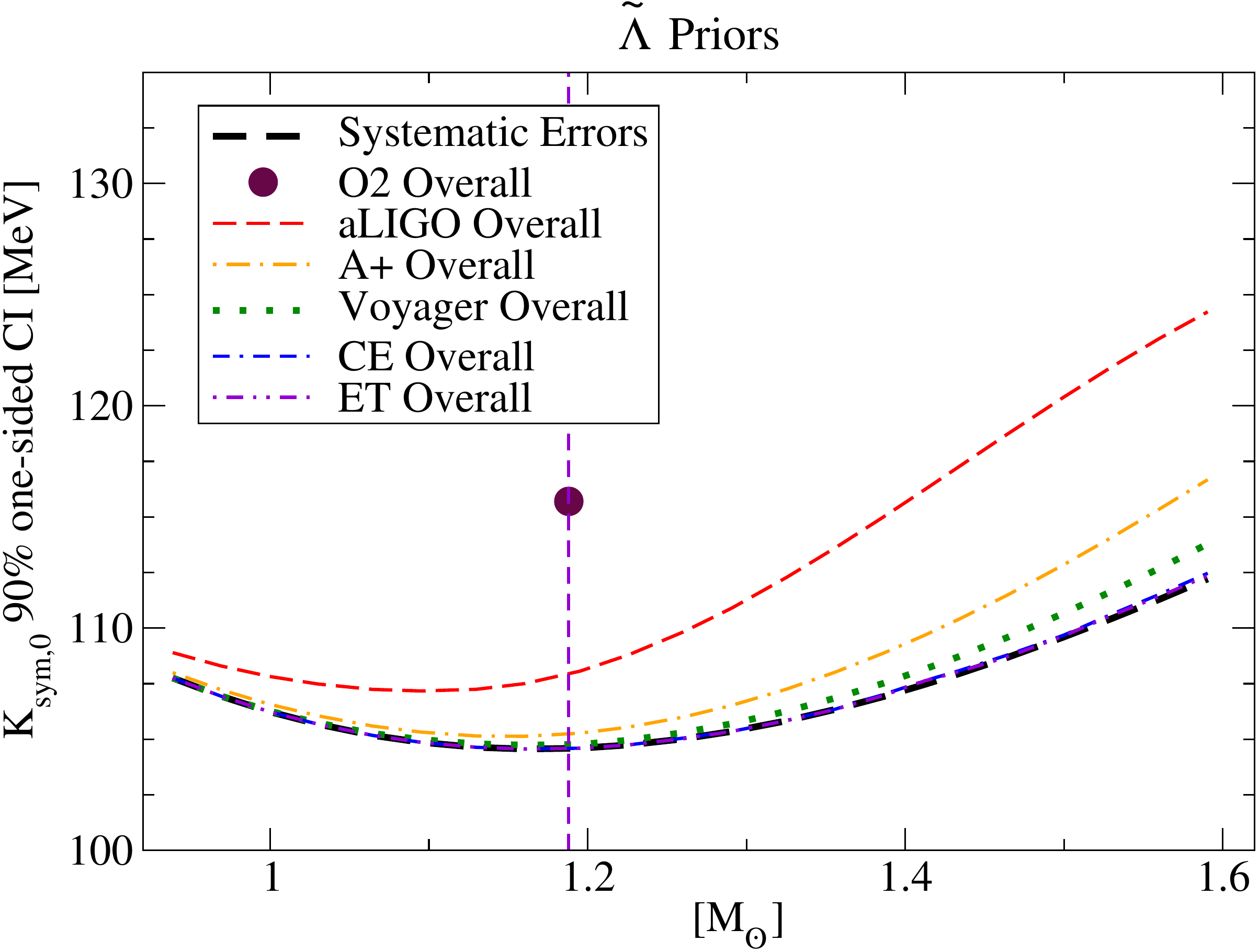}
\put(119,3.5){\small$\mathcal{M}$}
\end{overpic}
\end{center}
\caption{
The overall (statistical plus systematic) errors on $K_{\text{sym},0}$ using priors in $\tilde\Lambda$ for single-event measurements, plotted as a function of the binary systems' chirp mass -- applicable to any future binary NS merger. 
We fix the mass ratio as $q=0.9$ consistent with GW170817 and choose the distance and sky location of the binaries such that it gives the signal-to-noise ratio (SNR) of 32.4 for the O2 run, again corresponding to GW170817.
This is repeated for 6 different interferometers (O2, aLIGO, A\texttt{+}, Voyager, CE, ET).
Note that the error on O2 appears as a single point, corresponding to GW170817 - the single event observed on O2.
We also present systematic errors due to scattering in correlations between $K_{\text{sym},0}$ and $\tilde \Lambda$.
Observe how as one improves the detector sensitivity, the statistical errors become subdominant, and the overall errors approach the systematic uncertainties' ``wall".
This indicates the need to further reduce the EoS variation in the scattering that is the origin of systematic uncertainties before stronger constraints on $K_{\text{sym},0}$ can be derived. Additionally shown by the dashed vertical line is the chirp mass $\mathcal{M}=1.188\text{ M}_{\odot}$ corresponding to GW170817.
}
\label{fig:OverallVsSystematic}
\end{figure}

The corresponding one-sided 90\% confidence intervals on $K_{\text{sym},0}$ for single-event detections, along with the calculated systematic errors, are plotted in Fig.~\ref{fig:OverallVsSystematic} as a function of chirp mass for 6 different GW interferometers.
In this figure, we observe that as the detector sensitivity is increased, the statistical errors become subdominant rather quickly.
This in turn forces the overall errors to approach the
systematic error ``wall" at $\sim104$ MeV, caused partially by uncertainties in the EoS at low-density which are less sensitive to neutron star tidal deformabilities.
For this reason, the curves corresponding to the 3rd-generation detectors CE and ET become indistinguishable from the systematic errors -- indicating the necessity to reduce such errors in order for the further constraint of $K_{\text{sym},0}$ to become possible.

Following this, we offer a method to further decrease the statistical errors in the measurement of $K_{\text{sym},0}$.
This is accomplished by repeating the same analysis for the fixed chirp mass of $1.188\text{ M}_\odot$ with the first coefficient, $\Lambda_{1.4}$, of the Taylor expanded tidal deformability $\Lambda \approx \Lambda_{1.4}+\Lambda_{1.4}'(1-\frac{m}{m_0})$ about $m_0=1.4\text{ M}_{\odot}$\footnote{$m_0$ will remain fixed for the remainder of the analysis, with the exception of Sec.~\ref{sec:systematics}, where we consider the effect of variations in $m_0$.}, rather than the mass-weighted tidal deformability $\tilde\Lambda$.
$\Lambda_{1.4}$ (or the tidal deformability at $1.4\text{ M}_{\odot}$) is mass-independent, and thus, it is identical for all future GW events.
This way, we can combine the uncertainties for multiple detected events when it becomes applicable for future detectors.
As was observed in Fig.~\ref{fig:OverallVsSystematic}, we found that the uncertainties in $K_{\text{sym},0}$ became dominated by systematics for the single-event analyses on Voyager-era detectors and beyond.
By combining GW170817-like events detected on aLIGO and A\texttt{+}, we find that one can further reduce the statistical errors in $K_{\text{sym},0}$ such that the errors similarly become dominated by systematics.

\begin{figure}
\begin{center} 
\includegraphics[width=8.cm]{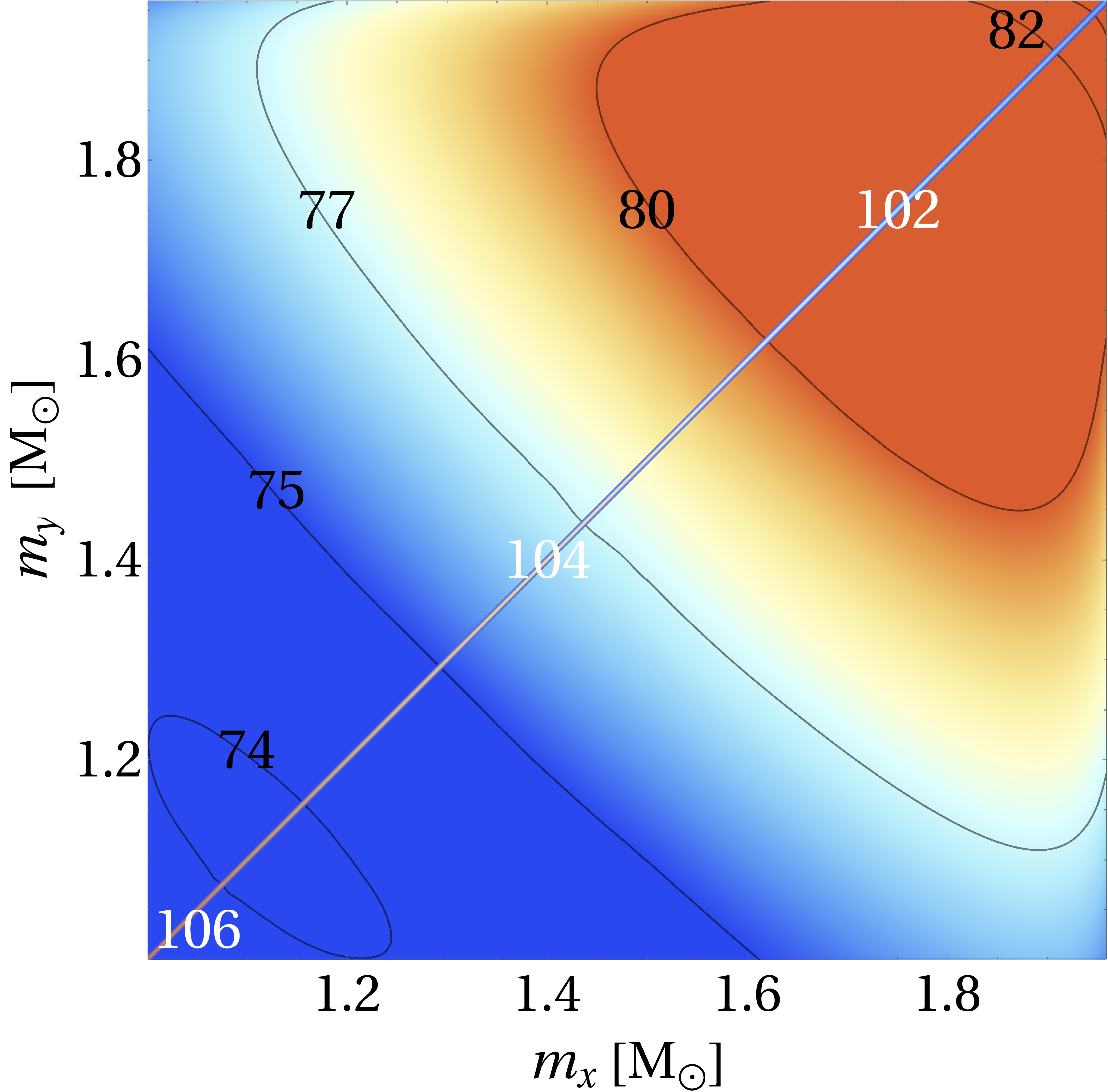}
\end{center}
\caption{
Contours displaying the systematic errors in $K_{\text{sym},0}$ [MeV] as a function of the masses $m_x$ and $m_y$ used to compute the four-dimensional probability distribution between $K_{\text{sym},0}$, $\Lambda_{1.5}$, $\Lambda_x$, and $\Lambda_y$.
The systematic errors are then computed by evaluating the probability distribution at the fiducial values of $\Lambda_{1.5}$, $\Lambda_x$, and $\Lambda_y$, and then taking the 90\% confidence interval of the resulting distribution in $K_{\text{sym},0}$.
Observe how a large reduction in systematic errors to $\sim74$ MeV can occur by including information about the tidal deformability at 3 different NS masses spread throughout their realistic range.
The diagonal contours at $m_x = m_y$ labeled in white correspond to the systematic errors obtained from the reduced three-dimensional probability distribution $P(K_{\text{sym},0},\Lambda_x,\Lambda_{1.5})$.
}
\label{fig:systematicContourFixly}
\end{figure}

Finally, we investigate the reduction of systematic errors by adding information about the tidal deformability at various different masses.
We begin by generating a four-dimensional Gaussian probability distribution $P(K_{\text{sym},0},\Lambda_{m_x},\Lambda_{m_y},\Lambda_{m_z})$.
The systematic error on $K_{\text{sym}}$ is obtained by first evaluating $P(K_{\text{sym},0},\Lambda_{m_x},\Lambda_{m_y},\Lambda_{m_z})$ at the fiducial values of $\Lambda_{m_x}$, $\Lambda_{m_y}$, and $\Lambda_{m_z}$ and estimating the 90\% confidence interval.
Figure~\ref{fig:systematicContourFixly} presents the systematic errors when $m_z$ is fixed to be $1.5\text{ M}_\odot$, and ($m_x,m_y$) are varied between $1.0\text{ M}_\odot$ and $2.0\text{ M}_\odot$.
Observe how the resulting systematic uncertainties are reduced to $\sim74$ MeV for certain combinations of $m_x$ and $m_y$.

The remainder of this paper is organized as follows.
We begin in Sec.~\ref{sec:theory} with a review on the tidal deformability of neutron stars, as well as the nuclear matter EoS and its constituent nuclear parameters.
We follow this up in Sec.~\ref{sec:correlations} with a study on the correlations between various nuclear parameters and the mass-weighted tidal deformability.
In Sec.~\ref{sec:chirpmass}, we analyze the measurement accuracy on such nuclear parameters as a function of the chirp mass, applicable to future binary NS mergers.
In Sec.~\ref{sec:systematics}, we discuss how one can further reduce systematic errors by considering correlations among nuclear parameters and multiple tidal deformabilities at different NS masses.
We conclude in Sec.~\ref{sec:conclusion} with a discussion on the implications of our results, as well as an outline into possible avenues of future work.
We have adopted geometric units such that $G = c = 1$ throughout.


\section{Background and theory}\label{sec:theory}

In this section we begin with a review on the NS tidal deformability in Sec.~\ref{sec:tidal}, followed up by a review on the NS equation of state and its constituent nuclear parameters in Secs.~\ref{sec:parameters} and~\ref{sec:eos}.


\subsection{Neutron star tidal deformability}\label{sec:tidal}

Here we offer a brief overview on how one can extract information on the internal structure of a NS by way of the GW observations of binary NS merger events, such as GW170817.
In the presence of a neighboring tidal field $\mathcal{E}_{ij}$ (such as a NS in a binary system with a compact companion), NSs will acquire a quadrupole moment $Q_{ij}$ characterized as the linear response to $\mathcal{E}_{ij}$:
\begin{equation}
Q_{ij}=-\lambda \mathcal{E}_{ij},
\end{equation}
with tidal deformability $\lambda$~\cite{Flanagan2008,hinderer-love,damour-nagar,Binnington:2009bb,Yagi2013}.
The tidal deformability characterizes the NSs corresponding deformation from sphericity, and can be made unitless by the following normalization:
\begin{equation}
\Lambda\equiv\frac{\lambda}{M^5}
\end{equation}
with stellar mass $M$.

Following Refs.~\cite{hinderer-love,damour-nagar,Yagi2013,Zack:nuclearConstraints}, the dimensionless tidal deformability $\Lambda$ can be computed by isolating different asymptotic limits of the gravitational potential in the buffer zone $R \ll r \ll \mathcal{L}$ given by
\begin{align}
\Phi (x^i)=&\frac{1+g_{tt}}{2} \nonumber \\
=&-\frac{M}{r} - \frac{3}{2}\frac{Q_{ij}}{r^3} \Bigg(\frac{x^i}{r} \frac{x^j}{r}-\frac{1}{3}\delta_{ij} \Bigg) + \mathcal{O} \Bigg( \frac{\mathcal{L}^4}{r^4} \Bigg) \nonumber \\
&+ \frac{1}{2} \mathcal{E}_{ij} x^i x^j + \mathcal{O} \Bigg( \frac{r^3}{R^3} \Bigg),
\end{align}
where $r = |x^i|$, $\mathcal{L}$ is the length scale of the companion-induced curvature, and $R$ is the stellar radius.
Here, $g_{tt}$ corresponds to the $tt$-component of the full spacetime metric:
\begin{equation}
g_{\alpha\beta}=g_{\alpha\beta}^{(0)}+h_{\alpha\beta},
\end{equation}
constructed via a non-spinning, spherically-symmetric background solution $g_{\alpha\beta}^{(0)}$ perturbed by the tidal deformation with metric components $h_{\alpha\beta}$.
The perturbed Einstein equations may then be solved in the interior of the NS and matched to the exterior solution at the surface of the star.
Further, the radius $R$ may then be determined from the condition $p(R)=0$.

In this investigation, we consider the scenario of two NSs orbiting each other in a binary system, like GW170817.
In this case, each NS individually obtains quadrupole moments from the neighboring tidal field, resulting in two highly-correlated tidal deformabilities $\Lambda_1$ and $\Lambda_2$.
Due to these correlations, individual tidal deformabilities are very difficult to extract from GW observations.
Typically, it is useful to reparameterize the waveform via independent linear combinations of $\Lambda_1$ and $\Lambda_2$ which enter the gravitational waveform at 5th post-Newtonian (PN) and 6PN orders\footnote{$n$PN order corrections enter the gravitational waveform at relative powers of $(v/c)^{2n}$.} respectively.
The dominant tidal effect in the resulting waveform is known as the \emph{mass-weighted tidal deformability}, and is given by~\cite{Flanagan2008}
\begin{equation}
\tilde{\Lambda} = \frac{16}{13} \frac{(1+12q) \Lambda_1+(12+q)q^4\Lambda_2}{(1+q)^5},
\end{equation}
with mass ratio $q \equiv m_2/m_1$ ($m_1 \geq m_2$).
Here we also define the \emph{chirp mass} of the binary system, which is the primary controlling factor of the merger inspiral defined by
\begin{equation}
\mathcal{M}\equiv\Bigg(\frac{q^3}{1+q}\Bigg)^{1/5}m_1.
\end{equation}
Similarly to $\tilde\Lambda$, this quantity can be measured with much higher accuracy than either of the individual masses $m_1$, $m_2$, or the mass ratio $q$.
For this reason, we consider the binary chirp mass $\mathcal{M}$ to be the dominant dependent variable in this analysis, cataloging our various results as a function of $\mathcal{M}$ for any future GW event.


\subsection{The nuclear matter parameters}\label{sec:parameters}

While the NS EoS is not currently known, there are many methods one can use to restrict it using various observations. 
This is because the structure of a NS and many of its observables such as mass, radius, tidal deformability, etc. rely strongly on the underlying EoS of nuclear matter.
For example, GW observations may help constrain the EoS in the pressure-density plane~\cite{LIGO:posterior}.
In this paper, we show how GW detections can aid in the constraint of various characteristics of the EoS, known as the nuclear matter parameters~\cite{Vidana2009}.

As originally considered in Ref.~\cite{Myers:1969zz} and followed up in Refs.~\cite{Vidana2009,Alam2016,Malik2018,Zack:nuclearConstraints}, we offer a generic method to parameterize NS EoSs.
This is done by first Taylor expanding the energy per nucleon $e$ of asymmetric nuclear matter about $\delta=0$ (symmetric nuclear matter), where $\delta \equiv (n_n-n_p)/n$ is the isospin symmetry parameter for nuclear matter with $n_n$ neutron density, $n_p$ proton density, and total density $n=n_n+n_p$:
\begin{equation}
e(n,\delta)=e(n,0)+S_2(n)\delta^2+\mathcal{O}(\delta^4).
\end{equation}
Here $e(n,0)$ and $S_2(n)$ are the symmetric and second-order asymmetric nuclear matter energy per nucleon, respectively.
Such energies are then further Taylor expanded about the \emph{nuclear saturation density} $n_0\approx2.3\times 10^{14}\text{ g/cm}^3$ as 
\begin{align}
\begin{split}
e(n,0)&=e_0+\frac{K_0}{2} y^2 + \frac{Q_0}{6}y^3 + \mathcal{O}(y^4),\\
S_2(n)&=J_0+L_0 y + \frac{K_\mathrm{sym,0}}{2} y^2 + \mathcal{O}(y^3),
\end{split}
\end{align}
with $y\equiv(n-n_0)/3n_0$.
The above coefficients, all evaluated at the nuclear saturation density, determine the NS EoS and are referred to in the literature~\cite{Myers:1969zz,Vidana2009} as the energy per particle $e_0$; incompressibility coefficient $K_0$; third derivative of symmetric matter $Q_0$; the slope of the incompressibility $M_0\equiv Q_0+12K_0$~\cite{Alam2016,Malik2018}; symmetry energy $J_0$; its slope $L_0$; and its curvature $K_\mathrm{sym,0}$.

In this paper, we expand upon previous works~\cite{Alam2016,Malik2018,Zack:nuclearConstraints} and investigate the correlations between the mass-weighted tidal deformability $\tilde\Lambda$ and various nuclear parameters $L_0$, $K_0$, $M_0$, and $K_{\text{sym},0}$ in order to derive constraints on such parameters.
Specifically, we focus on the curvature of the symmetry energy $K_{\text{sym},0}$, shown to be one of the most uncertain features of the nuclear matter EoS, especially at supranuclear densities found primarily in NSs~\cite{Li:2019xxz}.


\subsection{The supranuclear equation of state}\label{sec:eos}

We now explain which EoSs we use in this paper.
In Ref.~\cite{Zack:nuclearConstraints}, we showed the importance of considering a wide range of physically valid EoSs when computing constraints on nuclear parameters, in order to more properly take into account the systematic errors.
In the current analysis, we employ a restricted set of the same EoSs as was used previously, taking into account the observed correlations between nuclear parameters $J_0$ and $L_0$.
Starting with the 121 nuclear EoS models found in Ref.~\cite{Zack:nuclearConstraints}, we further remove 63 EoS models which do not comply with the allowed regions shown in Fig. 8 of Tews \emph{et al.}~\cite{Tews2017}.
Here, they combined an exclusion region $J_0(L_0)$ with the ``accepted" 95.4\% correlation confidence bands between $J_0$ and $L_0$. See App.~\ref{app:EoS-comparison} for the impact of restricted EoSs on correlations between $K_\mathrm{sym,0}$ and $\tilde \Lambda$.

Taking the shared region between the above two exclusions results in 58 different nuclear EoS models, which can be classified into 3 distinct classes: 13 non-relativistic ``Skyrme-type" EoSs, 5 relativistic-mean-field (RMF) EoSs, and 40 EoS models developed with a phenomenological variation method (PEs).
The Skyrme-type models used consist of: SKa, Sly230a~\cite{Chabanat1997}, Sly2, Sly9~\cite{Chabanat1995}, Sly4~\cite{Chabanat1998}, SkOp~\cite{Reinhard1999}, SK255, SK272~\cite{Agrawal2003}, BSK20, BSK21~\cite{Goriely2010}, BSK22, BSK24, BSK26~\cite{Goriely2013}.
Further, the RMF models used are: BSR2, BSR6~\cite{Dhiman2007,Agrawal2010}, NL3$\omega \rho$~\cite{Carriere2003}, DD2~\cite{Typel2010}, DDH$\delta$~\cite{Gaitanos2004}.
All 18 of the above EoSs originate from the minimal set of EoSs used in Refs.~\cite{Alam2016,Malik2018}, now restricted by nuclear matter correlations.
Lastly, following Ref.~\cite{Read2009}, the high-density core-region of the EoSs used in our analysis are matched to the low-density crust-region of the SLy EoS model~\cite{Douchin:2001sv} at about half of the nuclear saturation density, $\rho_{\text{stitch}} \approx 1.3 \times 10^{14} \text{ g/cm}^3$.

One last class of EoSs indirectly used in our analysis can be found in Ref.~\cite{Zack:URrelations}, which we call ``LVC constrained" EoSs in this paper.
By sampling the full physical EoS parameter space, the LIGO and Virgo Collaboration~\cite{LIGO:posterior,Carney:2018sdv} derived a marginalized 90\% posterior region on the NS pressure as a function of mass density (EoS) from GW170817, as seen in Fig. 2 of~\cite{LIGO:posterior}.
By randomly sampling the EoS posteriors from this analysis, a set of 100 ``constrained" EoSs were obtained, restricted by the GW observation of GW170817. 
While we do not directly utilize these 100 EoSs in the current analysis, we use them to estimate the mean value of the mass-weighted tidal deformability $\tilde\Lambda$ in Sec.~\ref{sec:futureSingle}, seen by Fig.~\ref{fig:meanLt}.


\section{Correlations between $\tilde\Lambda$ and nuclear parameters}\label{sec:correlations}

\begin{figure}
\begin{center} 
\begin{overpic}[width=\linewidth]{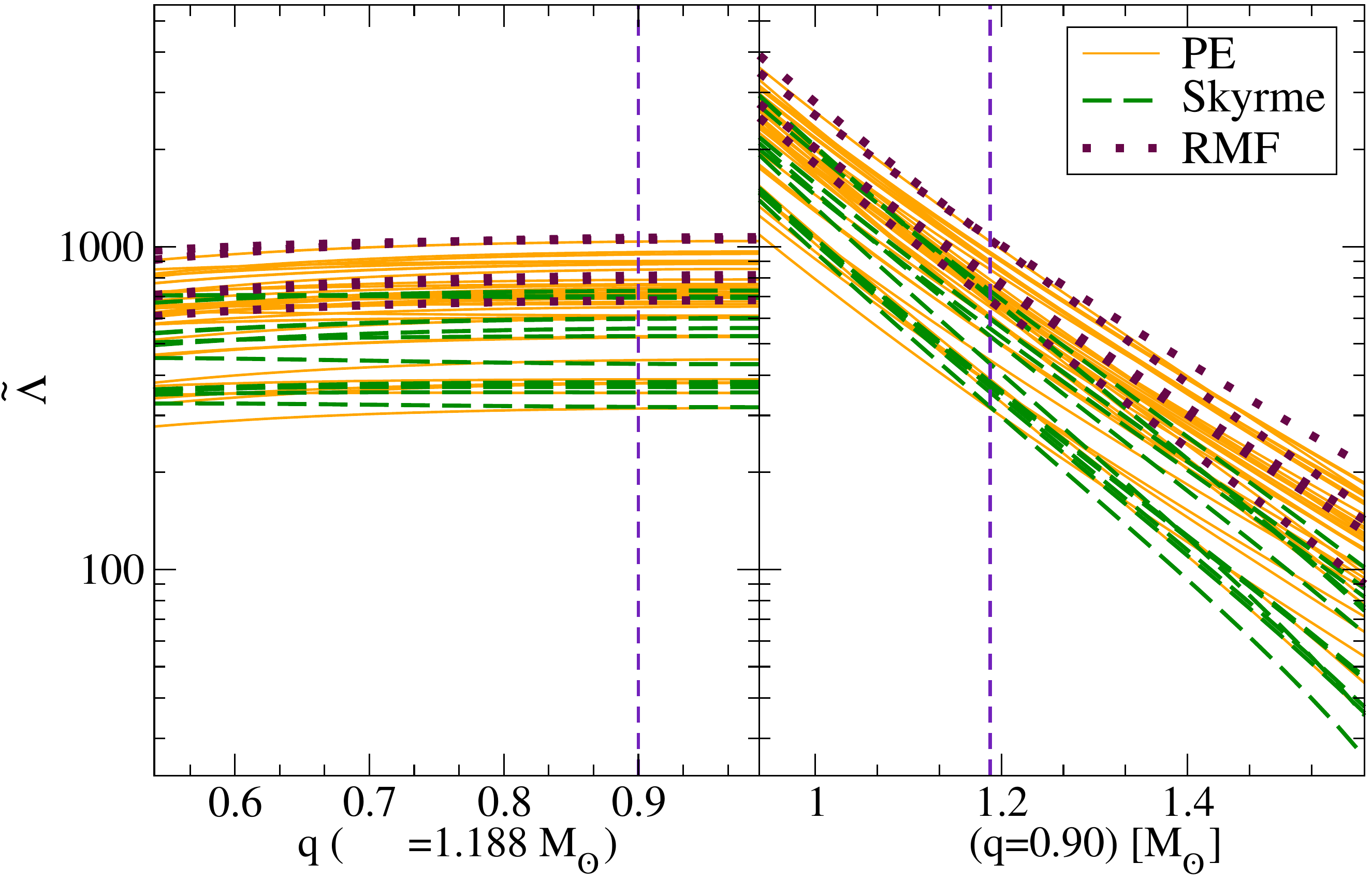}
\put(62,3){\small$\mathcal{M}$}
\put(162,3){\small$\mathcal{M}$}
\end{overpic}
\end{center}
\caption{
(left) Mass-weighted tidal deformability $\tilde\Lambda$ for each EoS model used in this analysis as a function of the mass ratio $q$ for a fixed chirp mass of $\mathcal{M}=1.188 \text{ M}_{\odot}$. The vertical dashed line at $q=0.90$ corresponds to GW170817. Observe that $\tilde\Lambda$ is insensitive to the choice of $q$, which justifies our method of keeping the mass ratio fixed. (right) Similar to the left panel but as a function of the chirp mass $\mathcal{M}$ for a fixed mass ratio of $q=0.90$. The vertical dashed line at $\mathcal{M}=1.188 \text{ M}_{\odot}$ corresponds to GW170817.
}
\label{fig:LT}
\end{figure}

We begin the current analysis by studying the correlations between the mass-weighted tidal deformability $\tilde\Lambda$ and various nuclear parameters $K_0$, $M_0$, and $K_{\text{sym},0}$ as was done in Ref.~\cite{Zack:nuclearConstraints}.
In the previous analysis, we studied the correlations and constraints as a function of the mass ratio $q \equiv m_1/m_2$ for the fixed chirp mass of $1.188\text{ M}_{\odot}$; corresponding to GW170817.
We here supplement this investigation by considering the correlations and constraints as a function of varying chirp mass at a fixed mass ratio, applicable to any number of future GW observations.
The left panel of Fig.~\ref{fig:LT} justifies the use of a fixed mass ratio by presenting $\tilde\Lambda$ as a function of $q$ for the various EoS models used in this analysis.
Observe how $\tilde\Lambda$ is insensitive to the choice of $q$. Such feature is absent in the right panel of Fig.~\ref{fig:LT}, where $\tilde\Lambda$ is plotted as a function of chirp mass for fixed $q$.
Thus, for the remainder of this analysis we fix the mass ratio to be $q=0.90$, corresponding to the center of $0.80 \leq q \leq 1.00$ derived in Ref.~\cite{Coughlin:2018fis} for GW170817.

We measure the amount of correlation between two observables $x$ and $y$ via
\begin{equation}
\mathcal{C}(x,y)=\frac{\Sigma_{xy}}{\sqrt{\Sigma_{xx}\Sigma_{yy}}},
\end{equation}
with covariances $\Sigma_{ab}$ $(a=(x,y), b=(x,y))$ given by
\begin{equation}\label{eq:covariance}
\Sigma_{ab}=\frac{1}{N}\sum^N_{i=0}a_ib_i-\frac{1}{N^2}\Big( \sum^N_{i=0}a_i \Big) \Big( \sum^N_{i=0}b_i \Big).
\end{equation}
Here $N$ represents the number of data points.
A correlation of $\mathcal{C}=1$ represents perfect correlation between observables $x$ and $y$, while $\mathcal{C}=0$ corresponds to no correlation. 

In Ref.~\cite{Zack:nuclearConstraints}, we studied the universal relations between $\tilde\Lambda$ and various nuclear parameters $K_0$, $M_0$, $K_{\text{sym},0}$ in order to derive constraints on such parameters.
Additionally, it was shown in Refs.~\cite{Alam2016,Malik2018,Zack:nuclearConstraints} that certain linear combinations of nuclear parameters, specifically $K_0+\alpha L_0$, $M_0+\beta L_0$, and $K_{\text{sym},0}+\gamma L_0$, exhibit heightened correlations, allowing one to derive more accurate constraints on the individual nuclear parameters.
However, it was found in Ref.~\cite{Zack:nuclearConstraints} that this came at the expense of additional sources of uncertainty which, if properly accounted for, enlarges the resulting constraints on the nuclear parameters $K_0$, $M_0$, and $K_{\text{sym},0}$.
It was also found that the single nuclear parameters as well as the linear combinations involving $K_0$ and $M_0$ observed poor correlations of $\mathcal{C} \lesssim 0.50$; indicating somewhat unreliable constraints on the nuclear parameters.

\begin{figure}
\begin{center} 
\begin{overpic}[width=\linewidth]{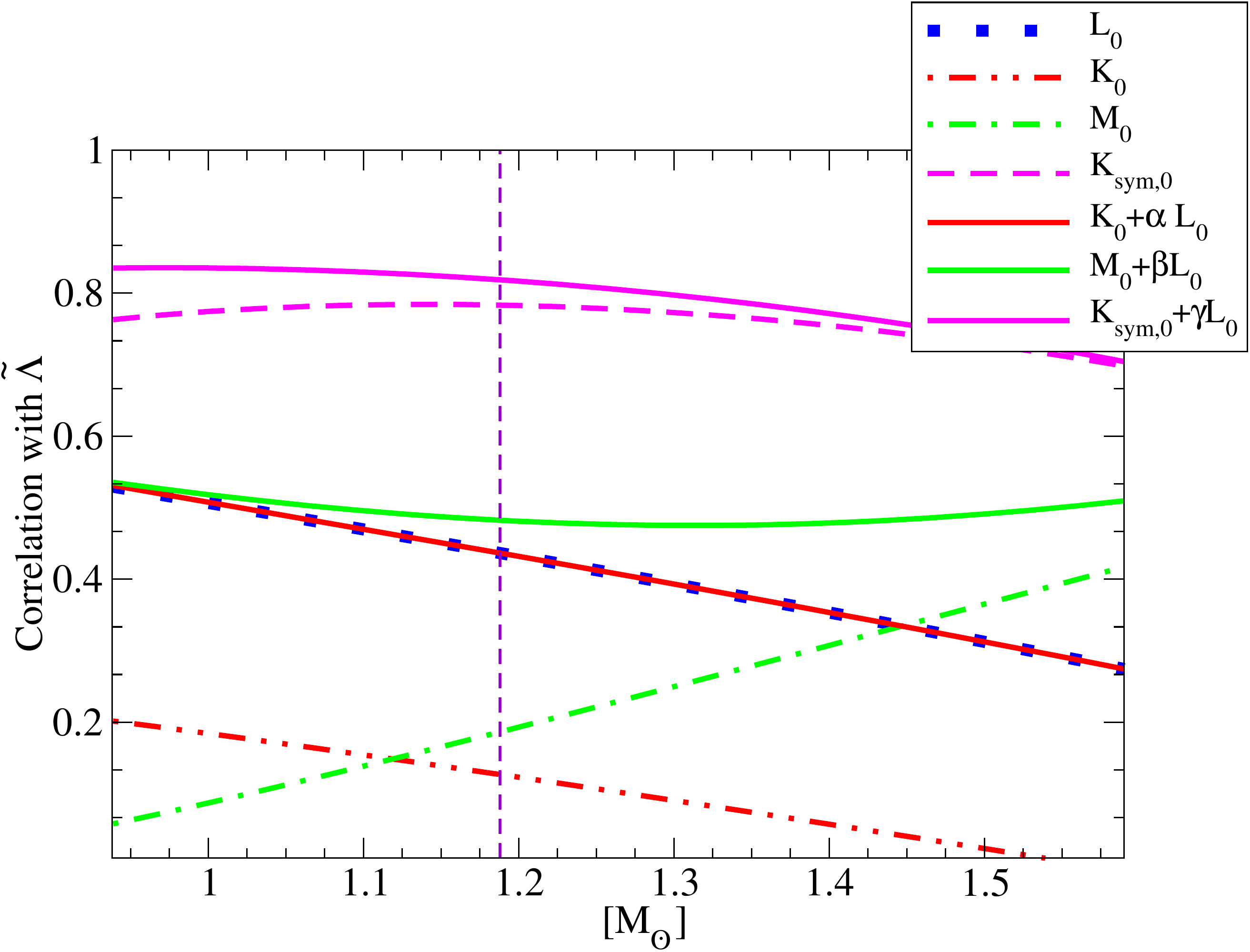}
\put(106,3){\small$\mathcal{M}$}
\end{overpic}
\end{center}
\caption{
Correlations between $\tilde\Lambda$ and various nuclear parameters as a function of the chirp mass.
Observe how low-order nuclear parameters $K_0$ and $M_0$ show poor correlations, while high-order parameter $K_{\text{sym},0}$ is highly correlated -- both with and without a linear combination with $L_0$.
Additionally shown by the dashed vertical line is the chirp mass of $1.188\text{ M}_{\odot}$ corresponding to GW170817, studied in detail by Ref.~\cite{Zack:nuclearConstraints}.
}
\label{fig:CorrOfM}
\end{figure}

Figure~\ref{fig:CorrOfM} similarly shows the above correlations as a function of chirp mass for a fixed mass ratio of $q=0.90$. 
Observe how, similar to what was found in Ref.~\cite{Zack:nuclearConstraints}, the correlations for $K_0$, $M_0$, $K_0+\alpha L_0$, and $M_0+\beta L_0$ are exceedingly poor for all values of chirp mass.
$K_{\text{sym},0}$ on the other hand, remains highly correlated with $\tilde\Lambda$ across the entire range of $\mathcal{M}$.
We also observe how correlations are not improved by much when considering linear combinations between $K_{\text{sym},0}$ and $L_0$.

Could other combinations of nuclear parameters give stronger correlations? To address this question, we further explore new combinations of nuclear parameters in App.~\ref{app:multiplicative}. 
In particular, we consider the ``multiplicative" combinations of $K_0 L_0^{\eta}$, $M_0 L_0^{\nu}$, and $K_{\text{sym},0} L_0^{\mu}$ that is motivated from Refs.~\cite{Sotani:2013dga,Silva:2016myw}. We found that such new combinations do not offer any advantages in terms of correlations and constraints.

For the above reasons, we consider only the curvature of the symmetry energy $K_{\text{sym},0}$ for the remainder of the investigation, without combinations with other parameters which would otherwise introduce additional uncertainties in the computation of constraints.

\section{Nuclear parameter constraints with future GW observations}\label{sec:chirpmass}

\begin{figure}
\begin{center} 
\includegraphics[width=\columnwidth]{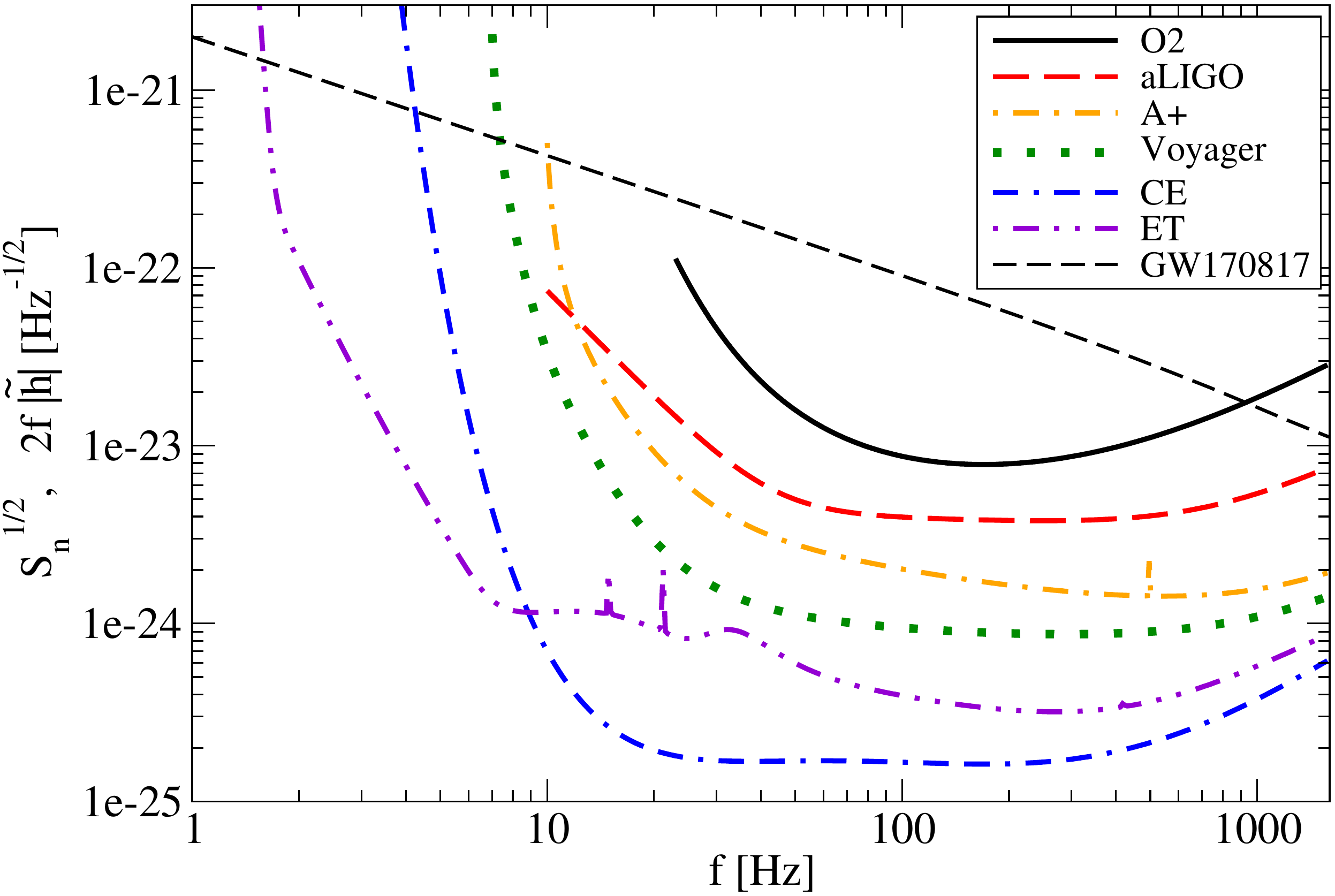}
\end{center}
\caption{
Spectral noise densities $\sqrt{S_n(f)}$ plotted for detectors: LIGO O2, aLIGO, A\texttt{+}, Voyager, CE, and ET-D as interpolated from publicly available data.
Spectral noise densities are plotted from $f_{\text{min}}=(23,10,10,7,1,1) \text{ Hz}$, respectively, to $f_{\text{max}}=1649 \text{ Hz}$.
Also shown is the frequency evolution of the characteristic amplitude $2 \sqrt{f} |\tilde{h}|$ for GW170817 using the IMRPhenomD~\cite{PhenomDI,PhenomDII} gravitational waveform template. The ratio between GW spectrum and signal roughly corresponds to signal-to-noise ratio.
}
\label{fig:sensitivities}
\end{figure}

Now that we have identified the high-correlation behavior of $K_{\text{sym},0}$, we proceed to compute projected bounds on the curvature of the symmetry energy as a function of chirp mass that is applicable to any future event.
Additionally we offer the same analysis repeated for 5 anticipated future detector sensitivities $S_n(f)$ for detectors O2~\cite{aLIGO}, aLIGO~\cite{aLIGO}, A\texttt{+}~\cite{Ap_Voyager_CE}, Voyager~\cite{Ap_Voyager_CE}, ET~\cite{ET} and CE~\cite{Ap_Voyager_CE} (see Fig.~\ref{fig:sensitivities}), which would allow one to compute the corresponding posterior distribution on $K_{\text{sym},0}$ given an events' chirp mass $\mathcal{M}$.

\subsection{Single events}\label{sec:futureSingle}

\begin{figure}
\begin{center}
\begin{overpic}[width=\linewidth]{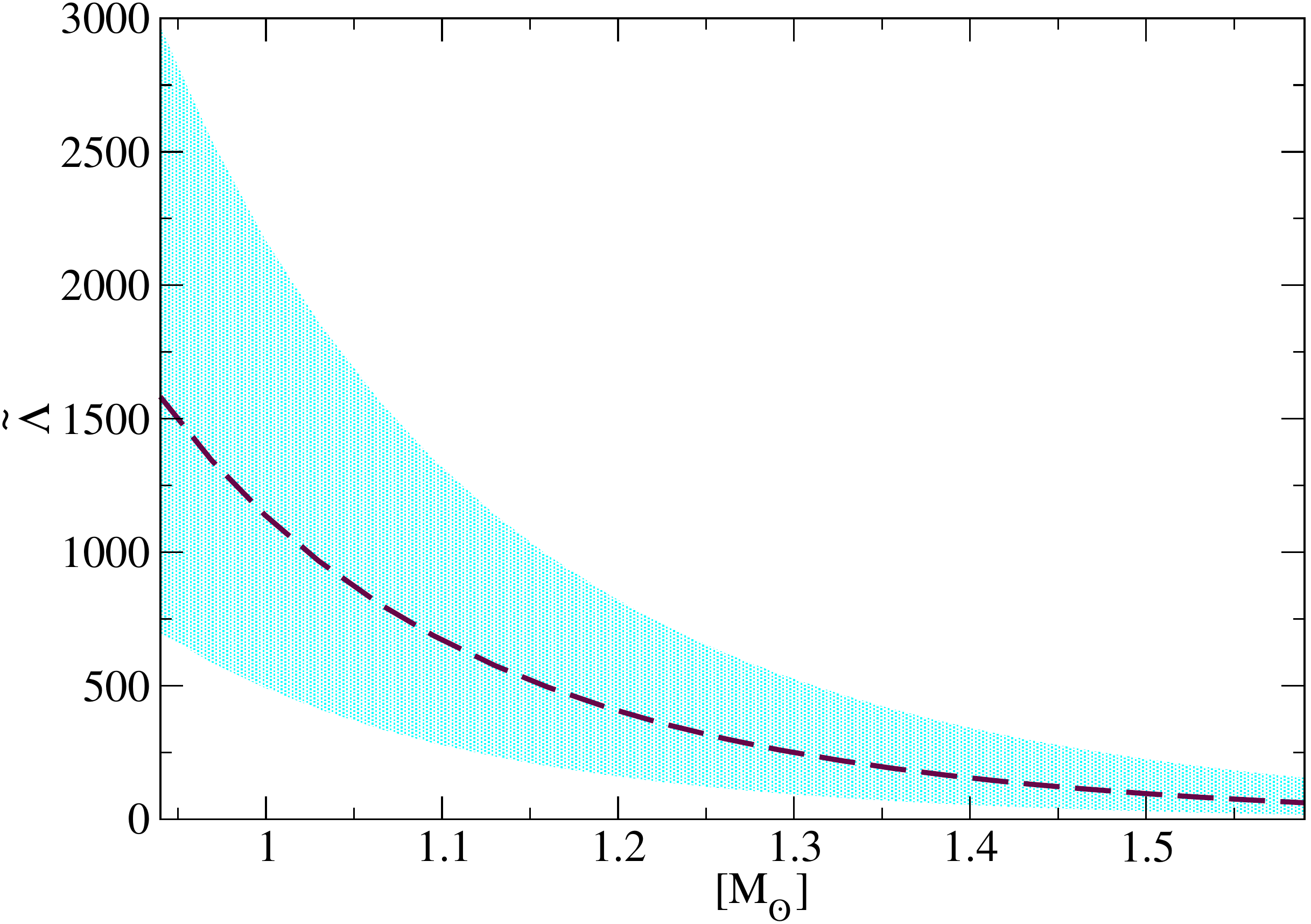}
\put(122,3){\small$\mathcal{M}$}
\end{overpic}
\end{center}
\caption{
Mean value of $\tilde\Lambda$ (dashed maroon curve) as a function of chirp mass $\mathcal{M}$, computed as the mean value of the ``LVC constrained" EoSs (cyan shaded region) from Ref.~\cite{Zack:URrelations} for each value of chirp mass.
This mean value corresponds to $\mu_{\tilde\Lambda}(\mathcal{M})$ used in the generation of the approximated $\tilde\Lambda$ probability distributions in Eq.~\eqref{eq:LtPrior} needed to compute constraints on $K_{\text{sym},0}$ in Eq.~\eqref{eq:P_Ksym}.
}
\label{fig:meanLt}
\end{figure}

\begin{figure}
\begin{center} 
\begin{overpic}[width=\linewidth]{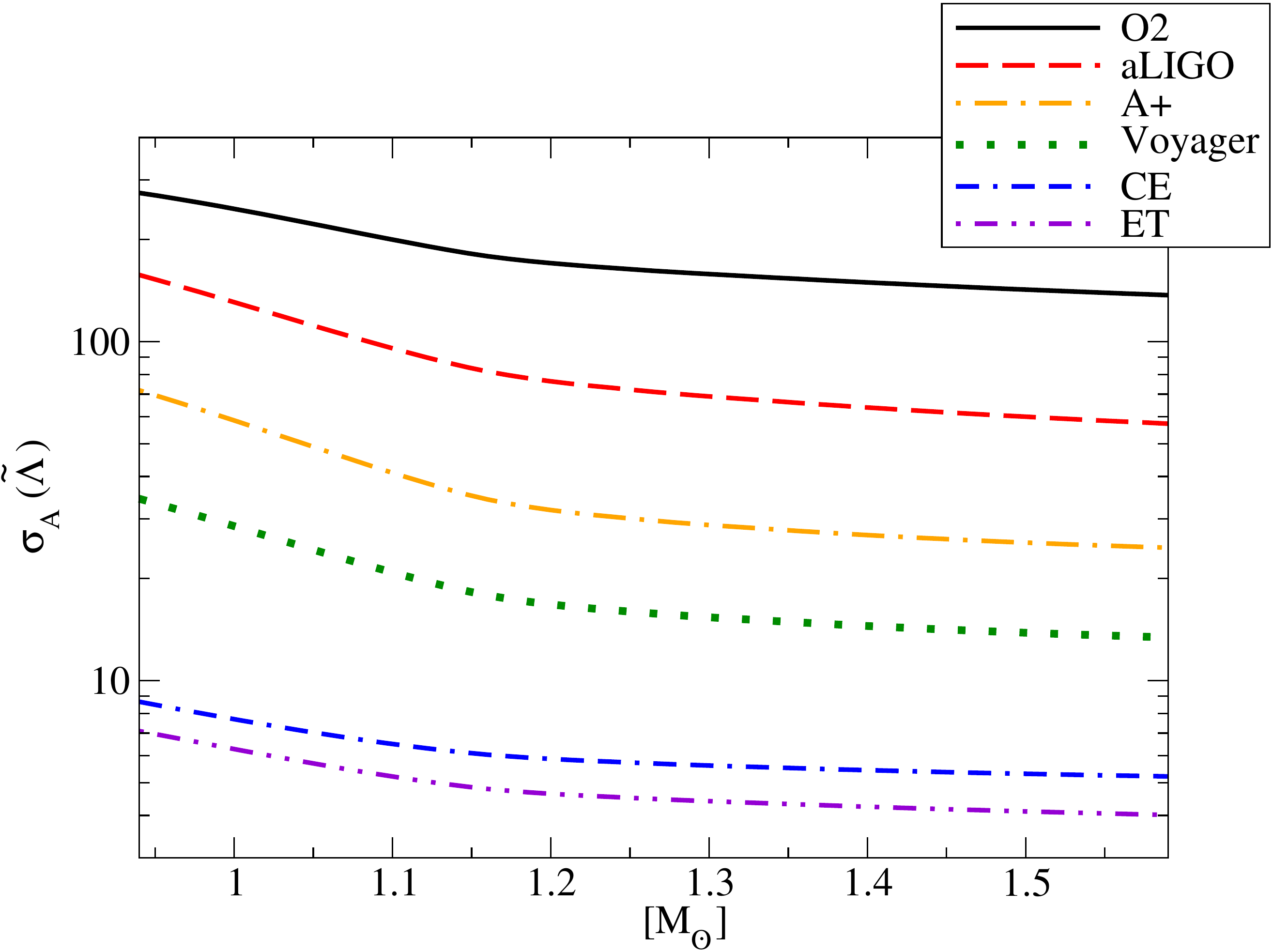}
\put(109,3){\small$\mathcal{M}$}
\end{overpic}
\end{center}
\caption{
Approximate 90\% measurement accuracies of $\tilde\Lambda$ on detectors $A=($O2, aLIGO, A\texttt{+}, Voyager, CE, ET$)$ as a function of chirp mass, computed via simple Fisher analyses.
These correspond to the standard deviations $\sigma_A$ used in the generation of the approximated $\tilde\Lambda$ probability distributions needed to compute constraints on $K_{\text{sym},0}$.
}
\label{fig:sigmaLt}
\end{figure}

Previously in Ref.~\cite{Zack:nuclearConstraints}, a posterior distribution on $\tilde\Lambda$ as derived from GW170817, was utilized in order to compute posterior distributions on the nuclear parameters.
In this analysis of future observations however, no such distribution is available.
To remedy this, we approximate the effective ``future" posterior distribution on $\tilde\Lambda$ as a Gaussian probability distribution given by
\begin{equation}\label{eq:LtPrior}
P_A(\tilde\Lambda)= \frac{1}{\sqrt{2\pi\sigma^2_{A}}} e^{-(\tilde\Lambda-\mu_{\tilde\Lambda})^2/2\sigma_{A}^2}
\end{equation}
for detector $A$.
Here, $\mu_{\tilde\Lambda}=\mu_{\tilde\Lambda}(\mathcal{M})$ is computed from the mean value of the ``LVC constrained" EoSs~\cite{Zack:URrelations} described in Sec.~\ref{sec:eos} for each value of chirp mass, as shown by Fig.~\ref{fig:meanLt}.
Further, $\sigma_{A}$ is approximated via simple Fisher analyses (described below), which estimates the measurement accuracy on $\tilde\Lambda$ under the assumption of detector sensitivity $A$. Figure~\ref{fig:sigmaLt} presents $\sigma_{A}$ for all 6 detectors.

The Fisher analysis method~\cite{Cutler:Fisher,Berti:Fisher,Poisson:Fisher} estimates the accuracy with which one can extract best-fit parameters $\theta^a$ given prior probability distributions $\sigma_{\theta^a}$ and a template waveform $h$.
For this analysis, we consider the sky-averaged ``IMRPhenomD" gravitational waveform template $h$~\cite{PhenomDI,PhenomDII} for point particles, modified by the 5PN and 6PN tidal corrections given in Ref.~\cite{Wade:tidalCorrections}.
We utilize a template parameter vector $\theta^a$ consisting of
\begin{equation}\label{eq:template}
\theta^a=(\ln{\mathcal A},\phi_c,t_c,\ln{\mathcal{M}_z},\ln{\mathcal{\eta}},\chi_s,\chi_a,\tilde{\Lambda}, \delta\tilde{\Lambda}),
\end{equation}
where $\mathcal A=\frac{\mathcal{M}_z^{5/6}}{\sqrt{30}\pi^{2/3}D_L}$ is a normalized amplitude factor, $D_L$ is the luminosity distance to the event, $\mathcal M_z = (1+z) \mathcal M$ is the redshifted chirp mass, $\eta = m_1 m_2/(m_1+m_2)^2$ is the symmetric mass ratio, $\chi_{s,a}=\frac{1}{2}(\chi_1 \pm \chi_2)$ are the symmetric and anti-symmetric combinations of individual spins $\chi_{1,2}$, and $\delta\tilde\Lambda$ is a tidal parameter entering first at 6PN order. 
In this investigation, we utilize fiducial parameter values of $\phi_c=0$, $t_c=0$, $\chi_s=\chi_a=0$, $\delta\tilde\Lambda=0$, $\mathcal{M}$ and $\eta$ are chosen from the current chirp mass iteration with a mass ratio of $q=0.9$, and $\tilde\Lambda$ is computed from the mean value of the ``LVC constrained" EoSs.
Additionally, we impose Gaussian spin priors of $|\chi_{s,a}|<1$, and tidal priors of $0<\tilde\Lambda<3000$, and $|\delta\tilde\Lambda|<500$~\cite{Wade:tidalCorrections}.

To obtain a posterior distribution, one needs to know both the likelihood and prior distributions.
Assuming the prior distributions on template parameters $\theta^a$ are Gaussian\footnote{A more comprehensive Bayesian analysis assumes the more valid choice of uniform prior distributions.}, the resulting posterior distributions are also Gaussian with root-mean-squares of
\begin{equation}
\Delta\theta^a=\sqrt{\Big( \tilde{\Gamma}^{-1} \Big)^{aa}}.
\end{equation}
The Fisher matrix $\tilde\Gamma_{ab}$ is defined as
\begin{equation}
\tilde{\Gamma}_{ab} \equiv \Big( \frac{\partial h}{\partial \theta^a} \Big| \frac{\partial h}{\partial \theta^b} \Big) + \frac{1}{\sigma^2_{\theta^a}}\delta_{ab},
\end{equation}
where $\sigma_{\theta^a}$ are the prior root-mean-square estimates of parameters $\theta^a$, and the inner product $(a|b)$ is given by
\begin{equation}
(a|b) \equiv 2 \int\limits^{\infty}_0 \frac{\tilde{a}^*\tilde{b}+\tilde{b}^*\tilde{a}}{S_n(f)}df.
\end{equation}

Now that all of the tools are prepared, we next compute the posterior distributions on $K_{\text{sym},0}$ using the Gaussian prior distributions on $\tilde\Lambda$ computed above as a function of chirp mass, for future detectors.
Following the process used in Ref.~\cite{Zack:nuclearConstraints}, this is accomplished by first generating a two-dimensional Gaussian probability distribution between $K_{\text{sym},0}$ and $\tilde\Lambda$, taking into account the covariances between the two as
\begin{equation}\label{eq:2dPDF}
P(\tilde{\Lambda},K_{\text{sym},0})=\frac{1}{2\pi\sqrt{|\bm{\Sigma}|}}e^{-\frac{1}{2}(\bm{x}-\bm{\mu})^T\bm{\Sigma}^{-1}(\bm{x}-\bm{\mu})}.
\end{equation}
Here $\bm{x}$ and $\bm{\mu}$ are the 2D vectors containing $(\tilde\Lambda, K_{\text{sym},0})$ and their means respectively, and $\bm{\Sigma}$ is the covariance matrix with elements given by Eq.~\eqref{eq:covariance}.

\begin{figure}
\begin{center} 
\begin{overpic}[width=\linewidth]{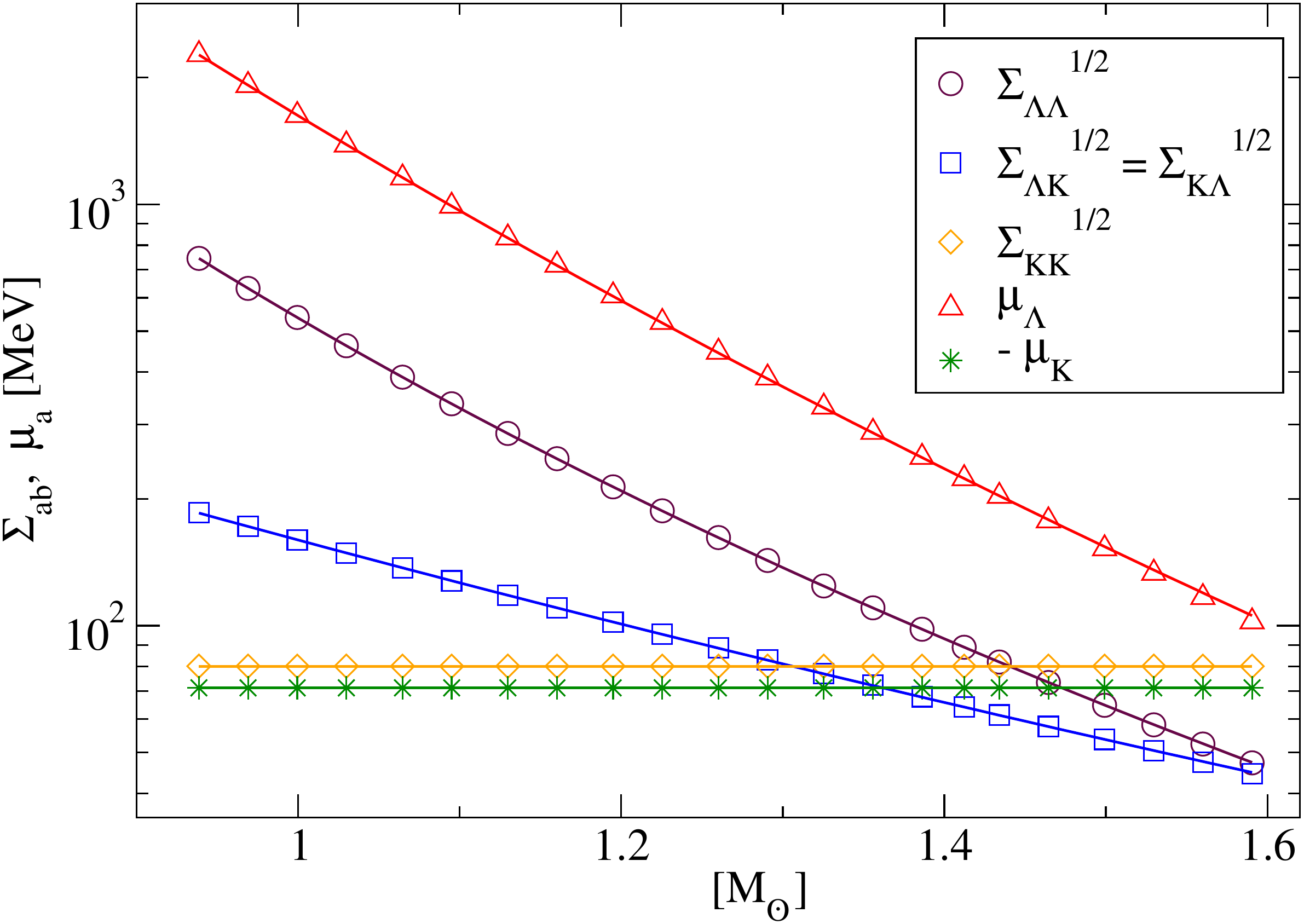}
\put(120,3.5){\small$\mathcal{M}$}
\end{overpic}
\end{center}
\caption{
Values of $\Sigma_{ab}$ and $\mu_a$ in Eq.~\eqref{eq:2dPDF} along with their respective fits (tabulated in Table~\ref{tab:CovMusLt}), necessary for the full reconstruction of the two-dimensional probability distributions between $\tilde\Lambda$ and $K_{\text{sym},0}$.
}
\label{fig:CovMusLt}
\end{figure}

Let us now offer readers the means to fully reproduce the results of the above analysis for any future event by constructing a fit for $\bm \mu$ and $\bm \Sigma$ in terms of chirp mass $\mathcal{M}$. Based on the relations between the former and the latter as shown in Fig.~\ref{fig:CovMusLt}, we create a fit in a logarithmic power expansion as
\begin{equation}
\log{y_i}=a_i+b_i \log{\mathcal{M}}+c_i (\log{\mathcal{M}})^2,
\end{equation}
with $y_i$ being the various parameters $\sqrt{\Sigma_{ab}}$ and $\mu_a$, and fitting coefficients $a_i$, $b_i$, and $c_i$ which are summarized in Table~\ref{tab:CovMusLt}.
Observe how well the fit agrees with the numerical data in Fig.~\ref{fig:CovMusLt}.

\begin{table}
\caption{
Respective fitting functions for the covariance matrix $\bm \Sigma$ and the mean vector $\bm \mu$ in Eq.~\eqref{eq:2dPDF} necessary for the full reconstruction of the two-dimensional probability distributions between $\tilde\Lambda$ and $K_{\text{sym},0}$.
Here, the values of $\Sigma_{\text{KK}}$ and $\mu_{\text{K}}$ correspond to the variance and mean of $K_{\text{sym},0}$, which are independent of chirp mass, thus require no fitting function.
}\label{tab:CovMusLt} 
\begin{tabular}{c || c} 
Parameter & Fitting Function\\
\hline
$\sqrt{\Sigma_{\Lambda\Lambda}}(\mathcal{M})$ [MeV] & $\text{Exp}\lbrack 6.287 - 11.86 \log{\mathcal{M}}-0.9803\log^2\mathcal{M} \rbrack$\\
$\sqrt{\Sigma_{\Lambda\text{K}}}(\mathcal{M})$ [MeV] & $\text{Exp}\lbrack 5.073 - 5.477 \log{\mathcal{M}}-4.103\log^2\mathcal{M} \rbrack$\\
$\sqrt{\Sigma_{\text{KK}}}(\mathcal{M})$ [MeV] & $80.11$\\
$\mu_{\Lambda}(\mathcal{M})$ [MeV] & $\text{Exp}\lbrack 7.394 - 12.27 \log{\mathcal{M}}-6.399\log^2\mathcal{M} \rbrack$\\
$\mu_{\text{K}}(\mathcal{M})$ [MeV] & $-71.6$
\end{tabular}
\end{table}

Constraints on $K_{\text{sym},0}$ are extracted by first computing the conditional probability distributions on $K_\text{sym,0}$ given a tidal deformability observation of $\tilde\Lambda_\text{obs}$.
By following Ref.~\cite{jensen_2007}, we can generate the one-dimensional conditional probability distribution on $K_\text{sym,0}$ by taking
\begin{widetext}
\begin{equation}\label{eq:conditional}
 P(K_\text{sym,0}|\tilde\Lambda_\text{obs})\sim \mathcal{N}\Bigg(\mu_{K_\text{sym,0}}+\frac{\sigma_{K_\text{sym,0}}}{\sigma_{\tilde\Lambda_\text{obs}}}C(\tilde\Lambda_\text{obs}-\mu_{{\tilde\Lambda_\text{obs}}}),(1-C^2)\sigma_{K_\text{sym,0}}^2\Bigg).
\end{equation}
\end{widetext}
Above, $\mathcal{N}(\mu,\sigma^2)$ is the normal distribution with mean and variance $\mu$ and $\sigma^2$, while $\mu_\A$ and $\sigma_\A^2$ are the mean and variances of $K_\text{sym,0}$ and $\tilde\Lambda_\text{obs}$.
Finally, we can combine the one-dimensional conditional probability distribution function of Eq.~\eqref{eq:conditional} with the one-dimensional prior distribution on $\tilde\Lambda$ of Eq.~\eqref{eq:LtPrior}.
Marginalizing over $\tilde\Lambda$ results in a posterior probability distribution on $K_{\text{sym},0}$
\begin{equation}
\label{eq:P_Ksym}
P_A(K_{\text{sym},0})=\int\limits^{\infty}_{-\infty}P(K_{\text{sym},0}|\tilde{\Lambda})P_A(\tilde\Lambda) \, d\tilde\Lambda,
\end{equation}
from which 90\% confidence intervals on the curvature of symmetry energy can be extracted.
This process is then repeated for $22$ values of chirp mass $\mathcal{M}$ across its feasible range, and then for each interferometer $A$.
Appendix~\ref{app:posteriorExample} exemplifies this by demonstrating the procedure for one value of chirp mass $\mathcal{M}=1.188 \text{ M}_{\odot}$ on interferometer O2, corresponding to GW170817.
The results found there are compared to those found in Ref.~\cite{Zack:nuclearConstraints} in order to demonstrate the accuracy of our approximated Gaussian $\tilde\Lambda$ priors, rather than the full posterior distribution found in Ref.~\cite{LIGO:posterior}. We found that we slightly \emph{underestimate} the errors in $K_{\text{sym},0}$ by using this method.

There is one important question to analyze here: how do the statistical errors on $K_{\text{sym},0}$ ($\sigma_A$ in $P_A(\tilde \Lambda)$ given in Eq.~\eqref{eq:LtPrior} that enters in Eq.~\eqref{eq:P_Ksym}) compare to the systematic errors (covariance $\bm \Sigma$ in $P(\tilde{\Lambda},K_{\text{sym},0})$ given in Eq.~\eqref{eq:2dPDF} that also enters in Eq.~\eqref{eq:P_Ksym})?
As more events are observed and the detector sensitivities $S_n(f)$ drop, the statistical errors on the measurement of $K_{\text{sym},0}$ approach zero, and the overall errors limit closer to the systematic error ``wall" introduced from the EoS variation in the universal relations.
We study this effect by first plotting the overall errors on $K_{\text{sym},0}$ as a function of chirp mass, defined to be the one-sided 90\% confidence interval on the posterior distribution of $K_{\text{sym},0}$.
Following this, we define the systematic errors to be the one-sided 90\% confidence interval of $K_{\text{sym},0}$ in the two-dimensional probability distribution evaluated at the central value $\mu_{\tilde\Lambda} (\mathcal M)$ of the $\tilde\Lambda$ prior distribution shown in Fig.~\ref{fig:meanLt}.
Equivalently, the fixed diagonal $K_{\text{sym},0}$ coefficient of the Gaussian argument $\exp \lbrack -\Sigma^{-1}_{\text{KK}}(K_{\text{sym},0}-\langle K_{\text{sym},0} \rangle)^2/2 + \dotsc \rbrack$ shows the systematic errors to be exactly equal to $(\Sigma_{\text{KK}}^{-1})^{-1/2}$.

Figure~\ref{fig:OverallVsSystematic} displays the results of the above described procedure; plotting the (one-sided 90\% confidence interval) overall and systematic errors on the measurement of $K_{\text{sym},0}$ as a function of chirp mass. 
We observe here the presence of a minimum in the uncertainties with respect to the chirp mass - a relic originating from the correlations between $K_{\text{sym},0}$ and $\tilde\Lambda$ seen in Fig.~\ref{fig:CorrOfM}, which similarly observe a maximum at the same chirp mass (and thus minimum EoS variation that generates systematic errors).
We do note, however, that while previous analyses by Refs.~\cite{Alam2016,Malik2018} required high correlations for the computation of constraints\footnote{Refs.~\cite{Alam2016,Malik2018} assumed the relationship between $\Lambda_{1.4}$ (the tidal deformability at $1.4\text{ M}_{\odot}$) and nuclear parameters to lay exactly on the best-fit line between the two. Thus, high degrees of correlation were absolutely necessary for accuracy on this claim.}, our analysis does not, as all covariances between $\tilde\Lambda$ and $K_{\text{sym},0}$ are taken into account by the two-dimensional probability distribution of Eq.~\eqref{eq:2dPDF}.

Observe also how, as predicted, the statistical errors drop as the more sensitive detectors are analyzed, reducing to almost zero as the overall errors limit to the fixed systematic error ``wall".
The overall errors on the highly-sensitive third generation interferometers CE and ET are indistinguishable from the systematic errors -- indicating that the error budget is highly dominated by systematics at this point.
Once the errors are dominated by systematics, improving detector sensitivities or observing new events will not aid in the further constraint of $K_{\text{sym},0}$.
This indicates the urgent need to reduce the systematic errors found in the EoS-variation of the universal relations for the Voyager-class detectors and beyond\footnote{A similar conclusion was reached in Ref.~\cite{Zack:URrelations}, where the detector statistical errors became comparable to the systematic errors from the binary Love universal relations for future detectors Voyager and beyond.}.


\subsection{Multiple events}\label{sec:futureMultiple}

The future of GW astronomy will become quite busy in terms of detected events. 
For example, future GW interferometer Cosmic Explorer will be detecting anywhere from $3\times 10^5$ to $4\times 10^6$~\cite{Zack:URrelations} binary NS merger events within its horizon distance per year - a staggering number which will certainly help reduce the statistical errors on tidal measurements.
How does one account for this effect when studying the uncertainties in future, undetected, events?
The dominant tidal parameter in the gravitational waveform, $\tilde\Lambda$, depends strongly on the subsequent masses in the binary system, something difficult to predict beforehand.
Ultimately, this prevents one from combining the uncertainties on $\tilde\Lambda$ for multiple events.

Fortunately, this can be remedied by following in the footsteps of Ref.~\cite{Zack:URrelations}, where we reparameterized the gravitational waveform to instead consider the $\Lambda_{1.4}$ and $\Lambda_{1.4}'$ tidal coefficients, generated by Taylor expanding the tidal deformability $\Lambda$ about the reference mass of $m_0=1.4 \text{ M}_{\odot}$~\cite{delPozzo:TaylorTidal,Yagi:binLove}:\footnote{We note here that a linear truncation of this Taylor series is valid for our purposes. By taking into account an additional quadratic term identified by $\Lambda_{1.4}''$, we found a reduction in measurement accuracy in $\Lambda_{1.4}$ by only $\leq5\%$, across various detectors and values of chirp mass.}
\begin{equation}
\Lambda \approx \Lambda_{1.4}+\Lambda_{1.4}'\left( 1-\frac{m}{m_0} \right).
\end{equation}
Here, $\Lambda_{1.4} \equiv \Lambda|_{1.4M_\odot}$ and $\Lambda_{1.4}' \equiv -d\Lambda/d\ln m|_{1.4M_\odot}$ are the dimensionless tidal deformability and its slope at $1.4 \text{ M}_{\odot}$, and they do not depend on the individual NS masses $m$ or any combination thereof (however they do depend on the fiducial mass value $m_0$ chosen).
Therefore, they are identical for every future binary NS merger event, and may be combined in uncertainty.

Similar to the correlation between $\tilde \Lambda$ and $K_\mathrm{sym,0}$, we also find a correlation between $\Lambda_{1.4}$ and $K_\mathrm{sym,0}$. We constructed a 2D Gaussian distribution $P(\Lambda_{1.4},K_{\text{sym},0})$ similar to $P(\tilde{\Lambda},K_{\text{sym},0})$ in Eq.~\eqref{eq:2dPDF} and find
\begin{equation}
\sqrt{\bm \Sigma} = 
\begin{pmatrix}
193.6 & 97.10 \\
97.10  & 80.11
\end{pmatrix} \ [\mathrm{MeV}]\,,
\quad
\bm \mu = \begin{pmatrix}
543.2 \\
 -71.164
\end{pmatrix} \ [\mathrm{MeV}]\,,
\end{equation}
for $\bm x = (\Lambda_{1.4}, K_\mathrm{sym,0})$. 
Notice that both $\bm \Sigma$ and $\bm \mu$ are independent of $\mathcal M$ in this case.

\begin{figure}
\begin{center} 
\includegraphics[width=1\linewidth]{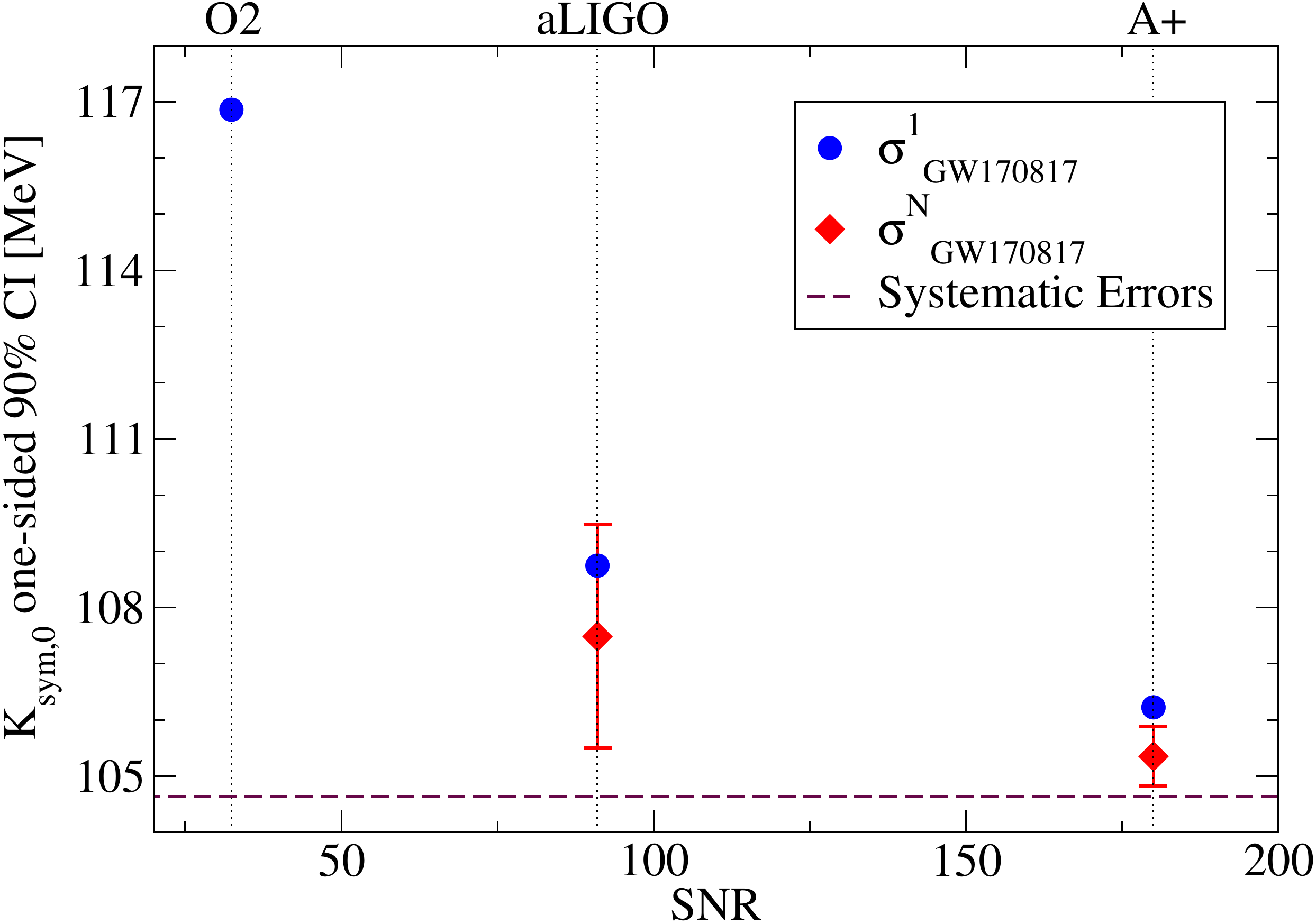}
\end{center}
\caption{
The overall errors on $K_{\text{sym},0}$ using priors on the combined $\Lambda_{1.4}$ (tidal deformability at $1.4\text{ M}_{\odot}$) uncertainty of multiple events (described in App.~\ref{app:multiple}), evaluated at the chirp mass $\mathcal{M}=1.188\text{ M}_{\odot}$.
The 90\% uncertainties on $K_{\text{sym},0}$ are shown as a function of the SNR of GW170817 as detected on each interferometer.
$\sigma^1_{\text{GW170817}}$ corresponds to the constraint formed with 1 GW170817-like observation, while $\sigma^N_{\text{GW170817}}$ forms the range bounded by the optimistic and pessimistic local binary NS coalescence rates.
While the single-event analysis of Fig.~\ref{fig:OverallVsSystematic} shows that single detections are nearly saturated by systematic uncertainties for Voyager-class detectors and beyond, here we show the effect stacking events can have on the aLIGO and A\texttt{+} analyses.
We observe that by combining multiple detections, even the aLIGO and A\texttt{+} interferometers approach the systematic error ``wall" (dashed horizontal line) with an optimistic number of detections.
}
\label{fig:stackedFisher}
\end{figure} 

In this section, we repeat the analysis performed in Sec.~\ref{sec:futureSingle} using the combined uncertainties on $\Lambda_{1.4}$ from $N_A$ unique events with chirp mass $1.188\text{ M}_\odot$, corresponding to the number of observed binary NS mergers within one observing year on detector $A$. We refer to App.~\ref{app:multiple} for details on how to combine information from multiple events, which closely follows App.~A of~\cite{Zack:URrelations}.
Fiducial values of $\Lambda_{1.4}$ and $\Lambda'_{1.4}$ were computed to be the mean values of $\Lambda|_{1.4M_\odot}$ and $-d\Lambda/d\ln m|_{1.4M_\odot}$ from the ``LVC constrained" EoSs.
Figure~\ref{fig:stackedFisher} shows how the combined-event uncertainties on $K_{\text{sym},0}$ for the fixed chirp mass of $1.188\text{ M}_\odot$ further become saturated on the aLIGO, and A\texttt{+} detectors as well.
As was shown in Fig.~\ref{fig:OverallVsSystematic}, the single-event uncertainties on $K_{\text{sym},0}$ become dominated by systematic errors for Voyager-class detectors and beyond, and thus there is not much point in stacking multiple events for these detectors to further reduce statistical errors on $\Lambda_{1.4}$.


\section{Reducing systematic errors via multidimensional correlations}\label{sec:systematics}

Let us now consider how we can reduce the systematic ``walls'' present in Fig.~\ref{fig:OverallVsSystematic}.
In Sec.~\ref{sec:futureMultiple}, this was computed by evaluating the two-dimensional probability distribution between $K_{\text{sym},0}$ and $\Lambda_{1.4}$ at the fiducial value of $\Lambda_{1.4}$, and then finding the 90\% confidence interval of the resulting probability distribution of $K_{\text{sym},0}$ to yield $\sim104$ MeV.
We here construct multidimensional correlations among $K_\mathrm{sym,0}$ and $\Lambda_{m_x}$ at a few different masses $m_x$ (since we expect to detect GWs from binary NSs with different masses with future observations) to see how adding information of the tidal deformability at multiple different masses may help us to reduce the systematic errors on $K_\mathrm{sym,0}$.

\begin{figure}
\begin{center} 
\includegraphics[width=8.cm]{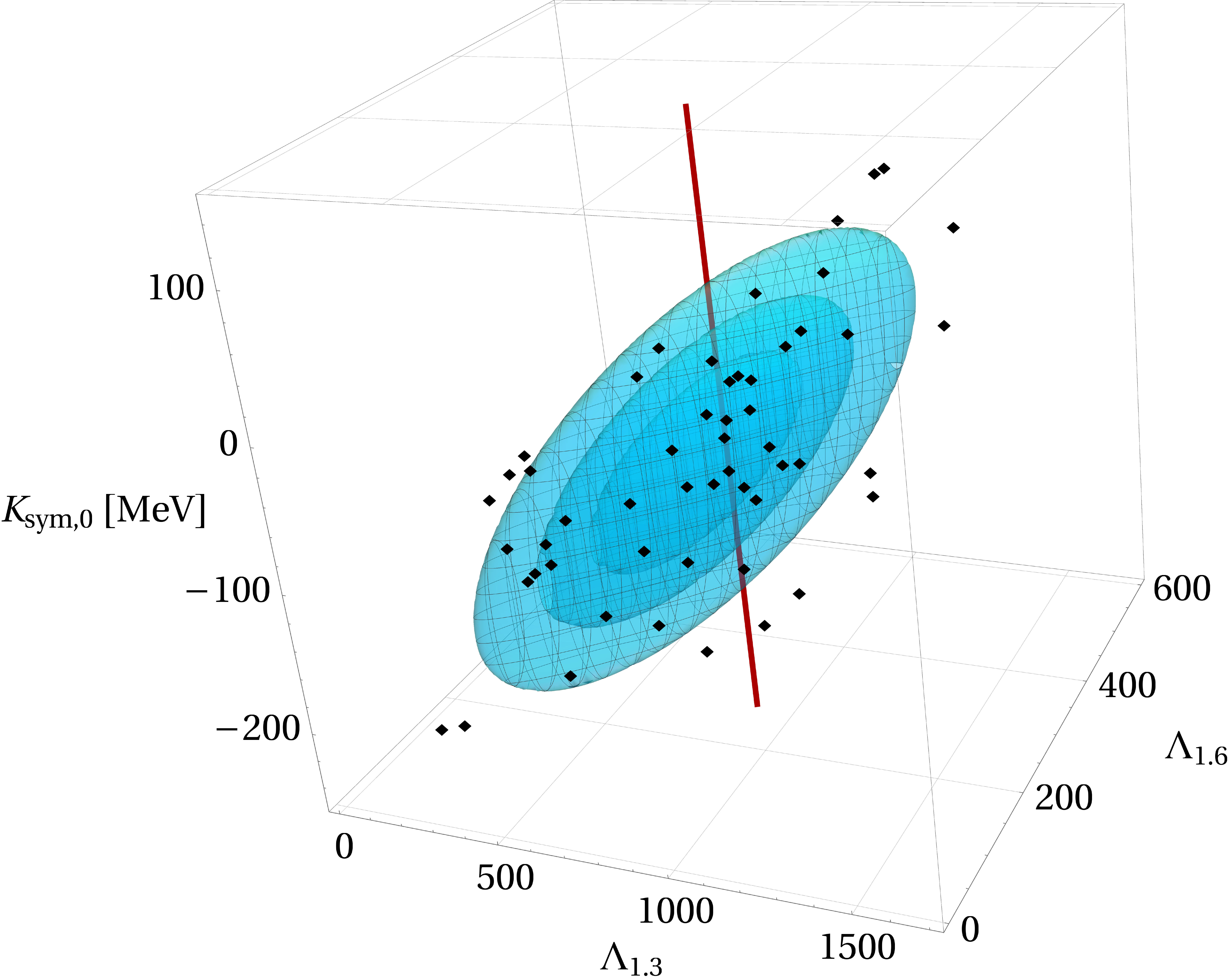}
\end{center}
\caption{
Example three-dimensional probability distribution between $\Lambda_{1.3}$, $\Lambda_{1.6}$, and $K_{\text{sym},0}$ (blue density contour).
Each EoS yields one point (black dot) in this space.
The resulting systematic errors in $K_{\text{sym},0}$ are computed by evaluating the probability distribution at the fiducial values of $\Lambda_{1.3}=886.8$ and $\Lambda_{1.6}=269.4$ (maroon line), at the 90\% confidence level.
}
\label{fig:3dPlot}
\end{figure}

Let us begin by using $\Lambda$ at two different masses $m_x$ and $m_y$. This requires us to find a three-dimensional correlation among $K_{\text{sym},0}$, $\Lambda_x (\equiv \Lambda_{m_x})$, and $\Lambda_y (\equiv \Lambda_{m_y})$, and construct a three-dimensional Gaussian distribution $P(K_{\text{sym},0}, \Lambda_x,\Lambda_y)$.
Figure~\ref{fig:3dPlot} shows an example of such a distribution for the case of $m_x=1.3\text{ M}_\odot$ and $m_y=1.6\text{ M}_\odot$.
The systematic error is then computed by evaluating the three-dimensional distribution at the fiducial values of $\Lambda_x$ and $\Lambda_y$, and then evaluating the resulting one-dimensional $K_{\text{sym},0}$ probability distribution at the 90\% confidence interval.

\begin{figure}
\begin{center} 
\includegraphics[width=8.cm]{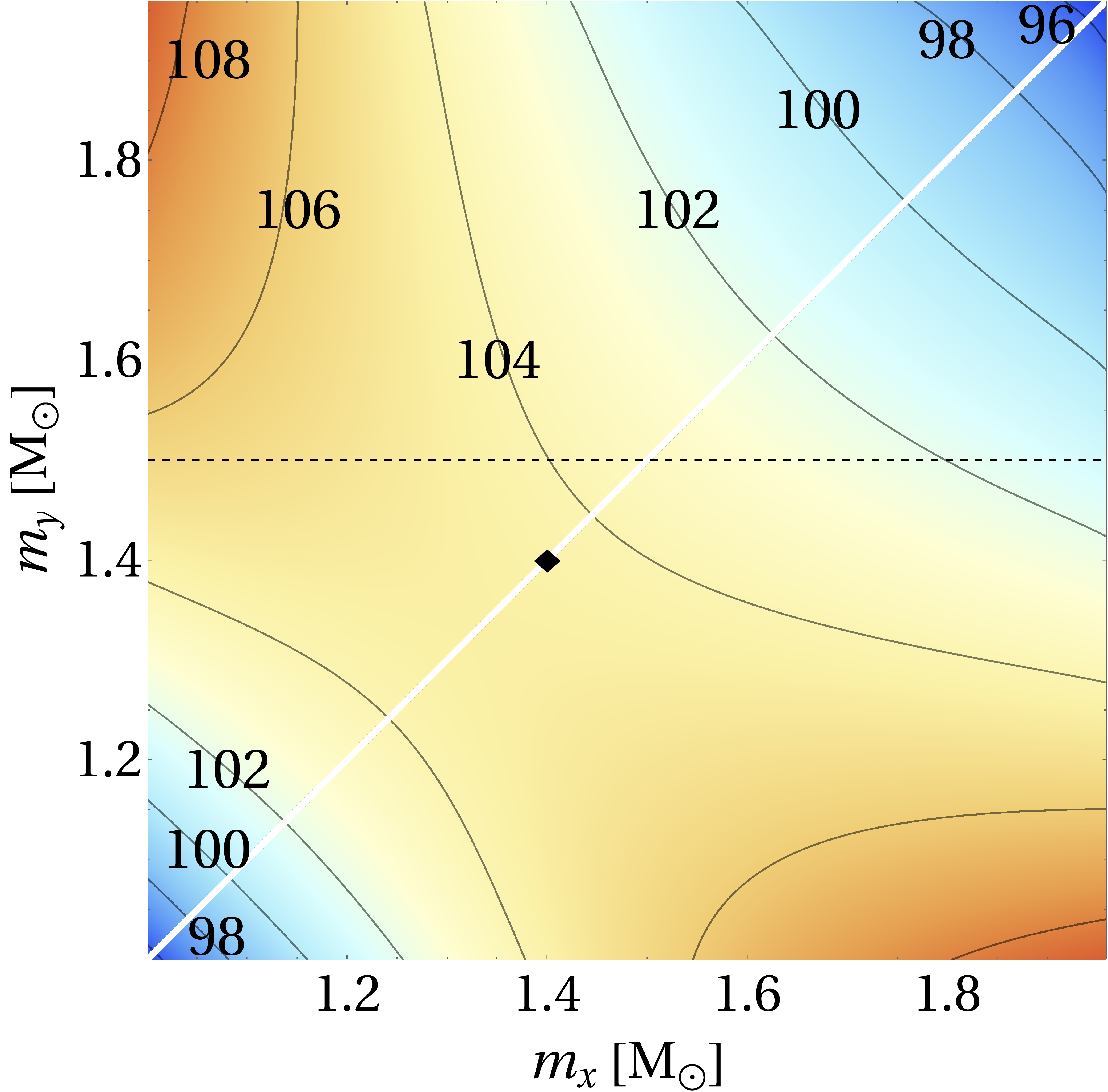}
\end{center}
\caption{
Similar to Fig.~\ref{fig:systematicContourFixly}, but computed from the three-dimensional probability distribution between $K_{\text{sym},0}$, $\Lambda_x$, and $\Lambda_y$, and evaluated at the fiducial values of $\Lambda_x$ and $\Lambda_y$.
The white diagonal line at $m_x=m_y$ corresponds to the systematic errors obtained from the reduced two-dimensional probability distribution $P(K_{\text{sym},0},\Lambda_x)$.
In particular, the black diamond represents the systematic error obtained with such a function with $\Lambda_{1.4}$ (the horizontal dashed line of Fig.~\ref{fig:stackedFisher}). 
The systematic errors along the horizontal dashed line at $m_y = 1.5M_\odot$ corresponds to $P(K_{\text{sym},0},\Lambda_x,\Lambda_{1.5})$, which is equivalent to $P(K_{\text{sym},0},\Lambda_x,\Lambda_x,\Lambda_{1.5})$ along the diagonal line in Fig.~\ref{fig:systematicContourFixly}.
Observe that having the information of additional $\Lambda$ values does not help in this case, and what matters is to have $m_x$ and $m_y$ to be both small or large.
}
\label{fig:systematicContour}
\end{figure}

Figure~\ref{fig:systematicContour} displays the resulting systematic uncertainties on $K_{\text{sym},0}$ using canonical masses $m_x$ and $m_y$ between $1\text{ M}_\odot$ and $2\text{ M}_\odot$.
Observe that the systematic errors can be reduced by setting both $m_x$ and $m_y$ to be large or small. However, this means that $m_x \approx m_y$, which corresponds to effectively using a two dimensional probability distribution. Thus, in this case, having additional information on $\Lambda$ at a different mass does not help to reduce the systematic errors.

\begin{figure}
\begin{center} 
\includegraphics[width=8.5cm]{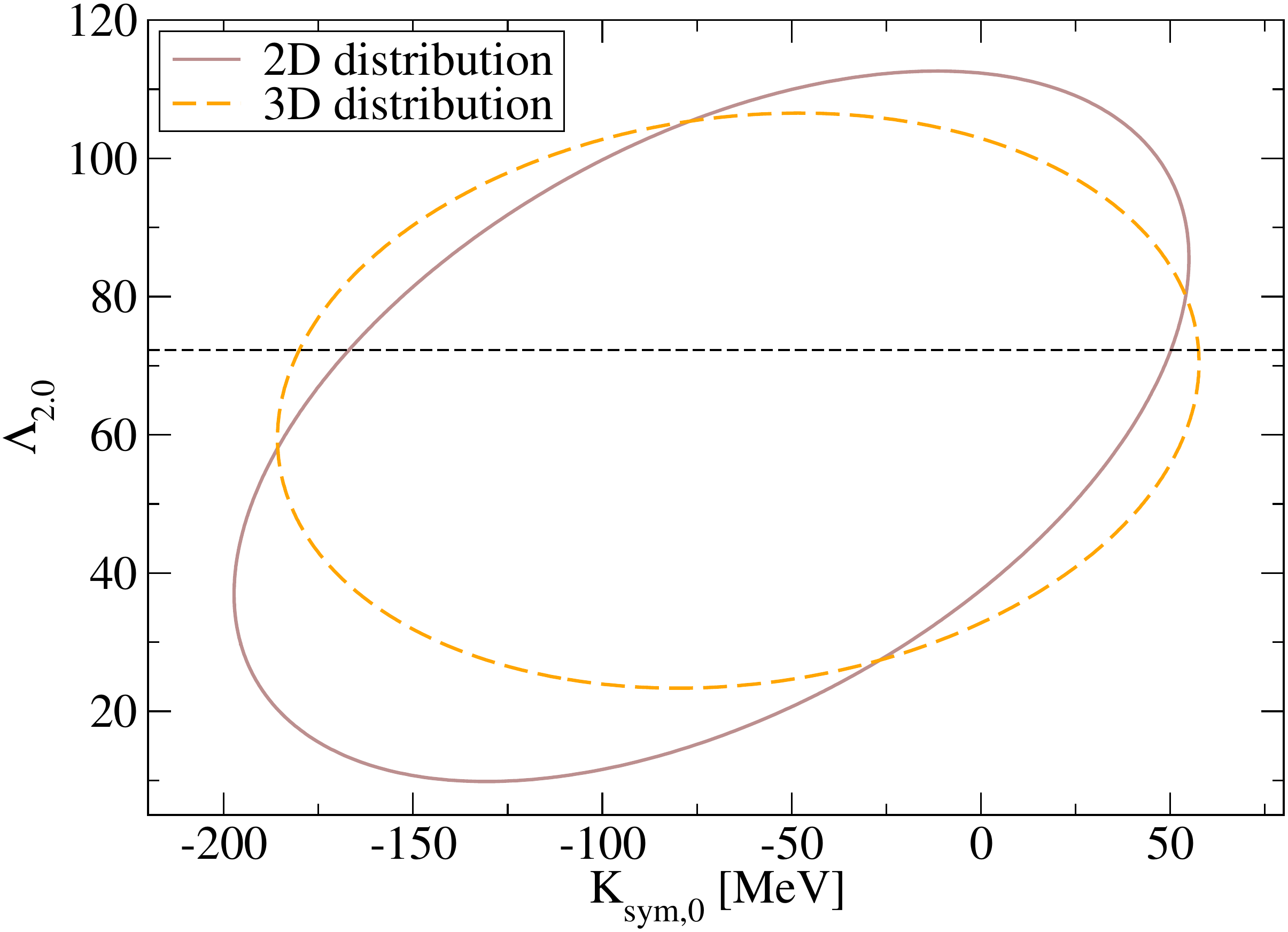}
\end{center}
\caption{90\% confidence interval contours of the two-dimensional probability distribution between $K_{\text{sym},0}$ and $\Lambda_{2.0}$ computed using two different methods:
(i) (brown) the two-dimensional probability distribution between $K_{\text{sym},0}$ and $\Lambda_{2.0}$, and (ii) (dashed orange) the three-dimensional probability distribution between $K_{\text{sym},0}$, $\Lambda_{2.0}$, and $\Lambda_{1.0}$. (We evaluate the latter at the fiducial value of $\Lambda_{1.0}$.)
To compute the systematic errors in $K_{\text{sym},0}$, one would evaluate such contours at the fiducial value of $\Lambda_{2.0}$, denoted by the horizontal line, and finding the 90\% confidence interval of the resulting one-dimensional probability distribution in $K_{\text{sym},0}$.
Observe that although the area of the brown contour is larger than that of the orange, the systematic error on $K_{\text{sym},0}$ from the former is smaller than that of the latter.
}
\label{fig:contourCases}
\end{figure}

It may sound strange that adding more pieces of information does not help to reduce the systematic errors. Let us explain why this is the case by comparing the systematic errors at $(m_x,m_y) = (1,2)M_\odot$ and $(m_x,m_y) = (2,2)M_\odot$.
Figure~\ref{fig:contourCases} compares the two-dimensional 90\% contours between $K_{\text{sym},0}$ and $\Lambda_{2.0}$ from two different methods by computing (i) directly the two-dimensional probability distribution from Eq.~\eqref{eq:2dPDF}, and (ii)  the three-dimensional probability distribution between $K_{\text{sym},0}$, $\Lambda_{2.0}$, and $\Lambda_{1.0}$, and then evaluating it at the fiducial value of $\Lambda_{1.0}$.
We observe that while the contour from the first case has a larger area (and value of $|\bm \Sigma|$) as expected due to the use of less information, it becomes distorted such that the systematic uncertainty (along the dashed horizontal line corresponding to the fiducial value of $\Lambda_{2.0}$) becomes smaller than that from the first case.

Let us now consider using $\Lambda$ at three different masses, $m_x$, $m_y$ and $m_z$. 
This requires us to find a four-dimensional correlation and construct the four-dimensional Gaussian probability distribution $P(K_{\text{sym},0},\Lambda_x,\Lambda_y,\Lambda_z)$.
We fix $m_z=1.5\text{ M}_\odot$, and allow $m_x$ and $m_y$ to vary between $\lbrack 1.0, 2.0 \rbrack\text{ M}_\odot$.
Similar to the process used previously, this probability distribution is evaluated at the fiducial values of $\Lambda_x$, $\Lambda_y$ and $\Lambda_{1.5}$: 
\begin{equation}
P'''(K_{\text{sym},0})=P(K_{\text{sym},0}, \bar{\Lambda}_x,\bar{\Lambda}_y,\bar{\Lambda}_{1.5}).
\end{equation}
The resulting 90\% confidence intervals are presented in Fig.~\ref{fig:systematicContourFixly} for the entire range of $m_x$ and $m_y$ mass values.
We observe that by including information about binaries with large, medium, and small masses together, the systematic errors can be improved drastically, down to $\sim74$ MeV. 
We also see that along the diagonal line of $m_x=m_y$, the four-dimensional probability distribution $P(K_{\text{sym},0},\Lambda_x,\Lambda_x,\Lambda_{1.5})$ reduces to the three-dimensional case $P(K_{\text{sym},0},\Lambda_x,\Lambda_{1.5})$, with uncertainties $\sim104$ MeV approaching that of Fig.~\ref{fig:systematicContour} along the horizontal dashed line.

\section{Discussions}\label{sec:conclusion}

Are there any other ways to further improve the constraints on $K_{\text{sym},0}$ using observed GW events?
One might think that the constraint on $\tilde\Lambda$ with GW170817 may help in this direction.
However, the restriction of data in \emph{only} the $\tilde\Lambda$ dimension does not help as systematic errors are found by evaluating the scattering width in the $K_{\text{sym},0}$ direction.

Finally, we briefly discuss the possibility of NSs with strong first-order phase transitions from hadronic to quark matter in the core, as described in Ref.~\cite{Paschalidis2018}.
With high enough observed chirp masses $\mathcal{M}$, future binary NS merger events could potentially be composed of one or both \emph{hybrid stars} (HSs) with quark-matter cores. 
The tidal deformabilities and thus, the nuclear parameters, depend on such structure, and could potentially disagree between events with varying chirp masses and combinations of NS/HS~\cite{Montana:2018bkb}.
Thus, significant variations between nuclear parameter measurements with future GW observations with varying chirp masses could potentially present evidence of strong phase transitions at around $2-3$ times the nuclear saturation density. 
If such transitions are present at sufficiently low-densities, then the nuclear matter parameters will be further decoupled from the GW observations and thus our lower limit for the nuclear matter parameter uncertainties will increase.
Alternatively, similar measurements of nuclear parameters could either indicate a pure hadronic matter EoS, or phase transitions occurring at higher nuclear densities.
The structure of such high-density transitions could be probed by the GW post-merger oscillation signal. 

As we showed in the previous section, one can use multidimensional correlations to reduce the systematic errors. 
Instead of using tidal deformabilities from different NS masses obtained from GWs alone, one can consider combining information from multi-messenger observations.
For example, pulse profiling techniques by NASA's Neutron star Interior Composition Explorer (NICER) instrument may provide high-precision measurements on the NS radius down to $5\%$~\cite{NICER:nsradius,NICER:nsEoS}. 
Thus, one can construct multidimensional correlations among nuclear parameters, tidal deformabilities and radii of NSs. 
The work along this direction is currently in progress~\cite{Kristen}.


\section*{Acknowledgments}\label{acknowledgments}

We thank David Nichols for his illuminating advice on conditional probability distributions.
Z.C. and K.Y. acknowledge support from NSF Award PHY-1806776. 
K.Y. would like to also acknowledge networking support by the COST Action GWverse CA16104. 
A.W.S. was supported by NSF grant PHY 1554876 and by the U.S. DOE Office of Nuclear Physics.

\appendix


\section{Original versus restricted sets of EoS}
\label{app:EoS-comparison}

\begin{figure}
\begin{center} 
\includegraphics[width=\linewidth]{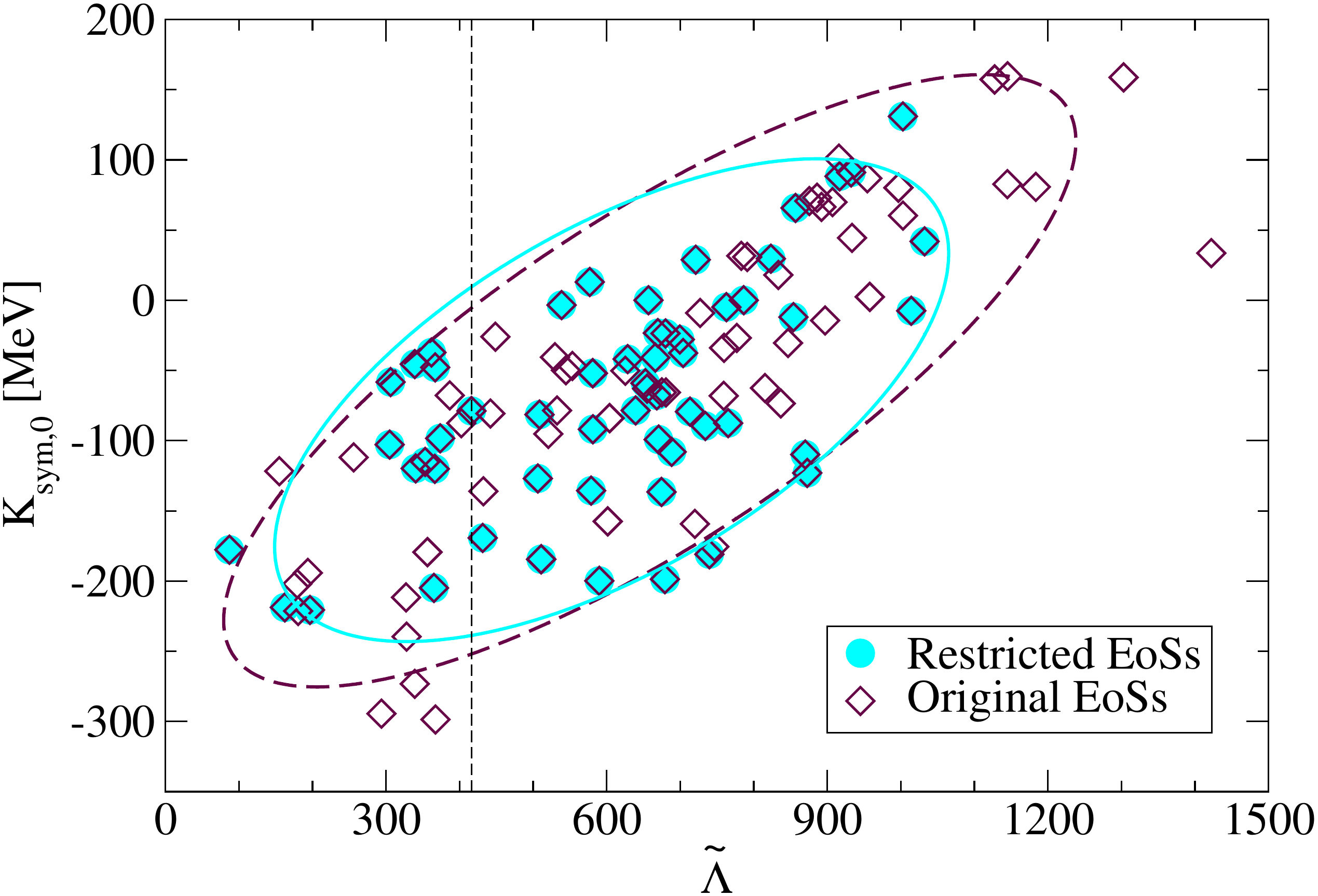}
\end{center}
\caption{Comparison of the two-dimensional $K_{\text{sym},0}-\tilde\Lambda$ correlations when using (i) the reduced set of EoSs taking into account the nuclear parameter correlations found in Ref.~\cite{Tews2017} (filled circle) and (ii) the original set of EoSs used in Ref.~\cite{Zack:nuclearConstraints} (open diamond).
Depicted by their respective 90\% confidence ellipses, we observe that the reduced set of EoSs shrinks the probability distribution in the direction of correlation, though such a set does not appreciably change the width.
The systematic errors, computed to be the 90\% width of the $K_{\text{sym},0}$ probability distribution evaluated at the fiducial value of $\tilde\Lambda$ (depicted by the vertical dashed line), are seen to be both $\sim104$ MeV, independent of which set of EoSs are used.
However, the overall errors \emph{are} observed to be reduced for the reduced set.
}
\label{fig:eosCompare}
\end{figure} 

In this appendix, we show how the restriction of EoSs described in Sec.~\ref{sec:eos} used in the current analysis impacts our observations, as compared to the original set of EoSs used in Ref.~\cite{Zack:nuclearConstraints}.
Figure~\ref{fig:eosCompare} shows a comparison between the two-dimensional probability distributions $P(K_{\text{sym},0},\tilde\Lambda)$ resulting from each set of EoSs.
We see that while the restriction to EoSs does indeed shrink the 90\% confidence intervals in the direction of correlation, the widths are approximately equal at the fiducial value of $\tilde\Lambda$ (where the systematic errors are analyzed).
This shows that while using a subset of EoSs may reduce the overall error\footnote{We indeed observed large reductions in the overall errors found in Fig.~\ref{fig:OverallVsSystematic} when using the restricted EoSs rather than the original ones, while the level of systematics stayed constant at $\sim104$ MeV.}, the level of systematic errors will remain mostly fixed.
Our result is consistent, for example, with Ref.~\cite{Carriere2003}, which found that the correlation between $L_0$ and the radius of a $1.4\text{ M}_\odot$ neutron star was weak because of the contribution of the high-density component of the EoS.


\section{Multiplicative combinations of nuclear parameters}\label{app:multiplicative}

In this appendix, we discuss the feasibility of using multiplicative combinations of nuclear parameters, such as $K_0 L_0^\eta$, rather than the linear combinations such as $K_0+\alpha L_0$ considered in Refs.~\cite{Alam2016,Malik2018,Zack:nuclearConstraints}.
Here we consider the following multiplicative combinations for comparison purposes: $K_0L_0^{\eta}$, $M_0L_0^{\mu}$, and $K_{\text{sym},0}L_0^{\nu}$, where coefficients $\eta$, $\mu$, and $\nu$ are similarly chosen to achieve maximal correlation. Such multiplicative combinations are similar to those considered in~\cite{Sotani:2013dga,Silva:2016myw}.

\begin{figure}
\begin{center} 
\begin{overpic}[width=\linewidth]{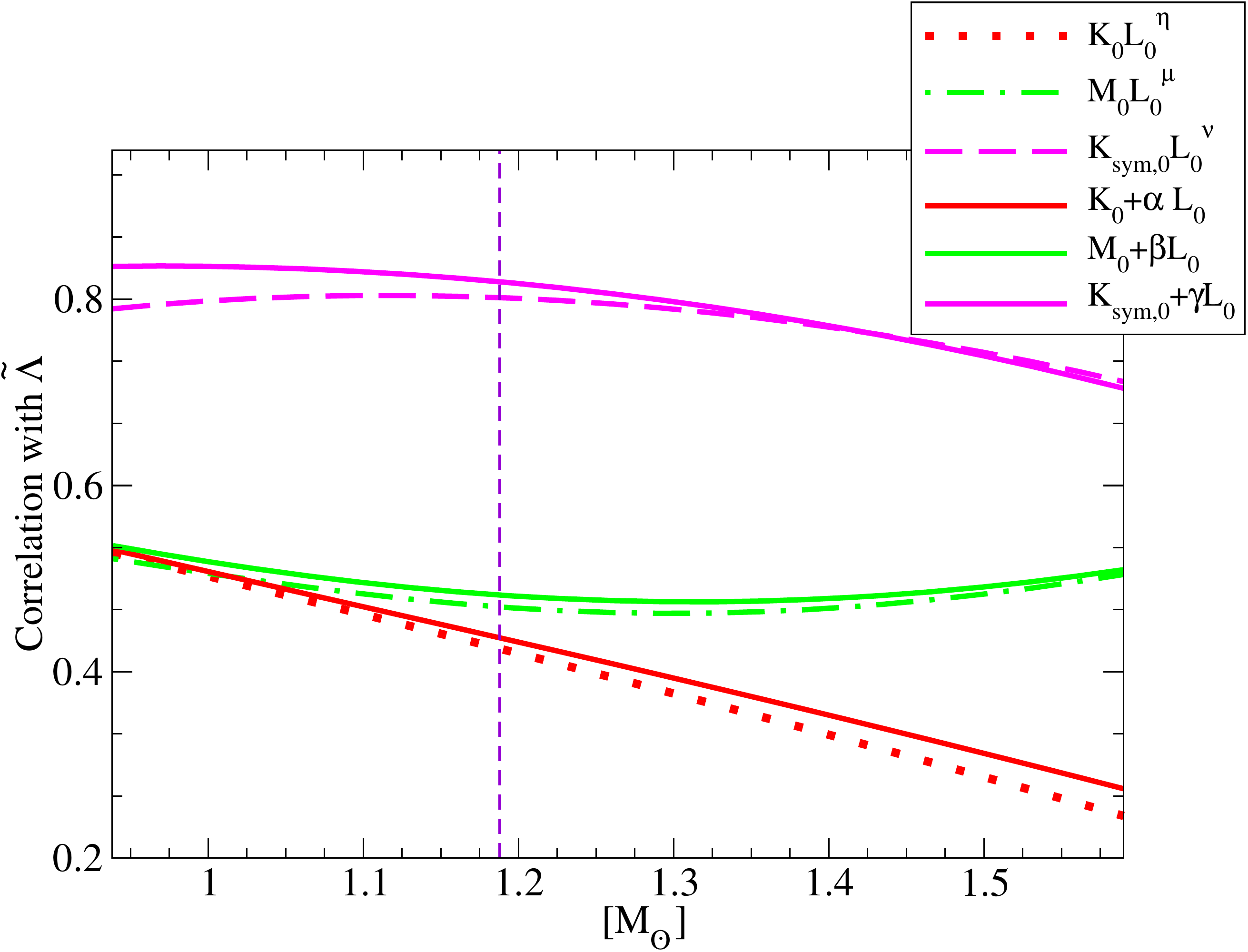}
\put(105,3){\small$\mathcal{M}$}
\end{overpic}
\end{center}
\caption{
Similar to Fig.~\ref{fig:CorrOfMmult}, but for the comparison between multiplicative and linear combinations of nuclear parameters: $K_0L_0^{\eta}$, $M_0L_0^{\mu}$, $K_{\text{sym},0}L_0^{\nu}$, $K_0+\alpha L_0$, $M_0+\beta L_0$, and $K_{\text{sym},0}+\gamma L_0$.
Here, parameters $\eta$, $\mu$, $\nu$, $\alpha$, $\beta$, and $\gamma$ are chosen such that the correlations with $\tilde\Lambda$ are maximal at each value of chirp mass.
Observe how both the linear and multiplicative combinations of nuclear parameters produce similar correlations with $\tilde\Lambda$, though the former outperforms the latter marginally for nearly all values of chirp mass.
}
\label{fig:CorrOfMmult}
\end{figure}

Figure~\ref{fig:CorrOfMmult} presents the correlations between $\tilde\Lambda$ and all 6 multiplicative and linear combinations of nuclear parameters considered in this analysis.
While the two classes of nuclear parameter combinations produce very similar correlations with $\tilde\Lambda$, we observe that the linear cases slightly outperform the multiplicative cases for nearly all values of chirp mass.
Similarly, repeating the analysis\footnote{Because the two-dimensional probability distribution is now between $\tilde\Lambda$ and $K_{\text{sym},0}L_0^{\nu}$, an additional marginalization over $L_0$: $\int^{\infty}_{-\infty}P(K_{\text{sym},0}L_0^\nu)P(L_0)dL_0$ must be performed ($P(L_0)$ is an additional prior distribution on $L_0$ given by Refs.~\cite{Lattimer2013,Lattimer2014,Tews2017,Oertel2017}) in order to extract the posterior distribution on $K_{\text{sym},0}$.} found in Sec.~\ref{sec:chirpmass} returns constraints on $K_{\text{sym},0}$ to be slightly worse than that considered in the the main analysis, due to the additional inclusion of uncertainties from nuclear parameter $L_0$. 
One arrives at a similar conclusion if one uses a linear combination with $\gamma \neq 0$~\cite{Zack:nuclearConstraints}.

From this evidence, we conclude with the remarks that the multiplicative combinations of nuclear parameters offer nothing new in terms of enhanced constraints on nuclear parameters.
The multiplicative combinations of nuclear parameters slightly under-perform their linear combination counterparts in terms of correlations with $\tilde\Lambda$.
Thus, we neglect their use and continue our analysis as was done previously in~\cite{Zack:nuclearConstraints}.


\section{Example computation of the $K_{\text{sym},0}$ posterior distribution}\label{app:posteriorExample}

\begin{figure}
\begin{center} 
\includegraphics[width=\columnwidth]{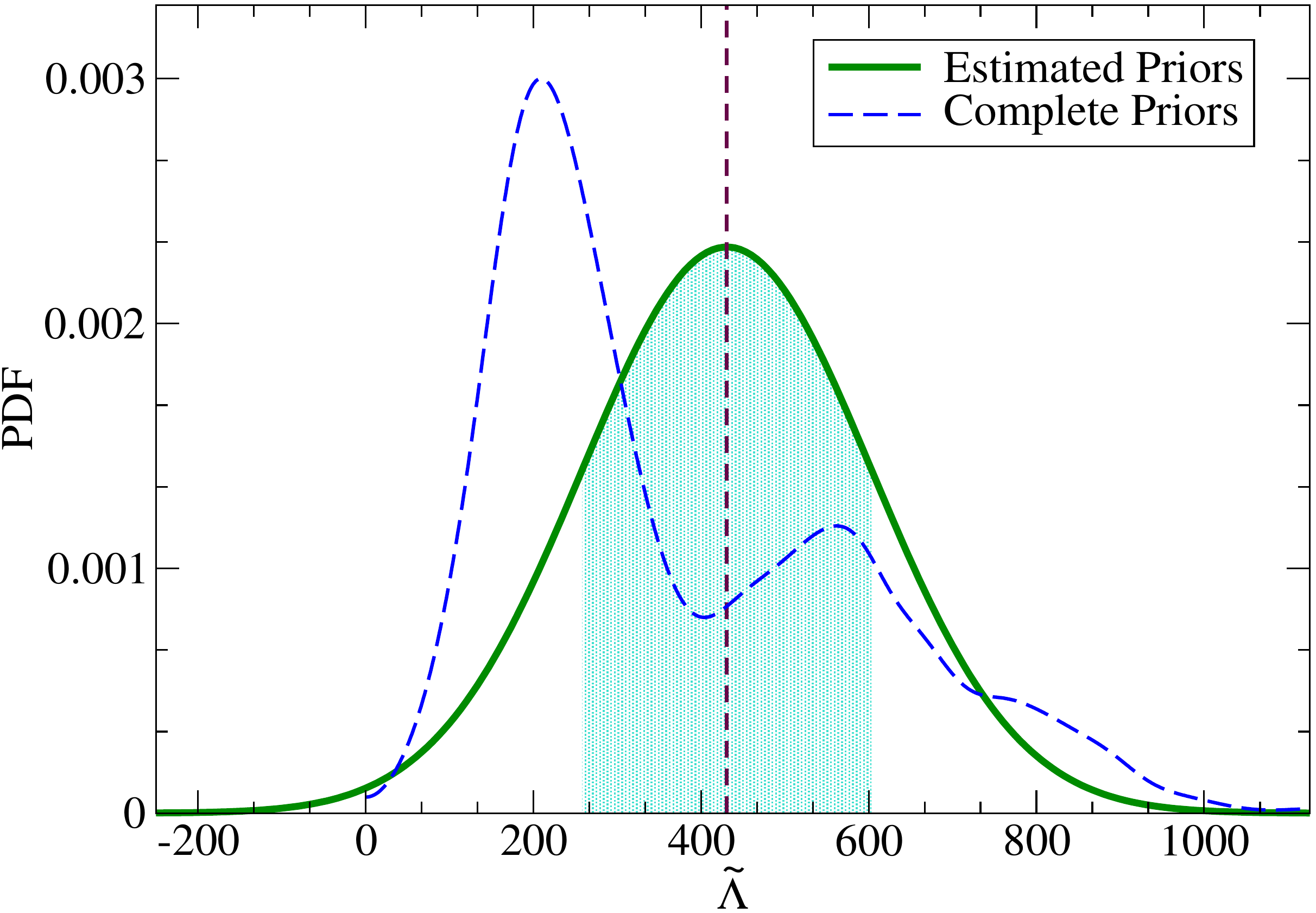}
\end{center}
\caption{
Prior distribution (solid green) on $\tilde\Lambda$ for O2 detector sensitivity with a chirp mass of $\mathcal{M}=1.188\text{ M}_{\odot}$.
This distribution, used to compute posteriors on $K_{\text{sym},0}$, is generated by assuming $\tilde\Lambda$ follows a Gaussian distribution in Eq.~\eqref{eq:PO2} with mean $\mu_{\tilde\Lambda}=430.8$ (dashed vertical line), and root-mean-square $\sigma_{\text{O2}}=172.5$ (cyan shaded region).
These are computed from the GW170817-constrained EoSs from Ref.~\cite{Zack:nuclearConstraints}, and from a simple Fisher analysis respectively.
Additionally shown in the figure is the posterior distribution on $\tilde\Lambda$ derived by the LIGO and Virgo Collaborations in Ref.~\cite{LIGO:posterior} (dashed blue).
}
\label{fig:LtPrior}
\end{figure}

In this appendix, we demonstrate the process of computing the posterior distribution on $K_{\text{sym},0}$ (used in Sec.~\ref{sec:chirpmass}) for one value of chirp mass, $\mathcal{M}=1.188\text{ M}_{\odot}$, corresponding to GW170817 on detector O2.
This case corresponds to the large dot in Fig.~\ref{fig:OverallVsSystematic}.
Referring to Figs.~\ref{fig:meanLt} and~\ref{fig:sigmaLt}, we observe that the mean and root-mean-square $\tilde\Lambda$ values for O2 detector sensitivity at $\mathcal{M}=1.188\text{ M}_{\odot}$ are given by $\mu_{\tilde\Lambda}=430.8$ and $\sigma_{\text{O2}}=172.5$, respectively.
This results in a prior distribution on $\tilde\Lambda$ shown in Fig.~\ref{fig:LtPrior}, given by
\begin{equation}
\label{eq:PO2}
P_{\text{O2}}(\tilde\Lambda)= \frac{1}{\sqrt{2\pi(172.5)^2}} e^{-(\tilde\Lambda-430.8)^2/2(172.5)^2}.
\end{equation}
We additionally show the true posterior distribution on $\tilde\Lambda$ derived in Ref.~\cite{LIGO:posterior}, which was used as a prior in our original analysis found in Ref.~\cite{Zack:nuclearConstraints} for comparison purposes.

\begin{figure}
\begin{center} 
\includegraphics[width=\columnwidth]{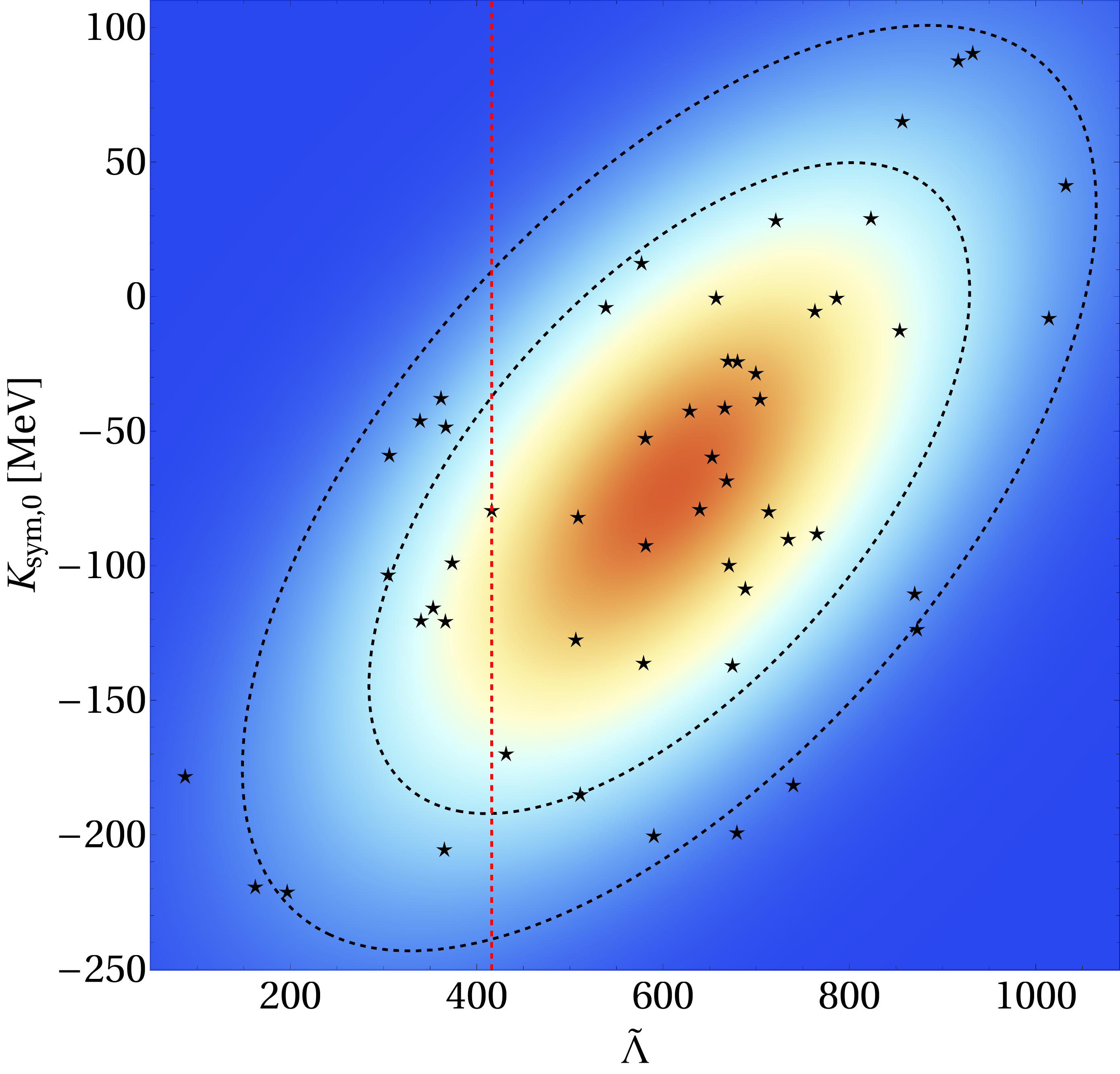}
\end{center}
\caption{
Two-dimensional normalized probability distribution between $K_{\text{sym},0}$ and $\tilde\Lambda$ as given by Eq.~\eqref{eq:2dPDF}, with the 68\% and 90\% confidence regions highlighted in black.
Overlaid on the distribution is the set of 58 data points corresponding to the various EoS models used in the analysis.
The vertical dashed line represents the mean of the prior $\tilde\Lambda$ distribution, at which the (one-side 90\% confidence interval) systematic error in $K_{\text{sym},0}$ is computed to be $\sigma_{\text{sys}}=104.6\text{ MeV}$.
}
\label{fig:probDistExample}
\end{figure}

Following along with Sec.~\ref{sec:futureSingle}, we generate the two-dimensional probability distribution between $K_{\text{sym},0}$ and $\tilde\Lambda$, given by Eq.~\eqref{eq:2dPDF}.
We find the covariance matrix and the mean vector to be
\begin{equation}
\mathbf{\Sigma}=\begin{pmatrix} 45610 & 10410 \\ 10410 & 6418 \end{pmatrix}, \quad
\bm{\mu}=\begin{pmatrix} 606.7 \\ -71.16 \end{pmatrix}\,,
\end{equation}
for $\bm x = (\tilde \Lambda, K_{\mathrm{sym,0}})$
This results in the two-dimensional probability distribution between $K_{\text{sym},0}$ and $\tilde\Lambda$ shown in Fig.~\ref{fig:probDistExample}.
The systematic error on $K_{\text{sym},0}$ is then computed by evaluating the 90\% confidence interval width of the distribution in the $K_{\text{sym},0}$-dimension at $\mu_{\tilde\Lambda}=430.8$, corresponding to the mean of the prior distribution in $\tilde\Lambda$.
The resulting (one-sided 90\% confidence level) systematic errors for this case are found to be $\sigma_{\text{sys}}=104.6\text{ MeV}$.

\begin{figure}
\begin{center} 
\includegraphics[width=\columnwidth]{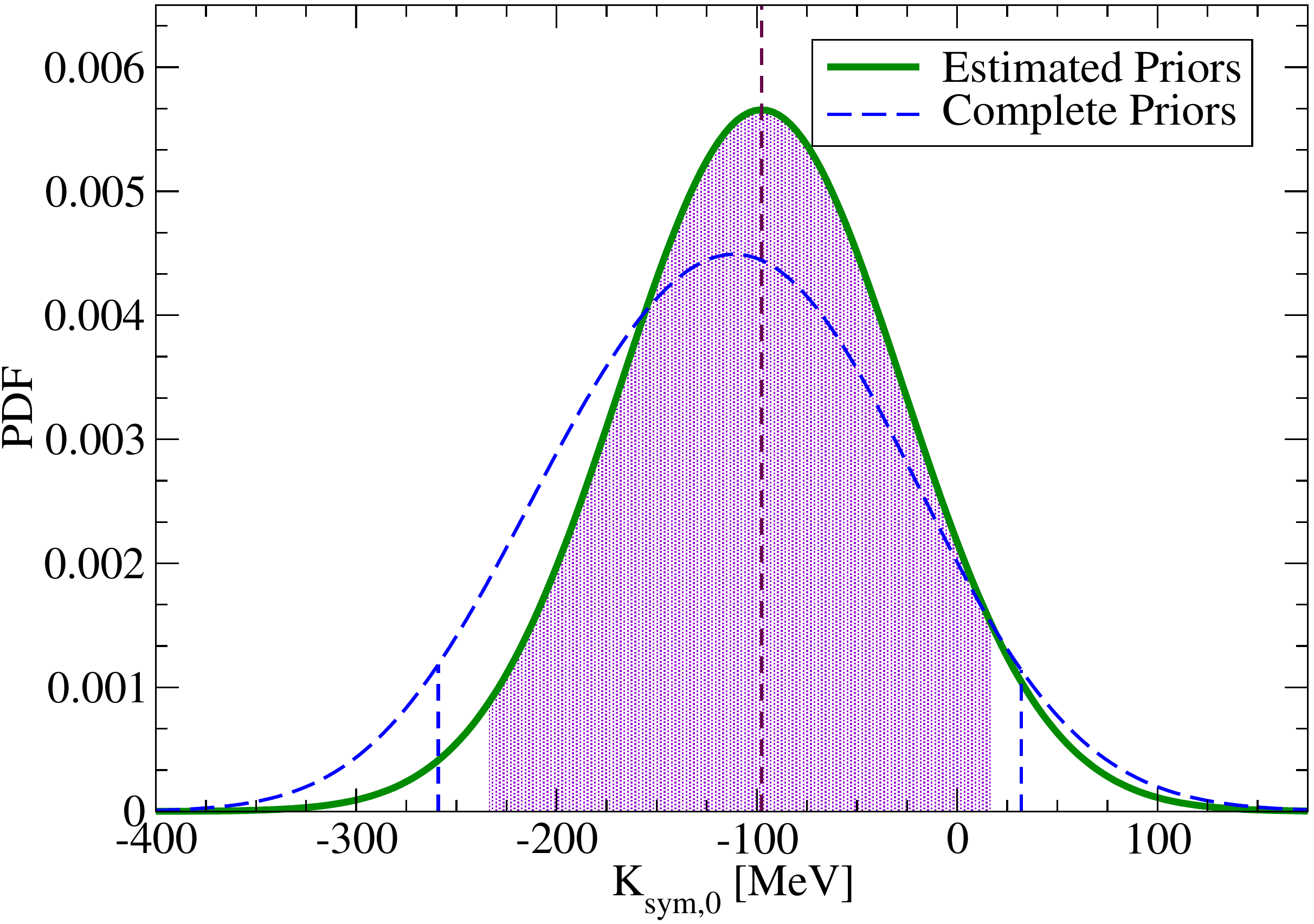}
\end{center}
\caption{
Resulting posterior distribution on $K_{\text{sym},0}$ (solid green), displaying a mean of $-97.63\text{ MeV}$ (dashed magenta) and a one-sided 90\% confidence interval of $116.0\text{ MeV}$ (shaded magenta).
Additionally shown (dashed blue) are the results found in Ref.~\cite{Zack:nuclearConstraints}, when using the full set of 121 EoSs and the full prior distribution in $\tilde\Lambda$, rather than the Gaussian estimation and restricted set of EoSs used here.
Observe that the approximation of Gaussian $\tilde\Lambda$ prior distributions slightly underestimates the uncertainties in $K_{\text{sym},0}$.
}
\label{fig:KsymPosterior}
\end{figure}

Next, we obtain the one-dimensional conditional probability distributions on $K_\text{sym,0}$ given tidal deformability observations of $\tilde\Lambda_\text{obs}$.
By following Eq.~\eqref{eq:conditional}, this is simply given by
\begin{equation}
P(K_\text{sym,0}|\tilde\Lambda_\text{obs})= \frac{\text{Exp}\left\lbrack-\frac{\left(K_\text{sym,0}-(0.228(\tilde\Lambda_\text{obs}-606.7)-71.16)\right)^2}{2(63.59)^2}\right\rbrack}{\sqrt{2\pi(63.59)^2}}.
\end{equation}
Finally, the posterior distribution on $K_{\text{sym},0}$ can be computed by combining the one-dimensional conditional probability distributions with the prior distribution on $\tilde\Lambda$, and then integrating over all observations of $\tilde\Lambda$:
\begin{equation}
P(K_{\text{sym},0})=\int\limits^{\infty}_{-\infty}P(K_{\text{sym},0}|\tilde{\Lambda})P_{\text{O2}}(\tilde\Lambda)d\tilde\Lambda.
\end{equation}
Figure~\ref{fig:KsymPosterior} finally displays the resulting posterior distribution on $K_{\text{sym},0}$, with a mean value of $-115^{+75}_{-73} \text{ MeV}$, giving a 90\% confidence interval of $-227\text{ MeV}\leq K_{\text{sym},0}\leq 7\text{ MeV}$ (or a one-sided $90\%$ confidence interval of $117$ MeV corresponding to the maroon dot in Fig.~\ref{fig:OverallVsSystematic}).
Comparing this to the resulting posterior distribution (also shown in Fig.~\ref{fig:KsymPosterior}) found in Ref.~\cite{Zack:nuclearConstraints} giving a 90\% confidence interval of $-285\text{ MeV}\leq K_{\text{sym},0}\leq 7\text{ MeV}$, we find that this approximation of $\tilde\Lambda$ priors slightly \textit{underestimates} the errors in $K_{\text{sym},0}$, but otherwise works quite well.
We also see that this approximation of $\tilde\Lambda$ priors skews the distribution less so than the LVC posterior, making it more normally distributed. 
We also note that here, we utilize a restricted set of EoSs compared to that done in Ref.~\cite{Zack:nuclearConstraints}, resulting in a slightly more accurate posterior distribution on $K_{\text{sym},0}$.

\section{Combining uncertainties from multiple events}
\label{app:multiple}

In this appendix, we explain how one can combine statistical uncertainties on certain parameters from multiple events. We closely follow App.~A of~\cite{Zack:URrelations}.
To estimate the number of detected binary NS merger events $N_A$ on interferometer $A$, we integrate the local binary NS merger rate across all redshift values within detector $A$'s horizon redshift $z_h$ with SNR thresholds of $\rho_{\text{th}}=8$.
Following the process used in Ref.~\cite{Zack:URrelations}, this is given by
\begin{equation}
N_A=\Delta \tau_0 \int\limits^{z_{h}}_0 4 \pi \lbrack  a_0r_1(z)\rbrack^2 \mathcal{R} r(z) \frac{d \tau}{dz} dz.
\end{equation}
In our cosmology, we choose $a_0r_1(z)$, $\frac{d\tau}{dz}$, and $r(z)$ to be
\begin{equation}
a_0r_1(z) = \frac{1}{H_0}\int\limits^z_0 \frac{dz'}{\sqrt{(1-\Omega_{\Lambda})(1+z')^3+\Omega_{\Lambda}}},
\end{equation}
\begin{equation}
\frac{d\tau}{dz} = \frac{1}{H_0} \frac{1}{1+z}\frac{1}{\sqrt{(1-\Omega_{\Lambda})(1+z')^3+\Omega_{\Lambda}}},
\end{equation}
\begin{equation}
r(z) = \left\{
\begin{array}{ll}
      1+2z & (z \leq 1) \\
      \frac{3}{4}(5-z) & (1\leq z\leq 5), \\
      0 & (z\geq 5)\\ 
\end{array} 
\right.
\end{equation}
where $H_0=70\text{ km s}^{-1}\text{ Mpc}^{-1}$ is the local Hubble constant, $\Omega_\Lambda=0.67$ is the universe's vacuum energy density, $\mathcal{R}=1540^{+3200}_{-1220}\text{ Gpc}^{-3}\text{ yr}^{-1}$ is the local binary NS coalescence rate density~\cite{Abbott2017}, and $\Delta\tau_0$ is chosen to be a one year observation period.
Doing so gives the rates found in Table VIII of Ref.~\cite{Zack:URrelations}. For demonstration purposes, we choose to use the lower and upper limits of these rates which are found to be $(2.0\times10^0,3.0\times10^2)$, $(1.6\times10^2,2.4\times10^3)$, $(2.2\times10^3,3.2\times10^4)$, $(7.2\times10^4,1.1\times10^6)$, and $(3.0\times10^5,4.4\times10^6)$ events per year for aLIGO, A\texttt{+}, Voyager, CE, and ET respectively.

Similarly following Ref.~\cite{Zack:URrelations}, the combined uncertainty $\sigma_{N_A}$ is computed by first simulating a population of $N_A$ events from the SNR ($\rho$) probability distribution:~\cite{Shutz:SNR,Chen:SNR}
\begin{equation}
f(\rho)=\frac{3\rho_{\text{th}}}{\rho^4},
\end{equation}
with an SNR threshold of $\rho_{\text{th}}=8$.
The combined population root-mean-square error on $\Lambda_{1.4}$ is then computed by integrating over all $N_A$ sources at various redshifts $z$:
\begin{equation}
\sigma_{N_A}^{-2}=\Delta \tau \int\limits^{z_h}_04 \pi \lbrack a_0 r_1(z)\rbrack^2 \mathcal{R}r(z)\frac{d\tau}{dz}\sigma^A(z)^{-2}dz.
\end{equation}
Here, $\sigma^A(z)$ is the redshift dependence of the root-mean-square error on detector $A$, evaluated via Fisher analyses at various redshifts up to $z_h$.
 

\bibliography{Zack}

 \newcommand{\noop}[1]{}
\begin{thebibliography}{76}%
\makeatletter
\providecommand \@ifxundefined [1]{%
 \@ifx{#1\undefined}
}%
\providecommand \@ifnum [1]{%
 \ifnum #1\expandafter \@firstoftwo
 \else \expandafter \@secondoftwo
 \fi
}%
\providecommand \@ifx [1]{%
 \ifx #1\expandafter \@firstoftwo
 \else \expandafter \@secondoftwo
 \fi
}%
\providecommand \natexlab [1]{#1}%
\providecommand \enquote  [1]{``#1''}%
\providecommand \bibnamefont  [1]{#1}%
\providecommand \bibfnamefont [1]{#1}%
\providecommand \citenamefont [1]{#1}%
\providecommand \href@noop [0]{\@secondoftwo}%
\providecommand \href [0]{\begingroup \@sanitize@url \@href}%
\providecommand \@href[1]{\@@startlink{#1}\@@href}%
\providecommand \@@href[1]{\endgroup#1\@@endlink}%
\providecommand \@sanitize@url [0]{\catcode `\\12\catcode `\$12\catcode
  `\&12\catcode `\#12\catcode `\^12\catcode `\_12\catcode `\%12\relax}%
\providecommand \@@startlink[1]{}%
\providecommand \@@endlink[0]{}%
\providecommand \url  [0]{\begingroup\@sanitize@url \@url }%
\providecommand \@url [1]{\endgroup\@href {#1}{\urlprefix }}%
\providecommand \urlprefix  [0]{URL }%
\providecommand \Eprint [0]{\href }%
\providecommand \doibase [0]{http://dx.doi.org/}%
\providecommand \selectlanguage [0]{\@gobble}%
\providecommand \bibinfo  [0]{\@secondoftwo}%
\providecommand \bibfield  [0]{\@secondoftwo}%
\providecommand \translation [1]{[#1]}%
\providecommand \BibitemOpen [0]{}%
\providecommand \bibitemStop [0]{}%
\providecommand \bibitemNoStop [0]{.\EOS\space}%
\providecommand \EOS [0]{\spacefactor3000\relax}%
\providecommand \BibitemShut  [1]{\csname bibitem#1\endcsname}%
\let\auto@bib@innerbib\@empty
\bibitem [{\citenamefont {Guver}\ and\ \citenamefont {Ozel}(2013)}]{guver}%
  \BibitemOpen
  \bibfield  {author} {\bibinfo {author} {\bibfnamefont {T.}~\bibnamefont
  {Guver}}\ and\ \bibinfo {author} {\bibfnamefont {F.}~\bibnamefont {Ozel}},\
  }\href {\doibase 10.1088/2041-8205/765/1/L1} {\bibfield  {journal} {\bibinfo
  {journal} {Astrophys. J.}\ }\textbf {\bibinfo {volume} {765}},\ \bibinfo
  {pages} {L1} (\bibinfo {year} {2013})},\ \Eprint
  {http://arxiv.org/abs/1301.0831} {arXiv:1301.0831 [astro-ph.HE]} \BibitemShut
  {NoStop}%
\bibitem [{\citenamefont {Ozel}\ \emph {et~al.}(2010)\citenamefont {Ozel},
  \citenamefont {Baym},\ and\ \citenamefont {Guver}}]{ozel-baym-guver}%
  \BibitemOpen
  \bibfield  {author} {\bibinfo {author} {\bibfnamefont {F.}~\bibnamefont
  {Ozel}}, \bibinfo {author} {\bibfnamefont {G.}~\bibnamefont {Baym}}, \ and\
  \bibinfo {author} {\bibfnamefont {T.}~\bibnamefont {Guver}},\ }\href
  {\doibase 10.1103/PhysRevD.82.101301} {\bibfield  {journal} {\bibinfo
  {journal} {Phys.Rev.}\ }\textbf {\bibinfo {volume} {D82}},\ \bibinfo {pages}
  {101301} (\bibinfo {year} {2010})},\ \Eprint {http://arxiv.org/abs/1002.3153}
  {arXiv:1002.3153 [astro-ph.HE]} \BibitemShut {NoStop}%
\bibitem [{\citenamefont {Steiner}\ \emph {et~al.}(2010)\citenamefont
  {Steiner}, \citenamefont {Lattimer},\ and\ \citenamefont
  {Brown}}]{steiner-lattimer-brown}%
  \BibitemOpen
  \bibfield  {author} {\bibinfo {author} {\bibfnamefont {A.~W.}\ \bibnamefont
  {Steiner}}, \bibinfo {author} {\bibfnamefont {J.~M.}\ \bibnamefont
  {Lattimer}}, \ and\ \bibinfo {author} {\bibfnamefont {E.~F.}\ \bibnamefont
  {Brown}},\ }\href {\doibase 10.1088/0004-637X/722/1/33} {\bibfield  {journal}
  {\bibinfo  {journal} {Astrophys.J.}\ }\textbf {\bibinfo {volume} {722}},\
  \bibinfo {pages} {33} (\bibinfo {year} {2010})}\BibitemShut {NoStop}%
\bibitem [{\citenamefont {Lattimer}\ and\ \citenamefont
  {Steiner}(2014)}]{Lattimer2014}%
  \BibitemOpen
  \bibfield  {author} {\bibinfo {author} {\bibfnamefont {J.~M.}\ \bibnamefont
  {Lattimer}}\ and\ \bibinfo {author} {\bibfnamefont {A.~W.}\ \bibnamefont
  {Steiner}},\ }\href {\doibase 10.1140/epja/i2014-14040-y} {\bibfield
  {journal} {\bibinfo  {journal} {The European Physical Journal A}\ }\textbf
  {\bibinfo {volume} {50}} (\bibinfo {year} {2014}),\
  10.1140/epja/i2014-14040-y}\BibitemShut {NoStop}%
\bibitem [{\citenamefont {Ozel}\ and\ \citenamefont
  {Freire}(2016)}]{Ozel:2016oaf}%
  \BibitemOpen
  \bibfield  {author} {\bibinfo {author} {\bibfnamefont {F.}~\bibnamefont
  {Ozel}}\ and\ \bibinfo {author} {\bibfnamefont {P.}~\bibnamefont {Freire}},\
  }\href {\doibase 10.1146/annurev-astro-081915-023322} {\bibfield  {journal}
  {\bibinfo  {journal} {Ann. Rev. Astron. Astrophys.}\ }\textbf {\bibinfo
  {volume} {54}},\ \bibinfo {pages} {401} (\bibinfo {year} {2016})},\ \Eprint
  {http://arxiv.org/abs/1603.02698} {arXiv:1603.02698 [astro-ph.HE]}
  \BibitemShut {NoStop}%
\bibitem [{\citenamefont {Abbott}\ \emph
  {et~al.}(2017{\natexlab{a}})\citenamefont {Abbott} \emph
  {et~al.}}]{TheLIGOScientific:2017qsa}%
  \BibitemOpen
  \bibfield  {author} {\bibinfo {author} {\bibfnamefont {B.~P.}\ \bibnamefont
  {Abbott}} \emph {et~al.} (\bibinfo {collaboration} {Virgo, LIGO
  Scientific}),\ }\href {\doibase 10.1103/PhysRevLett.119.161101} {\bibfield
  {journal} {\bibinfo  {journal} {Phys. Rev. Lett.}\ }\textbf {\bibinfo
  {volume} {119}},\ \bibinfo {pages} {161101} (\bibinfo {year}
  {2017}{\natexlab{a}})},\ \Eprint {http://arxiv.org/abs/1710.05832}
  {arXiv:1710.05832 [gr-qc]} \BibitemShut {NoStop}%
\bibitem [{\citenamefont {Abbott}\ \emph {et~al.}(2019)\citenamefont {Abbott}
  \emph {et~al.}}]{Abbott2018}%
  \BibitemOpen
  \bibfield  {author} {\bibinfo {author} {\bibfnamefont {B.~P.}\ \bibnamefont
  {Abbott}} \emph {et~al.} (\bibinfo {collaboration} {LIGO Scientific,
  Virgo}),\ }\href {\doibase 10.1103/PhysRevX.9.011001} {\bibfield  {journal}
  {\bibinfo  {journal} {Phys. Rev.}\ }\textbf {\bibinfo {volume} {X9}},\
  \bibinfo {pages} {011001} (\bibinfo {year} {2019})},\ \Eprint
  {http://arxiv.org/abs/1805.11579} {arXiv:1805.11579 [gr-qc]} \BibitemShut
  {NoStop}%
\bibitem [{\citenamefont {Abbott}\ \emph
  {et~al.}(2018{\natexlab{a}})\citenamefont {Abbott} \emph
  {et~al.}}]{Abbott:2018exr}%
  \BibitemOpen
  \bibfield  {author} {\bibinfo {author} {\bibfnamefont {B.~P.}\ \bibnamefont
  {Abbott}} \emph {et~al.} (\bibinfo {collaboration} {LIGO Scientific,
  Virgo}),\ }\href {\doibase 10.1103/PhysRevLett.121.161101} {\bibfield
  {journal} {\bibinfo  {journal} {Phys. Rev. Lett.}\ }\textbf {\bibinfo
  {volume} {121}},\ \bibinfo {pages} {161101} (\bibinfo {year}
  {2018}{\natexlab{a}})},\ \Eprint {http://arxiv.org/abs/1805.11581}
  {arXiv:1805.11581 [gr-qc]} \BibitemShut {NoStop}%
\bibitem [{\citenamefont {Paschalidis}\ \emph {et~al.}(2018)\citenamefont
  {Paschalidis}, \citenamefont {Yagi}, \citenamefont {Alvarez-Castillo},
  \citenamefont {Blaschke},\ and\ \citenamefont {Sedrakian}}]{Paschalidis2018}%
  \BibitemOpen
  \bibfield  {author} {\bibinfo {author} {\bibfnamefont {V.}~\bibnamefont
  {Paschalidis}}, \bibinfo {author} {\bibfnamefont {K.}~\bibnamefont {Yagi}},
  \bibinfo {author} {\bibfnamefont {D.}~\bibnamefont {Alvarez-Castillo}},
  \bibinfo {author} {\bibfnamefont {D.~B.}\ \bibnamefont {Blaschke}}, \ and\
  \bibinfo {author} {\bibfnamefont {A.}~\bibnamefont {Sedrakian}},\ }\href
  {\doibase 10.1103/physrevd.97.084038} {\bibfield  {journal} {\bibinfo
  {journal} {Physical Review D}\ }\textbf {\bibinfo {volume} {97}} (\bibinfo
  {year} {2018}),\ 10.1103/physrevd.97.084038}\BibitemShut {NoStop}%
\bibitem [{\citenamefont {Burgio}\ \emph {et~al.}(2018)\citenamefont {Burgio},
  \citenamefont {Drago}, \citenamefont {Pagliara}, \citenamefont {Schulze},\
  and\ \citenamefont {Wei}}]{Burgio2018}%
  \BibitemOpen
  \bibfield  {author} {\bibinfo {author} {\bibfnamefont {G.~F.}\ \bibnamefont
  {Burgio}}, \bibinfo {author} {\bibfnamefont {A.}~\bibnamefont {Drago}},
  \bibinfo {author} {\bibfnamefont {G.}~\bibnamefont {Pagliara}}, \bibinfo
  {author} {\bibfnamefont {H.~J.}\ \bibnamefont {Schulze}}, \ and\ \bibinfo
  {author} {\bibfnamefont {J.~B.}\ \bibnamefont {Wei}},\ }\href
  {http://arxiv.org/pdf/1803.09696v1} {\bibfield  {journal} {\bibinfo
  {journal} {Arxiv}\ } (\bibinfo {year} {2018})},\ \Eprint
  {http://arxiv.org/abs/1803.09696v1} {1803.09696v1} \BibitemShut {NoStop}%
\bibitem [{\citenamefont {Malik}\ \emph {et~al.}(2018)\citenamefont {Malik},
  \citenamefont {Alam}, \citenamefont {Fortin}, \citenamefont {ProvidÃªncia},
  \citenamefont {Agrawal}, \citenamefont {Jha}, \citenamefont {Kumar},\ and\
  \citenamefont {Patra}}]{Malik2018}%
  \BibitemOpen
  \bibfield  {author} {\bibinfo {author} {\bibfnamefont {T.}~\bibnamefont
  {Malik}}, \bibinfo {author} {\bibfnamefont {N.}~\bibnamefont {Alam}},
  \bibinfo {author} {\bibfnamefont {M.}~\bibnamefont {Fortin}}, \bibinfo
  {author} {\bibfnamefont {C.}~\bibnamefont {ProvidÃªncia}}, \bibinfo
  {author} {\bibfnamefont {B.~K.}\ \bibnamefont {Agrawal}}, \bibinfo {author}
  {\bibfnamefont {T.~K.}\ \bibnamefont {Jha}}, \bibinfo {author} {\bibfnamefont
  {B.}~\bibnamefont {Kumar}}, \ and\ \bibinfo {author} {\bibfnamefont {S.~K.}\
  \bibnamefont {Patra}},\ }\href {\doibase 10.1103/PhysRevC.98.035804}
  {\bibfield  {journal} {\bibinfo  {journal} {Phys. Rev.}\ }\textbf {\bibinfo
  {volume} {C98}},\ \bibinfo {pages} {035804} (\bibinfo {year} {2018})},\
  \Eprint {http://arxiv.org/abs/1805.11963} {arXiv:1805.11963 [nucl-th]}
  \BibitemShut {NoStop}%
\bibitem [{\citenamefont {Landry}\ and\ \citenamefont
  {Essick}(2019)}]{Landry:2018prl}%
  \BibitemOpen
  \bibfield  {author} {\bibinfo {author} {\bibfnamefont {P.}~\bibnamefont
  {Landry}}\ and\ \bibinfo {author} {\bibfnamefont {R.}~\bibnamefont
  {Essick}},\ }\href {\doibase 10.1103/PhysRevD.99.084049} {\bibfield
  {journal} {\bibinfo  {journal} {Phys. Rev.}\ }\textbf {\bibinfo {volume}
  {D99}},\ \bibinfo {pages} {084049} (\bibinfo {year} {2019})},\ \Eprint
  {http://arxiv.org/abs/1811.12529} {arXiv:1811.12529 [gr-qc]} \BibitemShut
  {NoStop}%
\bibitem [{\citenamefont {Flanagan}\ and\ \citenamefont
  {Hinderer}(2008)}]{Flanagan2008}%
  \BibitemOpen
  \bibfield  {author} {\bibinfo {author} {\bibfnamefont {{\'{E}}.~{\'{E}}.}\
  \bibnamefont {Flanagan}}\ and\ \bibinfo {author} {\bibfnamefont
  {T.}~\bibnamefont {Hinderer}},\ }\href {\doibase 10.1103/physrevd.77.021502}
  {\bibfield  {journal} {\bibinfo  {journal} {Physical Review D}\ }\textbf
  {\bibinfo {volume} {77}} (\bibinfo {year} {2008}),\
  10.1103/physrevd.77.021502}\BibitemShut {NoStop}%
\bibitem [{\citenamefont {De}\ \emph {et~al.}(2018)\citenamefont {De},
  \citenamefont {Finstad}, \citenamefont {Lattimer}, \citenamefont {Brown},
  \citenamefont {Berger},\ and\ \citenamefont {Biwer}}]{De:2018uhw}%
  \BibitemOpen
  \bibfield  {author} {\bibinfo {author} {\bibfnamefont {S.}~\bibnamefont
  {De}}, \bibinfo {author} {\bibfnamefont {D.}~\bibnamefont {Finstad}},
  \bibinfo {author} {\bibfnamefont {J.~M.}\ \bibnamefont {Lattimer}}, \bibinfo
  {author} {\bibfnamefont {D.~A.}\ \bibnamefont {Brown}}, \bibinfo {author}
  {\bibfnamefont {E.}~\bibnamefont {Berger}}, \ and\ \bibinfo {author}
  {\bibfnamefont {C.~M.}\ \bibnamefont {Biwer}},\ }\href {\doibase
  10.1103/PhysRevLett.121.259902, 10.1103/PhysRevLett.121.091102} {\bibfield
  {journal} {\bibinfo  {journal} {Phys. Rev. Lett.}\ }\textbf {\bibinfo
  {volume} {121}},\ \bibinfo {pages} {091102} (\bibinfo {year} {2018})},\
  \bibinfo {note} {[Erratum: Phys. Rev. Lett.121,no.25,259902(2018)]},\ \Eprint
  {http://arxiv.org/abs/1804.08583} {arXiv:1804.08583 [astro-ph.HE]}
  \BibitemShut {NoStop}%
\bibitem [{\citenamefont {Annala}\ \emph {et~al.}(2018)\citenamefont {Annala},
  \citenamefont {Gorda}, \citenamefont {Kurkela},\ and\ \citenamefont
  {Vuorinen}}]{Annala:2017llu}%
  \BibitemOpen
  \bibfield  {author} {\bibinfo {author} {\bibfnamefont {E.}~\bibnamefont
  {Annala}}, \bibinfo {author} {\bibfnamefont {T.}~\bibnamefont {Gorda}},
  \bibinfo {author} {\bibfnamefont {A.}~\bibnamefont {Kurkela}}, \ and\
  \bibinfo {author} {\bibfnamefont {A.}~\bibnamefont {Vuorinen}},\ }\href
  {\doibase 10.1103/PhysRevLett.120.172703} {\bibfield  {journal} {\bibinfo
  {journal} {Phys. Rev. Lett.}\ }\textbf {\bibinfo {volume} {120}},\ \bibinfo
  {pages} {172703} (\bibinfo {year} {2018})},\ \Eprint
  {http://arxiv.org/abs/1711.02644} {arXiv:1711.02644 [astro-ph.HE]}
  \BibitemShut {NoStop}%
\bibitem [{\citenamefont {Lim}\ and\ \citenamefont {Holt}(2018)}]{Lim:2018bkq}%
  \BibitemOpen
  \bibfield  {author} {\bibinfo {author} {\bibfnamefont {Y.}~\bibnamefont
  {Lim}}\ and\ \bibinfo {author} {\bibfnamefont {J.~W.}\ \bibnamefont {Holt}},\
  }\href {\doibase 10.1103/PhysRevLett.121.062701} {\bibfield  {journal}
  {\bibinfo  {journal} {Phys. Rev. Lett.}\ }\textbf {\bibinfo {volume} {121}},\
  \bibinfo {pages} {062701} (\bibinfo {year} {2018})},\ \Eprint
  {http://arxiv.org/abs/1803.02803} {arXiv:1803.02803 [nucl-th]} \BibitemShut
  {NoStop}%
\bibitem [{\citenamefont {Bauswein}\ \emph {et~al.}(2017)\citenamefont
  {Bauswein}, \citenamefont {Just}, \citenamefont {Janka},\ and\ \citenamefont
  {Stergioulas}}]{Bauswein:2017vtn}%
  \BibitemOpen
  \bibfield  {author} {\bibinfo {author} {\bibfnamefont {A.}~\bibnamefont
  {Bauswein}}, \bibinfo {author} {\bibfnamefont {O.}~\bibnamefont {Just}},
  \bibinfo {author} {\bibfnamefont {H.-T.}\ \bibnamefont {Janka}}, \ and\
  \bibinfo {author} {\bibfnamefont {N.}~\bibnamefont {Stergioulas}},\ }\href
  {\doibase 10.3847/2041-8213/aa9994} {\bibfield  {journal} {\bibinfo
  {journal} {Astrophys. J.}\ }\textbf {\bibinfo {volume} {850}},\ \bibinfo
  {pages} {L34} (\bibinfo {year} {2017})},\ \Eprint
  {http://arxiv.org/abs/1710.06843} {arXiv:1710.06843 [astro-ph.HE]}
  \BibitemShut {NoStop}%
\bibitem [{\citenamefont {Most}\ \emph {et~al.}(2018)\citenamefont {Most},
  \citenamefont {Weih}, \citenamefont {Rezzolla},\ and\ \citenamefont
  {Schaffner-Bielich}}]{Most:2018hfd}%
  \BibitemOpen
  \bibfield  {author} {\bibinfo {author} {\bibfnamefont {E.~R.}\ \bibnamefont
  {Most}}, \bibinfo {author} {\bibfnamefont {L.~R.}\ \bibnamefont {Weih}},
  \bibinfo {author} {\bibfnamefont {L.}~\bibnamefont {Rezzolla}}, \ and\
  \bibinfo {author} {\bibfnamefont {J.}~\bibnamefont {Schaffner-Bielich}},\
  }\href {\doibase 10.1103/PhysRevLett.120.261103} {\bibfield  {journal}
  {\bibinfo  {journal} {Phys. Rev. Lett.}\ }\textbf {\bibinfo {volume} {120}},\
  \bibinfo {pages} {261103} (\bibinfo {year} {2018})},\ \Eprint
  {http://arxiv.org/abs/1803.00549} {arXiv:1803.00549 [gr-qc]} \BibitemShut
  {NoStop}%
\bibitem [{\citenamefont {Read}\ \emph {et~al.}(2009)\citenamefont {Read},
  \citenamefont {Lackey}, \citenamefont {Owen},\ and\ \citenamefont
  {Friedman}}]{Read2009}%
  \BibitemOpen
  \bibfield  {author} {\bibinfo {author} {\bibfnamefont {J.~S.}\ \bibnamefont
  {Read}}, \bibinfo {author} {\bibfnamefont {B.~D.}\ \bibnamefont {Lackey}},
  \bibinfo {author} {\bibfnamefont {B.~J.}\ \bibnamefont {Owen}}, \ and\
  \bibinfo {author} {\bibfnamefont {J.~L.}\ \bibnamefont {Friedman}},\ }\href
  {\doibase 10.1103/physrevd.79.124032} {\bibfield  {journal} {\bibinfo
  {journal} {Physical Review D}\ }\textbf {\bibinfo {volume} {79}} (\bibinfo
  {year} {2009}),\ 10.1103/physrevd.79.124032}\BibitemShut {NoStop}%
\bibitem [{\citenamefont {Lackey}\ and\ \citenamefont
  {Wade}(2015)}]{Lackey:2014fwa}%
  \BibitemOpen
  \bibfield  {author} {\bibinfo {author} {\bibfnamefont {B.~D.}\ \bibnamefont
  {Lackey}}\ and\ \bibinfo {author} {\bibfnamefont {L.}~\bibnamefont {Wade}},\
  }\href {\doibase 10.1103/PhysRevD.91.043002} {\bibfield  {journal} {\bibinfo
  {journal} {Phys. Rev.}\ }\textbf {\bibinfo {volume} {D91}},\ \bibinfo {pages}
  {043002} (\bibinfo {year} {2015})},\ \Eprint {http://arxiv.org/abs/1410.8866}
  {arXiv:1410.8866 [gr-qc]} \BibitemShut {NoStop}%
\bibitem [{\citenamefont {Carney}\ \emph {et~al.}(2018)\citenamefont {Carney},
  \citenamefont {Wade},\ and\ \citenamefont {Irwin}}]{Carney:2018sdv}%
  \BibitemOpen
  \bibfield  {author} {\bibinfo {author} {\bibfnamefont {M.~F.}\ \bibnamefont
  {Carney}}, \bibinfo {author} {\bibfnamefont {L.~E.}\ \bibnamefont {Wade}}, \
  and\ \bibinfo {author} {\bibfnamefont {B.~S.}\ \bibnamefont {Irwin}},\ }\href
  {\doibase 10.1103/PhysRevD.98.063004} {\bibfield  {journal} {\bibinfo
  {journal} {Phys. Rev.}\ }\textbf {\bibinfo {volume} {D98}},\ \bibinfo {pages}
  {063004} (\bibinfo {year} {2018})},\ \Eprint
  {http://arxiv.org/abs/1805.11217} {arXiv:1805.11217 [gr-qc]} \BibitemShut
  {NoStop}%
\bibitem [{\citenamefont {Lindblom}(2010)}]{Lindblom:2010bb}%
  \BibitemOpen
  \bibfield  {author} {\bibinfo {author} {\bibfnamefont {L.}~\bibnamefont
  {Lindblom}},\ }\href {\doibase 10.1103/PhysRevD.82.103011} {\bibfield
  {journal} {\bibinfo  {journal} {Phys. Rev.}\ }\textbf {\bibinfo {volume}
  {D82}},\ \bibinfo {pages} {103011} (\bibinfo {year} {2010})},\ \Eprint
  {http://arxiv.org/abs/1009.0738} {arXiv:1009.0738 [astro-ph.HE]} \BibitemShut
  {NoStop}%
\bibitem [{\citenamefont {Lindblom}\ and\ \citenamefont
  {Indik}(2012)}]{Lindblom:2012zi}%
  \BibitemOpen
  \bibfield  {author} {\bibinfo {author} {\bibfnamefont {L.}~\bibnamefont
  {Lindblom}}\ and\ \bibinfo {author} {\bibfnamefont {N.~M.}\ \bibnamefont
  {Indik}},\ }\href {\doibase 10.1103/PhysRevD.86.084003} {\bibfield  {journal}
  {\bibinfo  {journal} {Phys. Rev.}\ }\textbf {\bibinfo {volume} {D86}},\
  \bibinfo {pages} {084003} (\bibinfo {year} {2012})},\ \Eprint
  {http://arxiv.org/abs/1207.3744} {arXiv:1207.3744 [astro-ph.HE]} \BibitemShut
  {NoStop}%
\bibitem [{\citenamefont {Lindblom}\ and\ \citenamefont
  {Indik}(2014)}]{Lindblom:2013kra}%
  \BibitemOpen
  \bibfield  {author} {\bibinfo {author} {\bibfnamefont {L.}~\bibnamefont
  {Lindblom}}\ and\ \bibinfo {author} {\bibfnamefont {N.~M.}\ \bibnamefont
  {Indik}},\ }\href {\doibase 10.1103/PhysRevD.89.064003,
  10.1103/PhysRevD.93.129903} {\bibfield  {journal} {\bibinfo  {journal} {Phys.
  Rev.}\ }\textbf {\bibinfo {volume} {D89}},\ \bibinfo {pages} {064003}
  (\bibinfo {year} {2014})},\ \bibinfo {note} {[Erratum: Phys.
  Rev.D93,no.12,129903(2016)]},\ \Eprint {http://arxiv.org/abs/1310.0803}
  {arXiv:1310.0803 [astro-ph.HE]} \BibitemShut {NoStop}%
\bibitem [{\citenamefont {Lindblom}(2018)}]{Lindblom:2018rfr}%
  \BibitemOpen
  \bibfield  {author} {\bibinfo {author} {\bibfnamefont {L.}~\bibnamefont
  {Lindblom}},\ }\href {\doibase 10.1103/PhysRevD.97.123019} {\bibfield
  {journal} {\bibinfo  {journal} {Phys. Rev.}\ }\textbf {\bibinfo {volume}
  {D97}},\ \bibinfo {pages} {123019} (\bibinfo {year} {2018})},\ \Eprint
  {http://arxiv.org/abs/1804.04072} {arXiv:1804.04072 [astro-ph.HE]}
  \BibitemShut {NoStop}%
\bibitem [{\citenamefont {Kumar}\ and\ \citenamefont
  {Landry}(2019)}]{Kumar:2019xgp}%
  \BibitemOpen
  \bibfield  {author} {\bibinfo {author} {\bibfnamefont {B.}~\bibnamefont
  {Kumar}}\ and\ \bibinfo {author} {\bibfnamefont {P.}~\bibnamefont {Landry}},\
  }\href@noop {} {\  (\bibinfo {year} {2019})},\ \Eprint
  {http://arxiv.org/abs/1902.04557} {arXiv:1902.04557 [gr-qc]} \BibitemShut
  {NoStop}%
\bibitem [{\citenamefont {Myers}\ and\ \citenamefont
  {Swiatecki}(1969)}]{Myers:1969zz}%
  \BibitemOpen
  \bibfield  {author} {\bibinfo {author} {\bibfnamefont {W.~D.}\ \bibnamefont
  {Myers}}\ and\ \bibinfo {author} {\bibfnamefont {W.~J.}\ \bibnamefont
  {Swiatecki}},\ }\href {\doibase 10.1016/0003-4916(69)90202-4} {\bibfield
  {journal} {\bibinfo  {journal} {Annals Phys.}\ }\textbf {\bibinfo {volume}
  {55}},\ \bibinfo {pages} {395} (\bibinfo {year} {1969})}\BibitemShut
  {NoStop}%
\bibitem [{\citenamefont {Vida{\~{n}}a}\ \emph {et~al.}(2009)\citenamefont
  {Vida{\~{n}}a}, \citenamefont {Provid{\^{e}}ncia}, \citenamefont {Polls},\
  and\ \citenamefont {Rios}}]{Vidana2009}%
  \BibitemOpen
  \bibfield  {author} {\bibinfo {author} {\bibfnamefont {I.}~\bibnamefont
  {Vida{\~{n}}a}}, \bibinfo {author} {\bibfnamefont {C.}~\bibnamefont
  {Provid{\^{e}}ncia}}, \bibinfo {author} {\bibfnamefont {A.}~\bibnamefont
  {Polls}}, \ and\ \bibinfo {author} {\bibfnamefont {A.}~\bibnamefont {Rios}},\
  }\href {\doibase 10.1103/physrevc.80.045806} {\bibfield  {journal} {\bibinfo
  {journal} {Physical Review C}\ }\textbf {\bibinfo {volume} {80}} (\bibinfo
  {year} {2009}),\ 10.1103/physrevc.80.045806}\BibitemShut {NoStop}%
\bibitem [{\citenamefont {Alam}\ \emph {et~al.}(2016)\citenamefont {Alam},
  \citenamefont {Agrawal}, \citenamefont {Fortin}, \citenamefont {Pais},
  \citenamefont {Provid{\^{e}}ncia}, \citenamefont {Raduta},\ and\
  \citenamefont {Sulaksono}}]{Alam2016}%
  \BibitemOpen
  \bibfield  {author} {\bibinfo {author} {\bibfnamefont {N.}~\bibnamefont
  {Alam}}, \bibinfo {author} {\bibfnamefont {B.~K.}\ \bibnamefont {Agrawal}},
  \bibinfo {author} {\bibfnamefont {M.}~\bibnamefont {Fortin}}, \bibinfo
  {author} {\bibfnamefont {H.}~\bibnamefont {Pais}}, \bibinfo {author}
  {\bibfnamefont {C.}~\bibnamefont {Provid{\^{e}}ncia}}, \bibinfo {author}
  {\bibfnamefont {A.~R.}\ \bibnamefont {Raduta}}, \ and\ \bibinfo {author}
  {\bibfnamefont {A.}~\bibnamefont {Sulaksono}},\ }\href {\doibase
  10.1103/physrevc.94.052801} {\bibfield  {journal} {\bibinfo  {journal}
  {Physical Review C}\ }\textbf {\bibinfo {volume} {94}} (\bibinfo {year}
  {2016}),\ 10.1103/physrevc.94.052801}\BibitemShut {NoStop}%
\bibitem [{\citenamefont {Sotani}\ \emph {et~al.}(2014)\citenamefont {Sotani},
  \citenamefont {Iida}, \citenamefont {Oyamatsu},\ and\ \citenamefont
  {Ohnishi}}]{Sotani:2013dga}%
  \BibitemOpen
  \bibfield  {author} {\bibinfo {author} {\bibfnamefont {H.}~\bibnamefont
  {Sotani}}, \bibinfo {author} {\bibfnamefont {K.}~\bibnamefont {Iida}},
  \bibinfo {author} {\bibfnamefont {K.}~\bibnamefont {Oyamatsu}}, \ and\
  \bibinfo {author} {\bibfnamefont {A.}~\bibnamefont {Ohnishi}},\ }\href
  {\doibase 10.1093/ptep/ptu052} {\bibfield  {journal} {\bibinfo  {journal}
  {PTEP}\ }\textbf {\bibinfo {volume} {2014}},\ \bibinfo {pages} {051E01}
  (\bibinfo {year} {2014})},\ \Eprint {http://arxiv.org/abs/1401.0161}
  {arXiv:1401.0161 [astro-ph.HE]} \BibitemShut {NoStop}%
\bibitem [{\citenamefont {Silva}\ \emph {et~al.}(2016)\citenamefont {Silva},
  \citenamefont {Sotani},\ and\ \citenamefont {Berti}}]{Silva:2016myw}%
  \BibitemOpen
  \bibfield  {author} {\bibinfo {author} {\bibfnamefont {H.~O.}\ \bibnamefont
  {Silva}}, \bibinfo {author} {\bibfnamefont {H.}~\bibnamefont {Sotani}}, \
  and\ \bibinfo {author} {\bibfnamefont {E.}~\bibnamefont {Berti}},\ }\href
  {\doibase 10.1093/mnras/stw969} {\bibfield  {journal} {\bibinfo  {journal}
  {Mon. Not. Roy. Astron. Soc.}\ }\textbf {\bibinfo {volume} {459}},\ \bibinfo
  {pages} {4378} (\bibinfo {year} {2016})},\ \Eprint
  {http://arxiv.org/abs/1601.03407} {arXiv:1601.03407 [astro-ph.HE]}
  \BibitemShut {NoStop}%
\bibitem [{\citenamefont {Abbott}\ \emph
  {et~al.}(2017{\natexlab{b}})\citenamefont {Abbott} \emph
  {et~al.}}]{Abbott2017}%
  \BibitemOpen
  \bibfield  {author} {\bibinfo {author} {\bibfnamefont {B.}~\bibnamefont
  {Abbott}} \emph {et~al.},\ }\href {\doibase 10.1103/physrevlett.119.161101}
  {\bibfield  {journal} {\bibinfo  {journal} {Physical Review Letters}\
  }\textbf {\bibinfo {volume} {119}} (\bibinfo {year} {2017}{\natexlab{b}}),\
  10.1103/physrevlett.119.161101}\BibitemShut {NoStop}%
\bibitem [{\citenamefont {Radice}\ \emph {et~al.}(2018)\citenamefont {Radice},
  \citenamefont {Perego}, \citenamefont {Zappa},\ and\ \citenamefont
  {Bernuzzi}}]{Radice2018}%
  \BibitemOpen
  \bibfield  {author} {\bibinfo {author} {\bibfnamefont {D.}~\bibnamefont
  {Radice}}, \bibinfo {author} {\bibfnamefont {A.}~\bibnamefont {Perego}},
  \bibinfo {author} {\bibfnamefont {F.}~\bibnamefont {Zappa}}, \ and\ \bibinfo
  {author} {\bibfnamefont {S.}~\bibnamefont {Bernuzzi}},\ }\href {\doibase
  10.3847/2041-8213/aaa402} {\bibfield  {journal} {\bibinfo  {journal} {The
  Astrophysical Journal}\ }\textbf {\bibinfo {volume} {852}},\ \bibinfo {pages}
  {L29} (\bibinfo {year} {2018})}\BibitemShut {NoStop}%
\bibitem [{\citenamefont {Oertel}\ \emph {et~al.}(2017)\citenamefont {Oertel},
  \citenamefont {Hempel}, \citenamefont {Kl\"ahn},\ and\ \citenamefont
  {Typel}}]{Oertel2017}%
  \BibitemOpen
  \bibfield  {author} {\bibinfo {author} {\bibfnamefont {M.}~\bibnamefont
  {Oertel}}, \bibinfo {author} {\bibfnamefont {M.}~\bibnamefont {Hempel}},
  \bibinfo {author} {\bibfnamefont {T.}~\bibnamefont {Kl\"ahn}}, \ and\
  \bibinfo {author} {\bibfnamefont {S.}~\bibnamefont {Typel}},\ }\href
  {\doibase 10.1103/RevModPhys.89.015007} {\bibfield  {journal} {\bibinfo
  {journal} {Rev. Mod. Phys.}\ }\textbf {\bibinfo {volume} {89}},\ \bibinfo
  {pages} {015007} (\bibinfo {year} {2017})}\BibitemShut {NoStop}%
\bibitem [{\citenamefont {Carson}\ \emph
  {et~al.}(2019{\natexlab{a}})\citenamefont {Carson}, \citenamefont {Steiner},\
  and\ \citenamefont {Yagi}}]{Zack:nuclearConstraints}%
  \BibitemOpen
  \bibfield  {author} {\bibinfo {author} {\bibfnamefont {Z.}~\bibnamefont
  {Carson}}, \bibinfo {author} {\bibfnamefont {A.~W.}\ \bibnamefont {Steiner}},
  \ and\ \bibinfo {author} {\bibfnamefont {K.}~\bibnamefont {Yagi}},\ }\href
  {\doibase 10.1103/PhysRevD.99.043010} {\bibfield  {journal} {\bibinfo
  {journal} {Phys. Rev.}\ }\textbf {\bibinfo {volume} {D99}},\ \bibinfo {pages}
  {043010} (\bibinfo {year} {2019}{\natexlab{a}})},\ \Eprint
  {http://arxiv.org/abs/1812.08910} {arXiv:1812.08910 [gr-qc]} \BibitemShut
  {NoStop}%
\bibitem [{\citenamefont {Tews}\ \emph {et~al.}(2017)\citenamefont {Tews},
  \citenamefont {Lattimer}, \citenamefont {Ohnishi},\ and\ \citenamefont
  {Kolomeitsev}}]{Tews2017}%
  \BibitemOpen
  \bibfield  {author} {\bibinfo {author} {\bibfnamefont {I.}~\bibnamefont
  {Tews}}, \bibinfo {author} {\bibfnamefont {J.~M.}\ \bibnamefont {Lattimer}},
  \bibinfo {author} {\bibfnamefont {A.}~\bibnamefont {Ohnishi}}, \ and\
  \bibinfo {author} {\bibfnamefont {E.~E.}\ \bibnamefont {Kolomeitsev}},\
  }\href {\doibase 10.3847/1538-4357/aa8db9} {\bibfield  {journal} {\bibinfo
  {journal} {The Astrophysical Journal}\ }\textbf {\bibinfo {volume} {848}},\
  \bibinfo {pages} {105} (\bibinfo {year} {2017})}\BibitemShut {NoStop}%
\bibitem [{aLI()}]{aLIGO}%
  \BibitemOpen
  \href {https://www.advancedligo.mit,.edu/} {\enquote {\bibinfo {title}
  {Advanced {LIGO}},}\ }\bibinfo {howpublished}
  {\url{https://www.advancedligo.mit.edu/}}\BibitemShut {NoStop}%
\bibitem [{Ap_()}]{Ap_Voyager_CE}%
  \BibitemOpen
  \href {https://dcc.ligo.org/ligo-T1400316/public} {\enquote {\bibinfo {title}
  {Ligo-t1400316-v4: Instrument science white paper},}\ }\bibinfo
  {howpublished} {\url{https://dcc.ligo.org/ligo-T1400316/public}}\BibitemShut
  {NoStop}%
\bibitem [{ET()}]{ET}%
  \BibitemOpen
  \href {http://www.et-gw.eu/} {\enquote {\bibinfo {title} {The {ET} project
  website},}\ }\bibinfo {howpublished} {\url{http://www.et-gw.eu/}}\BibitemShut
  {NoStop}%
\bibitem [{\citenamefont {Li}\ \emph {et~al.}(2019)\citenamefont {Li},
  \citenamefont {Krastev}, \citenamefont {Wen},\ and\ \citenamefont
  {Zhang}}]{Li:2019xxz}%
  \BibitemOpen
  \bibfield  {author} {\bibinfo {author} {\bibfnamefont {B.-A.}\ \bibnamefont
  {Li}}, \bibinfo {author} {\bibfnamefont {P.~G.}\ \bibnamefont {Krastev}},
  \bibinfo {author} {\bibfnamefont {D.-H.}\ \bibnamefont {Wen}}, \ and\
  \bibinfo {author} {\bibfnamefont {N.-B.}\ \bibnamefont {Zhang}},\ }\href@noop
  {} {\  (\bibinfo {year} {2019})},\ \Eprint {http://arxiv.org/abs/1905.13175}
  {arXiv:1905.13175 [nucl-th]} \BibitemShut {NoStop}%
\bibitem [{\citenamefont {Hinderer}(2008)}]{hinderer-love}%
  \BibitemOpen
  \bibfield  {author} {\bibinfo {author} {\bibfnamefont {T.}~\bibnamefont
  {Hinderer}},\ }\href {http://stacks.iop.org/0004-637X/677/i=2/a=1216}
  {\bibfield  {journal} {\bibinfo  {journal} {The Astrophysical Journal}\
  }\textbf {\bibinfo {volume} {677}},\ \bibinfo {pages} {1216} (\bibinfo {year}
  {2008})}\BibitemShut {NoStop}%
\bibitem [{\citenamefont {Damour}\ and\ \citenamefont
  {Nagar}(2009)}]{damour-nagar}%
  \BibitemOpen
  \bibfield  {author} {\bibinfo {author} {\bibfnamefont {T.}~\bibnamefont
  {Damour}}\ and\ \bibinfo {author} {\bibfnamefont {A.}~\bibnamefont {Nagar}},\
  }\href {\doibase 10.1103/PhysRevD.80.084035} {\bibfield  {journal} {\bibinfo
  {journal} {Phys.Rev.}\ }\textbf {\bibinfo {volume} {D80}},\ \bibinfo {pages}
  {084035} (\bibinfo {year} {2009})},\ \Eprint {http://arxiv.org/abs/0906.0096}
  {arXiv:0906.0096 [gr-qc]} \BibitemShut {NoStop}%
\bibitem [{\citenamefont {Binnington}\ and\ \citenamefont
  {Poisson}(2009)}]{Binnington:2009bb}%
  \BibitemOpen
  \bibfield  {author} {\bibinfo {author} {\bibfnamefont {T.}~\bibnamefont
  {Binnington}}\ and\ \bibinfo {author} {\bibfnamefont {E.}~\bibnamefont
  {Poisson}},\ }\href {\doibase 10.1103/PhysRevD.80.084018} {\bibfield
  {journal} {\bibinfo  {journal} {Phys. Rev.}\ }\textbf {\bibinfo {volume}
  {D80}},\ \bibinfo {pages} {084018} (\bibinfo {year} {2009})},\ \Eprint
  {http://arxiv.org/abs/0906.1366} {arXiv:0906.1366 [gr-qc]} \BibitemShut
  {NoStop}%
\bibitem [{\citenamefont {Yagi}\ and\ \citenamefont {Yunes}(2013)}]{Yagi2013}%
  \BibitemOpen
  \bibfield  {author} {\bibinfo {author} {\bibfnamefont {K.}~\bibnamefont
  {Yagi}}\ and\ \bibinfo {author} {\bibfnamefont {N.}~\bibnamefont {Yunes}},\
  }\href {\doibase 10.1103/physrevd.88.023009} {\bibfield  {journal} {\bibinfo
  {journal} {Physical Review D}\ }\textbf {\bibinfo {volume} {88}} (\bibinfo
  {year} {2013}),\ 10.1103/physrevd.88.023009}\BibitemShut {NoStop}%
\bibitem [{\citenamefont {Abbott}\ \emph
  {et~al.}(2018{\natexlab{b}})\citenamefont {Abbott} \emph
  {et~al.}}]{LIGO:posterior}%
  \BibitemOpen
  \bibfield  {author} {\bibinfo {author} {\bibfnamefont {B.~P.}\ \bibnamefont
  {Abbott}} \emph {et~al.} (\bibinfo {collaboration} {LIGO Scientific,
  Virgo}),\ }\href {\doibase 10.1103/PhysRevLett.121.161101} {\bibfield
  {journal} {\bibinfo  {journal} {Phys. Rev. Lett.}\ }\textbf {\bibinfo
  {volume} {121}},\ \bibinfo {pages} {161101} (\bibinfo {year}
  {2018}{\natexlab{b}})},\ \Eprint {http://arxiv.org/abs/1805.11581}
  {arXiv:1805.11581 [gr-qc]} \BibitemShut {NoStop}%
\bibitem [{\citenamefont {Chabanat}\ \emph {et~al.}(1997)\citenamefont
  {Chabanat}, \citenamefont {Bonche}, \citenamefont {Haensel}, \citenamefont
  {Meyer},\ and\ \citenamefont {Schaeffer}}]{Chabanat1997}%
  \BibitemOpen
  \bibfield  {author} {\bibinfo {author} {\bibfnamefont {E.}~\bibnamefont
  {Chabanat}}, \bibinfo {author} {\bibfnamefont {P.}~\bibnamefont {Bonche}},
  \bibinfo {author} {\bibfnamefont {P.}~\bibnamefont {Haensel}}, \bibinfo
  {author} {\bibfnamefont {J.}~\bibnamefont {Meyer}}, \ and\ \bibinfo {author}
  {\bibfnamefont {R.}~\bibnamefont {Schaeffer}},\ }\href {\doibase
  10.1016/s0375-9474(97)00596-4} {\bibfield  {journal} {\bibinfo  {journal}
  {Nuclear Physics A}\ }\textbf {\bibinfo {volume} {627}},\ \bibinfo {pages}
  {710} (\bibinfo {year} {1997})}\BibitemShut {NoStop}%
\bibitem [{\citenamefont {Chabanat}(1995)}]{Chabanat1995}%
  \BibitemOpen
  \bibfield  {author} {\bibinfo {author} {\bibfnamefont {E.}~\bibnamefont
  {Chabanat}},\ }\emph {\bibinfo {title} {Interactions effectives pour des
  conditions extremes d'isospin}},\ \href@noop {} {Ph.D. thesis},\ \bibinfo
  {school} {University Claude Bernard Lyon-I} (\bibinfo {year}
  {1995})\BibitemShut {NoStop}%
\bibitem [{\citenamefont {Chabanat}\ \emph {et~al.}(1998)\citenamefont
  {Chabanat}, \citenamefont {Bonche}, \citenamefont {Haensel}, \citenamefont
  {Meyer},\ and\ \citenamefont {Schaeffer}}]{Chabanat1998}%
  \BibitemOpen
  \bibfield  {author} {\bibinfo {author} {\bibfnamefont {E.}~\bibnamefont
  {Chabanat}}, \bibinfo {author} {\bibfnamefont {P.}~\bibnamefont {Bonche}},
  \bibinfo {author} {\bibfnamefont {P.}~\bibnamefont {Haensel}}, \bibinfo
  {author} {\bibfnamefont {J.}~\bibnamefont {Meyer}}, \ and\ \bibinfo {author}
  {\bibfnamefont {R.}~\bibnamefont {Schaeffer}},\ }\href {\doibase
  10.1016/s0375-9474(98)00180-8} {\bibfield  {journal} {\bibinfo  {journal}
  {Nuclear Physics A}\ }\textbf {\bibinfo {volume} {635}},\ \bibinfo {pages}
  {231} (\bibinfo {year} {1998})}\BibitemShut {NoStop}%
\bibitem [{\citenamefont {Reinhard}(1999)}]{Reinhard1999}%
  \BibitemOpen
  \bibfield  {author} {\bibinfo {author} {\bibfnamefont {P.-G.}\ \bibnamefont
  {Reinhard}},\ }\href {\doibase 10.1016/s0375-9474(99)00076-7} {\bibfield
  {journal} {\bibinfo  {journal} {Nuclear Physics A}\ }\textbf {\bibinfo
  {volume} {649}},\ \bibinfo {pages} {305} (\bibinfo {year}
  {1999})}\BibitemShut {NoStop}%
\bibitem [{\citenamefont {Agrawal}\ \emph {et~al.}(2003)\citenamefont
  {Agrawal}, \citenamefont {Shlomo},\ and\ \citenamefont {Au}}]{Agrawal2003}%
  \BibitemOpen
  \bibfield  {author} {\bibinfo {author} {\bibfnamefont {B.~K.}\ \bibnamefont
  {Agrawal}}, \bibinfo {author} {\bibfnamefont {S.}~\bibnamefont {Shlomo}}, \
  and\ \bibinfo {author} {\bibfnamefont {V.~K.}\ \bibnamefont {Au}},\ }\href
  {\doibase 10.1103/physrevc.68.031304} {\bibfield  {journal} {\bibinfo
  {journal} {Physical Review C}\ }\textbf {\bibinfo {volume} {68}} (\bibinfo
  {year} {2003}),\ 10.1103/physrevc.68.031304}\BibitemShut {NoStop}%
\bibitem [{\citenamefont {Goriely}\ \emph {et~al.}(2010)\citenamefont
  {Goriely}, \citenamefont {Chamel},\ and\ \citenamefont
  {Pearson}}]{Goriely2010}%
  \BibitemOpen
  \bibfield  {author} {\bibinfo {author} {\bibfnamefont {S.}~\bibnamefont
  {Goriely}}, \bibinfo {author} {\bibfnamefont {N.}~\bibnamefont {Chamel}}, \
  and\ \bibinfo {author} {\bibfnamefont {J.~M.}\ \bibnamefont {Pearson}},\
  }\href {\doibase 10.1103/physrevc.82.035804} {\bibfield  {journal} {\bibinfo
  {journal} {Physical Review C}\ }\textbf {\bibinfo {volume} {82}} (\bibinfo
  {year} {2010}),\ 10.1103/physrevc.82.035804}\BibitemShut {NoStop}%
\bibitem [{\citenamefont {Goriely}\ \emph {et~al.}(2013)\citenamefont
  {Goriely}, \citenamefont {Chamel},\ and\ \citenamefont
  {Pearson}}]{Goriely2013}%
  \BibitemOpen
  \bibfield  {author} {\bibinfo {author} {\bibfnamefont {S.}~\bibnamefont
  {Goriely}}, \bibinfo {author} {\bibfnamefont {N.}~\bibnamefont {Chamel}}, \
  and\ \bibinfo {author} {\bibfnamefont {J.~M.}\ \bibnamefont {Pearson}},\
  }\href {\doibase 10.1103/physrevc.88.024308} {\bibfield  {journal} {\bibinfo
  {journal} {Physical Review C}\ }\textbf {\bibinfo {volume} {88}} (\bibinfo
  {year} {2013}),\ 10.1103/physrevc.88.024308}\BibitemShut {NoStop}%
\bibitem [{\citenamefont {Dhiman}\ \emph {et~al.}(2007)\citenamefont {Dhiman},
  \citenamefont {Kumar},\ and\ \citenamefont {Agrawal}}]{Dhiman2007}%
  \BibitemOpen
  \bibfield  {author} {\bibinfo {author} {\bibfnamefont {S.~K.}\ \bibnamefont
  {Dhiman}}, \bibinfo {author} {\bibfnamefont {R.}~\bibnamefont {Kumar}}, \
  and\ \bibinfo {author} {\bibfnamefont {B.~K.}\ \bibnamefont {Agrawal}},\
  }\href {\doibase 10.1103/physrevc.76.045801} {\bibfield  {journal} {\bibinfo
  {journal} {Physical Review C}\ }\textbf {\bibinfo {volume} {76}} (\bibinfo
  {year} {2007}),\ 10.1103/physrevc.76.045801}\BibitemShut {NoStop}%
\bibitem [{\citenamefont {Agrawal}(2010)}]{Agrawal2010}%
  \BibitemOpen
  \bibfield  {author} {\bibinfo {author} {\bibfnamefont {B.~K.}\ \bibnamefont
  {Agrawal}},\ }\href {\doibase 10.1103/physrevc.81.034323} {\bibfield
  {journal} {\bibinfo  {journal} {Physical Review C}\ }\textbf {\bibinfo
  {volume} {81}} (\bibinfo {year} {2010}),\
  10.1103/physrevc.81.034323}\BibitemShut {NoStop}%
\bibitem [{\citenamefont {Carriere}\ \emph {et~al.}(2003)\citenamefont
  {Carriere}, \citenamefont {Horowitz},\ and\ \citenamefont
  {Piekarewicz}}]{Carriere2003}%
  \BibitemOpen
  \bibfield  {author} {\bibinfo {author} {\bibfnamefont {J.}~\bibnamefont
  {Carriere}}, \bibinfo {author} {\bibfnamefont {C.~J.}\ \bibnamefont
  {Horowitz}}, \ and\ \bibinfo {author} {\bibfnamefont {J.}~\bibnamefont
  {Piekarewicz}},\ }\href {\doibase 10.1086/376515} {\bibfield  {journal}
  {\bibinfo  {journal} {The Astrophysical Journal}\ }\textbf {\bibinfo {volume}
  {593}},\ \bibinfo {pages} {463} (\bibinfo {year} {2003})}\BibitemShut
  {NoStop}%
\bibitem [{\citenamefont {Typel}\ \emph {et~al.}(2010)\citenamefont {Typel},
  \citenamefont {RÃ¶pke}, \citenamefont {KlÃ€hn}, \citenamefont
  {Blaschke},\ and\ \citenamefont {Wolter}}]{Typel2010}%
  \BibitemOpen
  \bibfield  {author} {\bibinfo {author} {\bibfnamefont {S.}~\bibnamefont
  {Typel}}, \bibinfo {author} {\bibfnamefont {G.}~\bibnamefont {RÃ¶pke}},
  \bibinfo {author} {\bibfnamefont {T.}~\bibnamefont {KlÃ€hn}}, \bibinfo
  {author} {\bibfnamefont {D.}~\bibnamefont {Blaschke}}, \ and\ \bibinfo
  {author} {\bibfnamefont {H.~H.}\ \bibnamefont {Wolter}},\ }\href {\doibase
  10.1103/physrevc.81.015803} {\bibfield  {journal} {\bibinfo  {journal}
  {Physical Review C}\ }\textbf {\bibinfo {volume} {81}} (\bibinfo {year}
  {2010}),\ 10.1103/physrevc.81.015803}\BibitemShut {NoStop}%
\bibitem [{\citenamefont {Gaitanos}\ \emph {et~al.}(2004)\citenamefont
  {Gaitanos}, \citenamefont {Toro}, \citenamefont {Typel}, \citenamefont
  {Baran}, \citenamefont {Fuchs}, \citenamefont {Greco},\ and\ \citenamefont
  {Wolter}}]{Gaitanos2004}%
  \BibitemOpen
  \bibfield  {author} {\bibinfo {author} {\bibfnamefont {T.}~\bibnamefont
  {Gaitanos}}, \bibinfo {author} {\bibfnamefont {M.~D.}\ \bibnamefont {Toro}},
  \bibinfo {author} {\bibfnamefont {S.}~\bibnamefont {Typel}}, \bibinfo
  {author} {\bibfnamefont {V.}~\bibnamefont {Baran}}, \bibinfo {author}
  {\bibfnamefont {C.}~\bibnamefont {Fuchs}}, \bibinfo {author} {\bibfnamefont
  {V.}~\bibnamefont {Greco}}, \ and\ \bibinfo {author} {\bibfnamefont
  {H.}~\bibnamefont {Wolter}},\ }\href {\doibase
  10.1016/j.nuclphysa.2003.12.001} {\bibfield  {journal} {\bibinfo  {journal}
  {Nuclear Physics A}\ }\textbf {\bibinfo {volume} {732}},\ \bibinfo {pages}
  {24} (\bibinfo {year} {2004})}\BibitemShut {NoStop}%
\bibitem [{\citenamefont {Douchin}\ and\ \citenamefont
  {Haensel}(2001)}]{Douchin:2001sv}%
  \BibitemOpen
  \bibfield  {author} {\bibinfo {author} {\bibfnamefont {F.}~\bibnamefont
  {Douchin}}\ and\ \bibinfo {author} {\bibfnamefont {P.}~\bibnamefont
  {Haensel}},\ }\href {\doibase 10.1051/0004-6361:20011402} {\bibfield
  {journal} {\bibinfo  {journal} {Astron. Astrophys.}\ }\textbf {\bibinfo
  {volume} {380}},\ \bibinfo {pages} {151} (\bibinfo {year} {2001})},\ \Eprint
  {http://arxiv.org/abs/astro-ph/0111092} {arXiv:astro-ph/0111092 [astro-ph]}
  \BibitemShut {NoStop}%
\bibitem [{\citenamefont {Carson}\ \emph
  {et~al.}(2019{\natexlab{b}})\citenamefont {Carson}, \citenamefont
  {Chatziioannou}, \citenamefont {Haster}, \citenamefont {Yagi},\ and\
  \citenamefont {Yunes}}]{Zack:URrelations}%
  \BibitemOpen
  \bibfield  {author} {\bibinfo {author} {\bibfnamefont {Z.}~\bibnamefont
  {Carson}}, \bibinfo {author} {\bibfnamefont {K.}~\bibnamefont
  {Chatziioannou}}, \bibinfo {author} {\bibfnamefont {C.-J.}\ \bibnamefont
  {Haster}}, \bibinfo {author} {\bibfnamefont {K.}~\bibnamefont {Yagi}}, \ and\
  \bibinfo {author} {\bibfnamefont {N.}~\bibnamefont {Yunes}},\ }\href
  {\doibase 10.1103/PhysRevD.99.083016} {\bibfield  {journal} {\bibinfo
  {journal} {Phys. Rev.}\ }\textbf {\bibinfo {volume} {D99}},\ \bibinfo {pages}
  {083016} (\bibinfo {year} {2019}{\natexlab{b}})},\ \Eprint
  {http://arxiv.org/abs/1903.03909} {arXiv:1903.03909 [gr-qc]} \BibitemShut
  {NoStop}%
\bibitem [{\citenamefont {Coughlin}\ \emph {et~al.}(2019)\citenamefont
  {Coughlin}, \citenamefont {Dietrich}, \citenamefont {Margalit},\ and\
  \citenamefont {Metzger}}]{Coughlin:2018fis}%
  \BibitemOpen
  \bibfield  {author} {\bibinfo {author} {\bibfnamefont {M.~W.}\ \bibnamefont
  {Coughlin}}, \bibinfo {author} {\bibfnamefont {T.}~\bibnamefont {Dietrich}},
  \bibinfo {author} {\bibfnamefont {B.}~\bibnamefont {Margalit}}, \ and\
  \bibinfo {author} {\bibfnamefont {B.~D.}\ \bibnamefont {Metzger}},\ }\href
  {\doibase 10.1093/mnrasl/slz133} {\bibfield  {journal} {\bibinfo  {journal}
  {Mon. Not. Roy. Astron. Soc.}\ }\textbf {\bibinfo {volume} {489}},\ \bibinfo
  {pages} {L91} (\bibinfo {year} {2019})},\ \Eprint
  {http://arxiv.org/abs/1812.04803} {arXiv:1812.04803 [astro-ph.HE]}
  \BibitemShut {NoStop}%
\bibitem [{\citenamefont {Khan}\ \emph {et~al.}(2016)\citenamefont {Khan},
  \citenamefont {Husa}, \citenamefont {Hannam}, \citenamefont {Ohme},
  \citenamefont {P\"urrer}, \citenamefont {Forteza},\ and\ \citenamefont
  {Boh\'e}}]{PhenomDI}%
  \BibitemOpen
  \bibfield  {author} {\bibinfo {author} {\bibfnamefont {S.}~\bibnamefont
  {Khan}}, \bibinfo {author} {\bibfnamefont {S.}~\bibnamefont {Husa}}, \bibinfo
  {author} {\bibfnamefont {M.}~\bibnamefont {Hannam}}, \bibinfo {author}
  {\bibfnamefont {F.}~\bibnamefont {Ohme}}, \bibinfo {author} {\bibfnamefont
  {M.}~\bibnamefont {P\"urrer}}, \bibinfo {author} {\bibfnamefont {X.~J.}\
  \bibnamefont {Forteza}}, \ and\ \bibinfo {author} {\bibfnamefont
  {A.}~\bibnamefont {Boh\'e}},\ }\href {\doibase 10.1103/PhysRevD.93.044007}
  {\bibfield  {journal} {\bibinfo  {journal} {Phys. Rev. D}\ }\textbf {\bibinfo
  {volume} {93}},\ \bibinfo {pages} {044007} (\bibinfo {year}
  {2016})}\BibitemShut {NoStop}%
\bibitem [{\citenamefont {Husa}\ \emph {et~al.}(2016)\citenamefont {Husa},
  \citenamefont {Khan}, \citenamefont {Hannam}, \citenamefont {P\"urrer},
  \citenamefont {Ohme}, \citenamefont {Forteza},\ and\ \citenamefont
  {Boh\'e}}]{PhenomDII}%
  \BibitemOpen
  \bibfield  {author} {\bibinfo {author} {\bibfnamefont {S.}~\bibnamefont
  {Husa}}, \bibinfo {author} {\bibfnamefont {S.}~\bibnamefont {Khan}}, \bibinfo
  {author} {\bibfnamefont {M.}~\bibnamefont {Hannam}}, \bibinfo {author}
  {\bibfnamefont {M.}~\bibnamefont {P\"urrer}}, \bibinfo {author}
  {\bibfnamefont {F.}~\bibnamefont {Ohme}}, \bibinfo {author} {\bibfnamefont
  {X.~J.}\ \bibnamefont {Forteza}}, \ and\ \bibinfo {author} {\bibfnamefont
  {A.}~\bibnamefont {Boh\'e}},\ }\href {\doibase 10.1103/PhysRevD.93.044006}
  {\bibfield  {journal} {\bibinfo  {journal} {Phys. Rev. D}\ }\textbf {\bibinfo
  {volume} {93}},\ \bibinfo {pages} {044006} (\bibinfo {year}
  {2016})}\BibitemShut {NoStop}%
\bibitem [{\citenamefont {Cutler}\ and\ \citenamefont
  {Flanagan}(1994)}]{Cutler:Fisher}%
  \BibitemOpen
  \bibfield  {author} {\bibinfo {author} {\bibfnamefont {C.}~\bibnamefont
  {Cutler}}\ and\ \bibinfo {author} {\bibfnamefont {E.~E.}\ \bibnamefont
  {Flanagan}},\ }\href {\doibase 10.1103/PhysRevD.49.2658} {\bibfield
  {journal} {\bibinfo  {journal} {Phys. Rev. D}\ }\textbf {\bibinfo {volume}
  {49}},\ \bibinfo {pages} {2658} (\bibinfo {year} {1994})}\BibitemShut
  {NoStop}%
\bibitem [{\citenamefont {Berti}\ \emph {et~al.}(2005)\citenamefont {Berti},
  \citenamefont {Buonanno},\ and\ \citenamefont {Will}}]{Berti:Fisher}%
  \BibitemOpen
  \bibfield  {author} {\bibinfo {author} {\bibfnamefont {E.}~\bibnamefont
  {Berti}}, \bibinfo {author} {\bibfnamefont {A.}~\bibnamefont {Buonanno}}, \
  and\ \bibinfo {author} {\bibfnamefont {C.~M.}\ \bibnamefont {Will}},\ }\href
  {\doibase 10.1103/PhysRevD.71.084025} {\bibfield  {journal} {\bibinfo
  {journal} {Phys. Rev.}\ }\textbf {\bibinfo {volume} {D71}},\ \bibinfo {pages}
  {084025} (\bibinfo {year} {2005})},\ \Eprint
  {http://arxiv.org/abs/gr-qc/0411129} {arXiv:gr-qc/0411129 [gr-qc]}
  \BibitemShut {NoStop}%
\bibitem [{\citenamefont {Poisson}\ and\ \citenamefont
  {Will}(1995)}]{Poisson:Fisher}%
  \BibitemOpen
  \bibfield  {author} {\bibinfo {author} {\bibfnamefont {E.}~\bibnamefont
  {Poisson}}\ and\ \bibinfo {author} {\bibfnamefont {C.~M.}\ \bibnamefont
  {Will}},\ }\href {\doibase 10.1103/PhysRevD.52.848} {\bibfield  {journal}
  {\bibinfo  {journal} {Phys. Rev. D}\ }\textbf {\bibinfo {volume} {52}},\
  \bibinfo {pages} {848} (\bibinfo {year} {1995})}\BibitemShut {NoStop}%
\bibitem [{\citenamefont {Wade}\ \emph {et~al.}(2014)\citenamefont {Wade},
  \citenamefont {Creighton}, \citenamefont {Ochsner}, \citenamefont {Lackey},
  \citenamefont {Farr}, \citenamefont {Littenberg},\ and\ \citenamefont
  {Raymond}}]{Wade:tidalCorrections}%
  \BibitemOpen
  \bibfield  {author} {\bibinfo {author} {\bibfnamefont {L.}~\bibnamefont
  {Wade}}, \bibinfo {author} {\bibfnamefont {J.~D.~E.}\ \bibnamefont
  {Creighton}}, \bibinfo {author} {\bibfnamefont {E.}~\bibnamefont {Ochsner}},
  \bibinfo {author} {\bibfnamefont {B.~D.}\ \bibnamefont {Lackey}}, \bibinfo
  {author} {\bibfnamefont {B.~F.}\ \bibnamefont {Farr}}, \bibinfo {author}
  {\bibfnamefont {T.~B.}\ \bibnamefont {Littenberg}}, \ and\ \bibinfo {author}
  {\bibfnamefont {V.}~\bibnamefont {Raymond}},\ }\href {\doibase
  10.1103/PhysRevD.89.103012} {\bibfield  {journal} {\bibinfo  {journal} {Phys.
  Rev. D}\ }\textbf {\bibinfo {volume} {89}},\ \bibinfo {pages} {103012}
  (\bibinfo {year} {2014})}\BibitemShut {NoStop}%
\bibitem [{\citenamefont {Jensen}(2007)}]{jensen_2007}%
  \BibitemOpen
  \bibfield  {author} {\bibinfo {author} {\bibfnamefont {J.~L.}\ \bibnamefont
  {Jensen}},\ }\href@noop {} {\emph {\bibinfo {title} {Statistics for petroleum
  engineers and geoscientists}}}\ (\bibinfo  {publisher} {Elsevier},\ \bibinfo
  {year} {2007})\BibitemShut {NoStop}%
\bibitem [{\citenamefont {Del~Pozzo}\ \emph {et~al.}(2013)\citenamefont
  {Del~Pozzo}, \citenamefont {Li}, \citenamefont {Agathos}, \citenamefont {Van
  Den~Broeck},\ and\ \citenamefont {Vitale}}]{delPozzo:TaylorTidal}%
  \BibitemOpen
  \bibfield  {author} {\bibinfo {author} {\bibfnamefont {W.}~\bibnamefont
  {Del~Pozzo}}, \bibinfo {author} {\bibfnamefont {T.~G.~F.}\ \bibnamefont
  {Li}}, \bibinfo {author} {\bibfnamefont {M.}~\bibnamefont {Agathos}},
  \bibinfo {author} {\bibfnamefont {C.}~\bibnamefont {Van Den~Broeck}}, \ and\
  \bibinfo {author} {\bibfnamefont {S.}~\bibnamefont {Vitale}},\ }\href
  {\doibase 10.1103/PhysRevLett.111.071101} {\bibfield  {journal} {\bibinfo
  {journal} {Phys. Rev. Lett.}\ }\textbf {\bibinfo {volume} {111}},\ \bibinfo
  {pages} {071101} (\bibinfo {year} {2013})},\ \Eprint
  {http://arxiv.org/abs/1307.8338} {arXiv:1307.8338 [gr-qc]} \BibitemShut
  {NoStop}%
\bibitem [{\citenamefont {Yagi}\ and\ \citenamefont
  {Yunes}(2016)}]{Yagi:binLove}%
  \BibitemOpen
  \bibfield  {author} {\bibinfo {author} {\bibfnamefont {K.}~\bibnamefont
  {Yagi}}\ and\ \bibinfo {author} {\bibfnamefont {N.}~\bibnamefont {Yunes}},\
  }\href {\doibase 10.1088/1361-6382/34/1/015006} {\bibfield  {journal}
  {\bibinfo  {journal} {Classical and Quantum Gravity}\ }\textbf {\bibinfo
  {volume} {34}},\ \bibinfo {pages} {015006} (\bibinfo {year}
  {2016})}\BibitemShut {NoStop}%
\bibitem [{\citenamefont {Montana}\ \emph {et~al.}(2019)\citenamefont
  {Montana}, \citenamefont {Tolos}, \citenamefont {Hanauske},\ and\
  \citenamefont {Rezzolla}}]{Montana:2018bkb}%
  \BibitemOpen
  \bibfield  {author} {\bibinfo {author} {\bibfnamefont {G.}~\bibnamefont
  {Montana}}, \bibinfo {author} {\bibfnamefont {L.}~\bibnamefont {Tolos}},
  \bibinfo {author} {\bibfnamefont {M.}~\bibnamefont {Hanauske}}, \ and\
  \bibinfo {author} {\bibfnamefont {L.}~\bibnamefont {Rezzolla}},\ }\href
  {\doibase 10.1103/PhysRevD.99.103009} {\bibfield  {journal} {\bibinfo
  {journal} {Phys. Rev.}\ }\textbf {\bibinfo {volume} {D99}},\ \bibinfo {pages}
  {103009} (\bibinfo {year} {2019})},\ \Eprint
  {http://arxiv.org/abs/1811.10929} {arXiv:1811.10929 [astro-ph.HE]}
  \BibitemShut {NoStop}%
\bibitem [{\citenamefont {Ozel}\ \emph {et~al.}(2016)\citenamefont {Ozel},
  \citenamefont {Psaltis}, \citenamefont {Arzoumanian}, \citenamefont
  {Morsink},\ and\ \citenamefont {Baubock}}]{NICER:nsradius}%
  \BibitemOpen
  \bibfield  {author} {\bibinfo {author} {\bibfnamefont {F.}~\bibnamefont
  {Ozel}}, \bibinfo {author} {\bibfnamefont {D.}~\bibnamefont {Psaltis}},
  \bibinfo {author} {\bibfnamefont {Z.}~\bibnamefont {Arzoumanian}}, \bibinfo
  {author} {\bibfnamefont {S.}~\bibnamefont {Morsink}}, \ and\ \bibinfo
  {author} {\bibfnamefont {M.}~\bibnamefont {Baubock}},\ }\href {\doibase
  10.3847/0004-637X/832/1/92} {\bibfield  {journal} {\bibinfo  {journal}
  {Astrophys. J.}\ }\textbf {\bibinfo {volume} {832}},\ \bibinfo {pages} {92}
  (\bibinfo {year} {2016})},\ \Eprint {http://arxiv.org/abs/1512.03067}
  {arXiv:1512.03067 [astro-ph.HE]} \BibitemShut {NoStop}%
\bibitem [{\citenamefont {Sieniawska}\ \emph {et~al.}(2018)\citenamefont
  {Sieniawska}, \citenamefont {Bejger},\ and\ \citenamefont
  {Haskell}}]{NICER:nsEoS}%
  \BibitemOpen
  \bibfield  {author} {\bibinfo {author} {\bibfnamefont {M.}~\bibnamefont
  {Sieniawska}}, \bibinfo {author} {\bibfnamefont {M.}~\bibnamefont {Bejger}},
  \ and\ \bibinfo {author} {\bibfnamefont {B.}~\bibnamefont {Haskell}},\ }\href
  {\doibase 10.1051/0004-6361/201833071} {\bibfield  {journal} {\bibinfo
  {journal} {Astron. Astrophys.}\ }\textbf {\bibinfo {volume} {616}},\ \bibinfo
  {pages} {A105} (\bibinfo {year} {2018})},\ \Eprint
  {http://arxiv.org/abs/1803.08813} {arXiv:1803.08813 [astro-ph.HE]}
  \BibitemShut {NoStop}%
\bibitem [{\citenamefont {Schumacher}\ \emph {et~al.}(tion)\citenamefont
  {Schumacher}, \citenamefont {Zimmerman}, \citenamefont {Carson},\ and\
  \citenamefont {Yagi}}]{Kristen}%
  \BibitemOpen
  \bibfield  {author} {\bibinfo {author} {\bibfnamefont {K.}~\bibnamefont
  {Schumacher}}, \bibinfo {author} {\bibfnamefont {J.}~\bibnamefont
  {Zimmerman}}, \bibinfo {author} {\bibfnamefont {Z.}~\bibnamefont {Carson}}, \
  and\ \bibinfo {author} {\bibfnamefont {K.}~\bibnamefont {Yagi}},\ }\href@noop
  {} {} (\bibinfo {year} {in preparation})\BibitemShut {NoStop}%
\bibitem [{\citenamefont {Lattimer}\ and\ \citenamefont
  {Lim}(2013)}]{Lattimer2013}%
  \BibitemOpen
  \bibfield  {author} {\bibinfo {author} {\bibfnamefont {J.~M.}\ \bibnamefont
  {Lattimer}}\ and\ \bibinfo {author} {\bibfnamefont {Y.}~\bibnamefont {Lim}},\
  }\href {\doibase 10.1088/0004-637x/771/1/51} {\bibfield  {journal} {\bibinfo
  {journal} {The Astrophysical Journal}\ }\textbf {\bibinfo {volume} {771}},\
  \bibinfo {pages} {51} (\bibinfo {year} {2013})}\BibitemShut {NoStop}%
\bibitem [{\citenamefont {Schutz}(2011)}]{Shutz:SNR}%
  \BibitemOpen
  \bibfield  {author} {\bibinfo {author} {\bibfnamefont {B.~F.}\ \bibnamefont
  {Schutz}},\ }\href {\doibase 10.1088/0264-9381/28/12/125023} {\bibfield
  {journal} {\bibinfo  {journal} {Classical and Quantum Gravity}\ }\textbf
  {\bibinfo {volume} {28}},\ \bibinfo {pages} {125023} (\bibinfo {year}
  {2011})}\BibitemShut {NoStop}%
\bibitem [{\citenamefont {Chen}\ and\ \citenamefont {Holz}(2014)}]{Chen:SNR}%
  \BibitemOpen
  \bibfield  {author} {\bibinfo {author} {\bibfnamefont {H.-Y.}\ \bibnamefont
  {Chen}}\ and\ \bibinfo {author} {\bibfnamefont {D.~E.}\ \bibnamefont
  {Holz}},\ }\href@noop {} {\  (\bibinfo {year} {2014})},\ \Eprint
  {http://arxiv.org/abs/1409.0522} {arXiv:1409.0522 [gr-qc]} \BibitemShut
  {NoStop}%
\end{thebibliography}%
\end{document}